\documentclass[twocolumn]{aastex63}
\usepackage{times}
\usepackage{amsmath}
\usepackage{textcomp}
\usepackage{amssymb}
\usepackage[flushleft]{threeparttable}
\usepackage{boldline}
\usepackage{tabularx}
\usepackage{lineno}
\usepackage{gensymb}
\usepackage{hyperref}
\usepackage{url}
\usepackage{afterpage}
\usepackage{float}
\usepackage{mathrsfs}
\usepackage{rotating}
\usepackage{longtable}

\newcommand{\Msun}{\ensuremath{M_{\odot}} }

\newcommand{\ergflux}{\mbox{${\rm \, erg \,\, cm^{-2} \, s^{-1}}$}}
\newcommand{\gm}{$\gamma$}

\shorttitle{The evolution of Swift-BAT blazars}
\shortauthors{Marcotulli et al.}

\begin{document}

\title{BASS XXXIII: {\it Swift}-BAT blazars and their jets through cosmic time}

\author[0000-0002-8472-3649]{L. Marcotulli}
\altaffiliation{NHFP Einstein Fellow}
\affil{Yale Center for Astronomy \& Astrophysics, 52 Hillhouse Avenue, New Haven, CT 06511, USA}
\affil{Department of Physics, Yale University, P.O. Box 208120, New Haven, CT 06520, USA}
\affil{Department of Physics and Astronomy, Clemson University, Kinard Lab of Physics, Clemson, SC 29634-0978, USA}
\email{lea.marcotulli@yale.edu}
\author[0000-0002-6584-1703]{M. Ajello}
\affil{Department of Physics and Astronomy, Clemson University, Kinard Lab of Physics, Clemson, SC 29634-0978, USA}
\author[0000-0002-0745-9792]{C. M. Urry}
\affil{Yale Center for Astronomy \& Astrophysics, 52 Hillhouse Avenue, New Haven, CT 06511, USA}
\affil{Department of Physics, Yale University, P.O. Box 208120, New Haven, CT 06520, USA}
\author[0000-0001-7774-5308]{V. S. Paliya}
\affil{Inter-University Centre for Astronomy and Astrophysics (IUCAA), SPPU Campus, 411007, Pune, India}
\author[0000-0002-7998-9581]{M. Koss}
\affil{Eureka Scientific, 2452 Delmer Street, Suite 100, Oakland, CA 94602-3017, USA}
\author[0000-0002-5037-951X]{K. Oh}
\altaffiliation{JSPS Fellow}
\affil{Korea Astronomy \& Space Science institute, 776, Daedeokdae-ro, Yuseong-gu, Daejeon 34055, Republic of Korea}
\affil{Department of Astronomy, Kyoto University, Kitashirakawa-Oiwake-cho, Sakyo-ku, Kyoto 606-8502, Japan}
\author[0000-0002-2114-5626]{G. Madejski}
\affil{Kavli Institute for Particle Astrophysics and Cosmology, SLAC National Accelerator Laboratory, Menlo Park, CA 94025, USA}
\author[0000-0001-7821-6715]{Y. Ueda}
\affiliation{Department of Astronomy, Kyoto University, Kitashirakawa-Oiwake-cho, Sakyo-ku, Kyoto 606-8502, Japan}
\author[0000-0003-0476-6647]{M. Balokovi\'c}
\affil{Yale Center for Astronomy \& Astrophysics, 52 Hillhouse Avenue, New Haven, CT 06511, USA}
\affil{Department of Physics, Yale University, P.O. Box 208120, New Haven, CT 06520, USA}
\author[0000-0002-3683-7297]{B. Trakhtenbrot}
 \affiliation{School of Physics and Astronomy, Tel Aviv University, Tel Aviv 69978, Israel}
\author[0000-0001-5742-5980]{F. Ricci}
\affil{Instituto de Astrof\'isica and Centro de Astroingenier\'ia, Facultad de
F\'isica, Pontificia Universidad Cat\'olica de Chile, Casilla 306,
Santiago 22, Chile}
\author[0000-0001-5231-2645]{C. Ricci}
\affiliation{N\'ucleo de Astronom\'ia de la Facultad de Ingenier\'ia, Universidad Diego Portales, Av. Ej\'ercito Libertador 441, Santiago 22, Chile}
\affiliation{Kavli Institute for Astronomy and Astrophysics, Peking University, Beijing 100871, People's Republic of China}
 \affiliation{George Mason University, Department of Physics \& Astronomy, MS 3F3, 4400 University Drive, Fairfax, VA 22030, USA}
\author[0000-0003-2686-9241]{D. Stern}
\affiliation{Jet Propulsion Laboratory, California Institute of Technology, 4800 Oak Grove Drive, MS 169-224, Pasadena, CA 91109, USA}
\author{F. Harrison}
\affiliation{Cahill Center for Astronomy and Astrophysics, California Institute of Technology, Pasadena, CA 91125, USA}
\author[0000-0003-2284-8603]{M. C. Powell}
\affiliation{Institute of Particle Astrophysics and Cosmology, Stanford University, 452 Lomita Mall, Stanford, CA 94305, USA}

\author{BASS Collaboration}

\begin{abstract}
	We derive the most up-to-date Swift-Burst Alert Telescope (BAT) blazar luminosity function in the $14-195\,\rm keV$ range, 
	making use of a clean sample of 118 blazars detected in the BAT 105-month survey catalog, with newly obtained redshifts from the BAT AGN Spectroscopic Survey (BASS).
We determine the best-fit X-ray luminosity function for the whole blazar population, as well as for Flat Spectrum Radio Quasars (FSRQs) alone.
	The main results are: 
    (1) at any redshift, BAT detects the most luminous blazars, above any possible break in their luminosity distribution, which means we cannot differentiate between density and luminosity evolution;
	(2) the whole blazar population, dominated by FSRQs, 
	evolves positively up to redshift 
	$z\sim4.3$, confirming earlier results and implying lower number densities of
	blazars at higher redshifts than previously estimated.
	The contribution of this source class to the Cosmic X-ray Background at $14-195\,\rm keV$
	can range from 5-18\%, while possibly accounting for 100\% of the MeV background. We also derived 
	the average $14\,\rm keV - 10\,\rm GeV$ SED for BAT blazars, which allows us to  
	predict the number counts of sources 
	in the MeV range, as well as  the expected number of high-energy ($>$100\,TeV) neutrinos. A mission like COSI, will detect 40 MeV blazars and 2 coincident neutrinos. 
	Finally, taking into account beaming selection effects, the distribution and properties 
	of the parent population of these extragalactic jets are derived. We find that the distribution of viewing angles 
	is quite narrow, with most sources aligned within $<5\degree$ of the line of sight. 
Moreover, the average Lorentz factor, $<\Gamma>= 8-12$, is lower than previously suggested for these powerful sources.
\end{abstract}

\section{Introduction}\label{sec:intro}
The extragalactic universe is permeated at every observable wavelength by a rather uniform glow \citep[e.g.,][]{Hauser_2001,Dwek_2013,Gilli_2013,Ackermann_2015,Mozdzen_2017,Ajello_2018,Desai_2019,Planck_2019}. Referred to as ``backgrounds", 
 in some cases they are attributed to the integrated emission of many unresolved sources.
In others they can carry the imprint of truly diffuse emission processes, as well as signatures of the cosmic web structure \citep[e.g.,][]{He_2018}.
At the highest energies, the so-called extragalactic cosmic X- \citep[CXB, $F_{\rm CXB}\sim 10^{-7}\,\rm erg~cm^{-2}~s^{-1}~sr^{-1}$,][]{Ajello_2008,Gilli_2013} and $\gamma$-ray background \citep[EGB, $F_{\rm EGB}\sim10^{-9}\,\rm erg~cm^{-2}~s^{-1}~sr^{-1}$,][]{Ackermann_2015} can 
be accounted for in terms of unresolved sources \citep[e.g.,][]{Churazov_2007,Ajello_2009, Ueda_2014, Aird_2015, Ajello_2015, Capelluti_2017, Tonima_2020,  Marcotulli_2020_b}. 

In particular, in the hard X-ray regime ($\rm E>10\,\rm keV$) the CXB is %thought to be 
dominated by 
supermassive black holes accreting gas at the centers of galaxies.
A fraction ($\sim 10\%$) of these active galactic nuclei (AGNs) powers relativistic jets which, when pointing close to our line of sight, 
are called blazars. 
Canonically, blazars are differentiated into two major sub-classes, the Flat-Spectrum Radio Quasars (FSRQs) and BL Lacertae objects (BL Lacs), %where the observational dividing line is set 
distinguished by the presence or weakness/absence of optical emission lines stronger than $5$\AA~in equivalent width \citep[e.g.,][]{Schmitt_1968, Stein_1976}. 
Fueled by the most massive black holes ($M_{\rm BH}>10^8\Msun$), the jets' peculiar orientation enhances their emission and renders them visible up to very high-redshifts ($z>4$, see e.g., \citealp{Romani_2006,Sbarrato_2015,An_2018, Marcotulli_2020_a, An_2020}). 
If we could understand how the blazar population evolved through cosmic time, this would enable us to trace both the jets \citep[e.g.,][]{Ajello_2012} and supermassive black holes \citep[e.g.,][]{Sbarrato_2015} formation and evolution into the early universe \citep[e.g.,][]{Berti_Volonteri_2008, Volonteri_2010}. 
Indeed, what triggers and powers jet activity, and what is its relation to supermassive black hole accretion are still open questions in astrophysics. The fact that these blazars usually reside in old, already evolved, massive elliptical galaxies \citep[e.g.,][]{Urry_1999, Falomo_2000, Scarpa_2000, Chiaberge_2011, Olguin_2016} has provided some evidence possibly linking major merger events (more frequent in the high-redshift universe) as triggers for jet activity, as well as being a preferred channel for rapid supermassive black hole accretion \citep[see][]{Mayer_2010, Chiaberge_2015, Paliya_2020_a}. 
Moreover, it has been proposed that these jets extract energy tapping onto the central highly spinning black hole \citep{B_Z_1977,  Maraschi_2012, Ghisellini_2014, Schultze_2017}. 
Therefore following blazar jets through cosmic times could also shine a light onto the evolution of black hole spins in the universe.

The hard X-ray window (from $>10\rm\,keV$ up to $\sim500\rm\,keV$) is key to study these powerful sources. In fact, blazar spectral energy distribution (SED) at these
energies is dominated by the jet emission and is successfully explained by inverse Compton (IC) scattering of the relativistic jet electrons off a low-energy photon field.  
In orbit for more than sixteen years, the Burst Alert Telescope (BAT, $14-195\,\rm keV$, \citealp{Swift_BAT}) onboard the Neil Gehrels Swift Observatory \citep{Swift_2004} provides the best uniform all-sky survey of the brightest hard X-ray emitters in the universe. 
The most recent catalog, the BAT 105-month catalog \citep[hereafter BAT 105,][]{BAT_105}, contains more than one thousand sources detected at fluxes $\gtrsim10^{-12}\ergflux$ over the full $14-195\,\rm keV$ energy range, with blazars making up about 10\% of the total.  
Almost tripling the number of blazars detected in previous catalogs \citep[see][]{BAT_70}, the BAT 105 allows for better statistics and constraints on cosmic evolution studies of this class of sources.
Previous works \citep{Ajello_2009,Toda_2020} have shown that blazars, and in particular the FSRQ subclass, evolve positively in this energy range (i.e., sources are more numerous and/or more luminous at earlier times). 
Nonetheless, due to the limited sample size, which kind of evolution these sources follow and at what redshift the peak in space density occurs are still a matter of debate. 

In this work, we 
construct the most up-to-date BAT-blazar luminosity function employing the large blazar sample from the BAT 105 catalog. In Section~\ref{sec:sample} we describe the sample selection, its associated incompleteness and the sky coverage of the instrument. Section~\ref{sec:lf} details the mathematical description and method used to derive the best-fit XLF, and Section~\ref{sec:res} highlights the main results. In particular, Section~\ref{sec:lf_all} describes the results for the total blazar sample, while Section~\ref{sec:lf_fsrq} focuses on the FSRQ subclass, which dominates the sample and comprises of the intrinsically more luminous and higher redshift sources.
In order to derive blazars contribution to the high-energy cosmic backgrounds, in Section~\ref{sec:ave_sed} we derive the average BAT-blazar SED from $14\,\rm keV$ to $10\,\rm GeV$ using BAT and LAT data. Knowledge of both the luminosity function and average SED enables us to derive their contribution to the CXB in the hard X-ray regime ($14-500\,\rm keV$, Section~\ref{sec:bkg}) as well make predictions for the MeV background ($0.5-30\rm\,MeV$) in light of planned MeV missions like COSI \citep{Tomsick_2019} or potentially forthcoming ones like ASTROGAM. Finally, the blazar luminosity function allows us to examine the properties of 
 the parent population of jetted AGNs (Section~\ref{sec:jets}). 
The main results are highlighted and discussed in Section~\ref{sec:disc}.
Throughout, the following cosmological parameters are adopted: $H_0=67.4\,\rm km~s^{-1}~Mpc^{-1}$ and $\Omega_{M}=1-\Omega_{\Lambda}=0.315$ \citep{Plank_2020}.

\section{The Clean Sample} \label{sec:sample}
The first task in deriving any luminosity function is to construct a clean sample.
 Indeed, studying the evolution of one particular source class in luminosity entails having both flux and exact redshift measurements for all objects belonging to the desired population.
Originally constructed to scan the sky with the intent of detecting $\gamma$-ray bursts (GRBs), the BAT instrument covers every day about 80\% of the whole celestial sphere in the $14-195\,\rm keV$ range, and thus provides a complete all-sky hard X-ray survey. 
The latest BAT catalog is derived using 105 months of data and contains 1635 sources \citep{BAT_105}.

In an extensive work focused on the properties of BAT-blazars, \citet{Paliya_2019} carefully identified 146 blazars from the full BAT 105 catalog. Moreover, thanks to ongoing meticulous follow-up efforts in the BAT AGN Spectroscopic Survey (BASS) collaboration\footnote{\url{http://www.bass-survey.com/}}, using the VLT in the south and Palomar in the north, more sources have reported spectroscopic redshifts and updated counterpart associations
\citep{Kossa_2022, Kossb_2022}.  At present, 
the latest BAT 105 contains 160 known beamed AGN (i.e.~blazars). Moreover,
99.8\% (857/858) of counterparts for the BAT 70-month catalog have spectroscopic redshifts, 
as do the majority (1183/1635) of the counterparts in the 105-month catalog used here.
The only beamed AGN without a known redshift is B3 0133+388.
This source shows faint Ca H+K lines at redshift zero in two different Palomar spectra (and also in a Keck/LRIS spectrum shown in \citealp{Aliu_2012}). However, given the radio and Fermi-Large Area Telescope (LAT) detection, the source is unlikely to be Galactic, but may be a distant blazar close in projection to a foreground star. Following the BASS optical spectroscopic classification \citep[see][]{Koss_2017_b}, source types identified as beamed AGN are divided intro four main types: BZQ (i.e., blazars hosting broad Balmer lines in optical spectroscopy, also known as FSRQs); BZB (i.e., continuum-dominated blazars, also known as BL Lacs); BZG (i.e., continuum-dominated blazars with clearly visible 
galaxy emission); and BZU (sources of uncertain type).
Here, we use 
the most up-to-date BAT 105 catalog with redshifts and associated counterparts provided by the BASS collaboration to construct our blazar sample.

\subsection{Incompleteness}\label{sec:cut}
To work with the cleanest possible sample, we included only sources with:
\begin{enumerate}
\item 
BAT detection significance above the $5\sigma$ threshold;
\item 
Galactic latitude at $|b|>10\degree$;
\item 
Time-averaged BAT flux greater than 
$F_{\rm 14-195\,keV}>5.4\times10^{-12}\ergflux$, the minimum flux of the sky coverage (see below, Section~\ref{sec:eff} and Figure~\ref{fig:sky_cov}). 
\end{enumerate}
These criteria were chosen in order to minimize source confusion or uncertainty arising from Galactic and sub-threshold sources, as well as to be consistent with respect to the sky coverage calculation.
Applied to the entire BAT catalog, these cuts return 1069 sources (65\% of the total), of which 118 are classified as `beamed AGN'.  We define the incompleteness of our sample as the fraction of objects with respect to the total that lack of classification (classified as \texttt{U1},\texttt{U2}, \texttt{U3} or \texttt{Unk AGN} in the BAT 105). This results in 15 sources, accounting for an incompleteness of $\sim 1\%$.

\subsection{Sky Coverage}\label{sec:eff}
\begin{figure} 
\centering
        \includegraphics[width=\columnwidth]{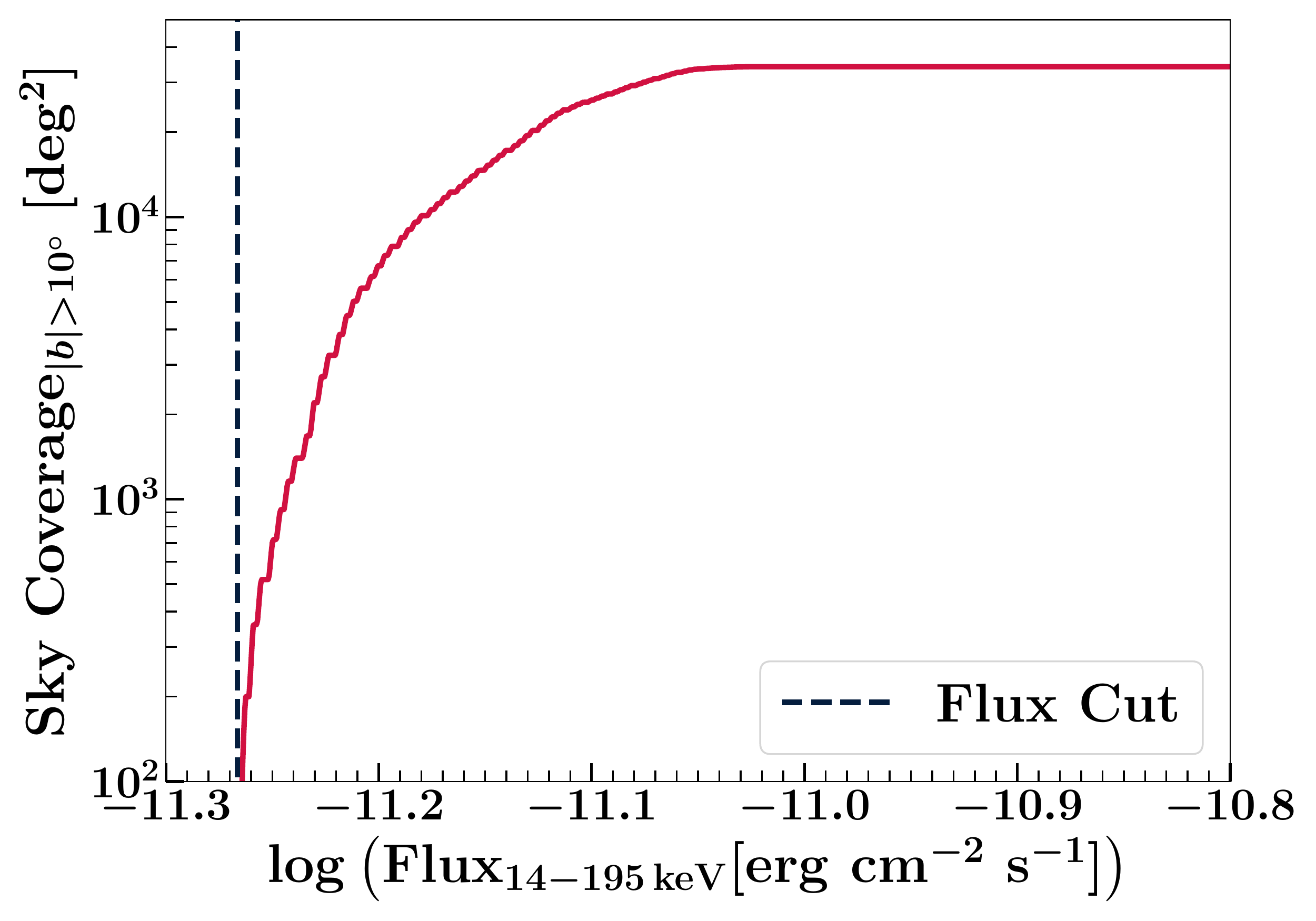}
        \caption{The Swift-BAT sky coverage for $|b|>10\degree$ and $5\sigma$ detection threshold, derived from \citet[][see Section~\ref{sec:eff}]{BAT_105}, rises from 300~deg$^2$ at the minimum sky coverage flux ({\it blue dashed line}) to 34,089~deg$^2$ above $F_{\rm 14-195\,keV}>10^{-11}\ergflux$.
         \label{fig:sky_cov}}
\end{figure} 
Despite the BAT survey averaging over 9 years of observations, the sky coverage of the survey (i.e., how much time the instrument has looked at a particular position in the sky) is not perfectly uniform \citep[see][]{BAT_105}. Hence the efficiency of the BAT instrument (i.e., the probability of detecting sources as a function of 
flux) changes depending on the chosen significance detection threshold. 
It is therefore paramount to know the sky coverage, $\omega(F)\equiv\omega(L,z)$, of the instrument, i.e., how the flux limit changes as a function of the surveyed sky solid angle ($\Omega$), given a significance threshold.
This function is reported in \citet{BAT_105} for the whole sky for a significance threshold of $5\sigma$.
In this work, in agreement with the chosen Galactic latitude and significance cuts 
(\S\,\ref{sec:cut}), 
we recalculate this function for $|b|>10\degree$ at $5\sigma$ threshold in the same way as described in \citet[][see Figure~\ref{fig:sky_cov}]{BAT_105}.
Sources with BAT flux $F_{\rm 14-195\,keV}>10^{-11}\ergflux$ are detected everywhere  at $|b|>10\degree$ ($\Omega_{\rm max, |b|>10\degree} \sim 34,089$~deg$^2$), while the available area decreases gradually toward lower fluxes, reaching $\sim 1\%$ of the surveyed area ($\Omega_{\rm min, |b|>10\degree}=300\,\rm deg^2$) at the limiting flux, $F_{\rm 14-195\,keV}\sim 5.4 \times 10^{-12}\ergflux$.

\begin{figure*} 
\centering
        \includegraphics[width=\columnwidth]{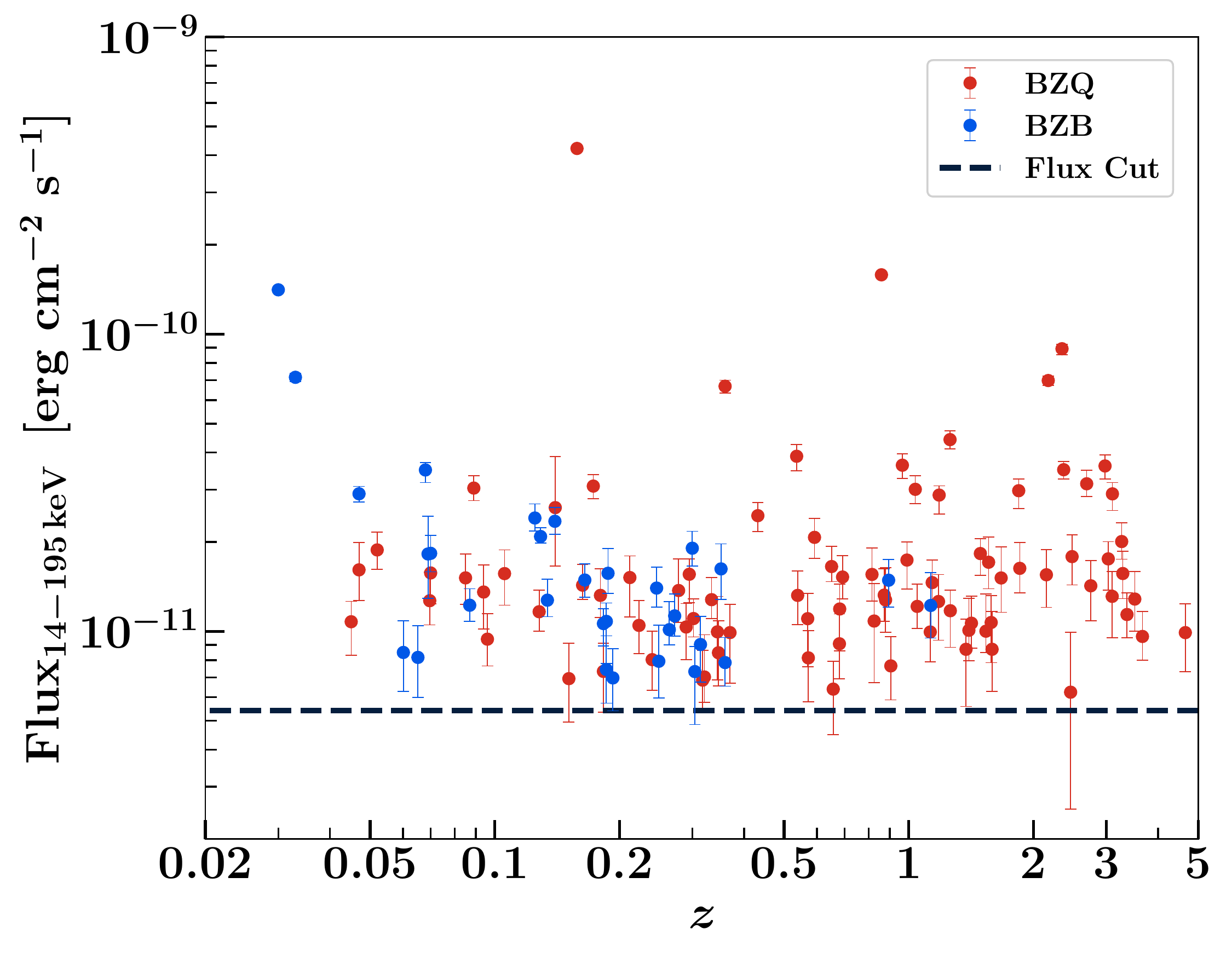}
        \includegraphics[width=\columnwidth]{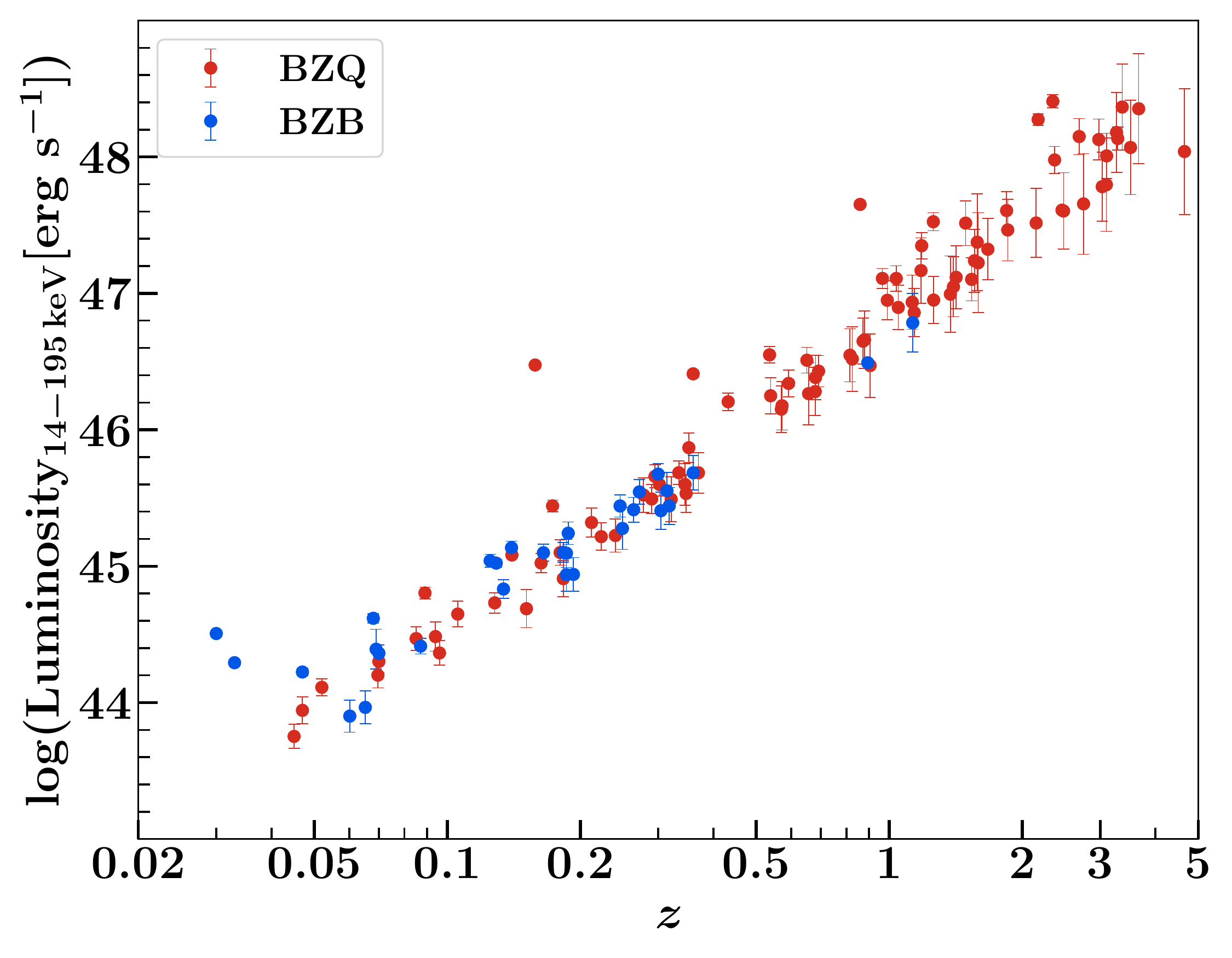}
        \caption{{\bf Left}: Hard X-ray ($14-195\,\rm keV$) flux versus redshift for our clean sample of 118 blazars (see Section~\ref{sec:eff}). The data points represent all identified blazars %the sources in our sample 
        at $|b|>10\degree$ detected above $5\sigma$ with $F_{\rm 14-195\,keV}>5.4 \times 10^{-12}\ergflux$ ({\it dashed line}) from the BAT 105 catalog: 
        ({\it red filled circles}) BZQs, ({\it blue filled circles}) BZBs, including 3 BZGs and 1 BZU. 
         {\bf Right}: K-corrected hard X-ray ($14-195\,\rm keV$) luminosity as function of redshift for our clean sample.
         \label{fig:blazar_sample}}
\end{figure*} 

\subsection{Constructing the Blazar Sample}
To avoid biases in the source selection, instead of relying directly on the BAT 105 classification,  we performed a standard positional cross match between 
the list of 1069 sources (with counterpart positions taken from the BASS DR2 multi-wavelength catalog\footnote{DR2 reference papers (XXI-XXX) are listed here: \url{https://www.bass-survey.com/publications.html}}) and already existing blazar catalogs. 
These consist of the Roma-BZCAT \citep{BZCAT}, the Combined Radio All-Sky Targeted Eight-GHz Survey \citep[CRATES,][]{CRATES}, the Candidate Gamma-Ray Blazar Survey Source Catalog \citep[CGRaBS,][]{CGRABS}, the WISE Blazar-like Radio-Loud Source catalog \citep[WIBRaLS,][]{WIBRaLS} and the Fermi-LAT Fourth Source Catalog \citep[4FGL,][]{4FGL}. Moreover, we checked the MOJAVE catalog \citep{MOJAVE} and a radio galaxy catalog \citep{Yuan_2012} for
sources with small reported jet viewing angles, $\theta<5\degree$, which can therefore be classified as blazars. 
The positional cross match was done using the \texttt{Sky with errors} algorithm in \texttt{TOPCAT}\footnote{\url{http://www.star.bris.ac.uk/~mbt/topcat}};  
source coordinates and positional errors were taken from the respective catalogs. 
We consider blazars sources that have positional crossmatches with one or several of 
the above mentioned catalogs (i.e.\ within the 1$\sigma$ positional uncertainty, there was overlap between the catalog counterpart and the BAT source position).
Our clean blazar sample contains 118 sources ($\sim 75\%$ of the total BAT 105 beamed AGN  sample), 114 classified as BZQ, 33 as BZB, 10 as BZG, and 2 as BZU \citep{Koss_2017_b}.
We emphasize that these are the same 118 `beamed AGN' found by applying Section~\ref{sec:eff} cuts to the total BAT-105 sample.
The main properties of the clean sample are listed in Table~\ref{tab:sample}.  Figure~\ref{fig:blazar_sample} shows the distribution of sources flux (left panel) and K-corrected $14-195\,\rm keV$ luminosity (right panel) as a function of redshift.

To evaluate the incompleteness of the sample, we employ the zeroth-order assumption that uncertain sources (15 out of 1069, see Section~\ref{sec:cut}) are distributed in type as the associated ones in the catalog. Since blazars represent 
10\% of the associated sources, we expect a 10\% of the unassociated ones to be blazars, adding only one or two extra objects to our list. This incompleteness is completely negligible and of no impact for our results.

\begingroup
\renewcommand*{\arraystretch}{1.2}
\begin{table*}[t!]
\centering
	\caption{Mean properties of clean blazar sample.\label{tab:sample}}
\begin{tabular}{ l | c  c c c c c }
       & Number$^{a}$ & $<\Gamma_{\rm 14-195\, keV}>^{b}$ &  $<F_{\rm 14-195\, keV}>^{b}$ & $<L_{\rm 14-195\, keV}>^{b}$ & $z_{\rm min}^{c}$ & $z_{\rm max}^{c}$ \\
       &  &   &  $[$\ergflux$]$ & $[\rm erg~s^{-1}]$ &  &  \\
\hline
 Total & 118 &  $1.94\pm0.47$  & $2.31\times10^{-11}$ & $1.22\times10^{46}$ & 0.03 & 4.65 \\
 BZQ   & 88  &  $1.82\pm0.38$  & $2.40\times10^{-11}$ & $2.75\times10^{46}$ & 0.04 & 4.65  \\
 BZB$^{d}$   & 30  &  $2.30\pm0.51$  & $2.06\times10^{-11}$ & $1.03\times10^{45}$ & 0.03 & 1.13 \\
 \hline 
 \multicolumn{7}{l}{\begin{minipage}{0.8\textwidth}
 \tablenotetext{a}{Number of sources in the sample.}
 \tablenotetext{b}{BAT 105 average spectral properties: spectral index ($\Gamma_{\rm 14-195\,keV}$); flux ($F_{\rm 14-195\,keV}$); luminosity($L_{\rm 14-195\,keV}$).}
 \tablenotetext{c}{Redshift statistics: minimum and maximum redshift.}
 \tablenotetext{d}{Included in the BZB classification are 4 BZG and 1 BZU.}
\end{minipage}%
}
\end{tabular}
\end{table*}
\endgroup

\section{The Luminosity Function}\label{sec:lf}
The luminosity function of a particular source class is defined as the number of objects per unit comoving volume ($dV$) and luminosity interval ($dL$). It can be written in its differential form as:
\begin{equation}
\phi(L,V(z)) = \frac{d^2 N}{dV dL}(L,V(z)).
\end{equation}
The above can be interpreted as a function of redshift ($z$), adopting the transformation of comoving volume per unit redshift and solid angle ($dV/dzd\Omega$, see \citealp{Hogg_1999}), as follows:
\begin{equation}\label{eq:pl}
\phi(L,z) = \frac{d^2 N}{dV dL}(L,V(z))\times\frac{dV}{dzd\Omega}.
\end{equation}
Throughout this work, the luminosity labeled as $L$ indicates the $14-195\,\rm keV$ X-ray luminosity ($L\equiv L_{\rm X}$) derived using the flux and redshift information in the BAT 105 catalog\footnote{In blazars there is very little obscuration, therefore 
$F_{\rm 14-195\rm keV, obs}=F_{\rm 14-195\rm keV, intr}$.}. 
To calculate its evolution, it is custom to factorize $\phi(L,V(z))$ into a local luminosity function, $\phi(L,V(z=0))$, accompanied by an evolutionary factor, $e(z)$. 
For the purpose of our analysis, we adopt the notations and conventions detailed in \citet[][hereafter A09]{Ajello_2009}, \citet[][hereafter A12]{Ajello_2012} and \citet[][hereafter A14]{Ajello_2014}, which are here summarized. 

The simplest parametrization for $\phi(L,V(z=0))$ is a straightforward power law
\begin{equation}\label{eq:pl_dn}
	\begin{split}
		&\phi(L,V(z=0)) = \frac{dN}{dL} = \frac{dN}{d\log L}\frac{d\log L}{dL}\\
		&               = \frac{A}{\ln(10)L} \left(\frac{L}{L_*}\right)^{-\gamma}
			        = \frac{A}{\ln(10)L_*} \left(\frac{L}{L_*}\right)^{-\gamma'},\\
        \end{split}
\end{equation}
where $L_*$ is a constant luminosity scale, fixed here to $10^{44}\,\rm erg~s^{-1}$, $A$ is the normalization factor and $\gamma$ the power-law index
of the $dN/d\log L$, while $\gamma'=\gamma+1$ is the power-law index of the $dN/dL$. 
Another envisaged scenario
is a distribution described by a break occurring at some luminosity $L_*$, and it can be represented by a smoothly-joint broken power law of the form 
\begin{equation}\label{eq:bpl_dn}
\begin{split}
	\phi(L,V(z=0))& = \frac{A}{\ln(10)L} \left[\left(\frac{L}{L_*}\right)^{\gamma_1}+\left(\frac{L}{L_*}\right)^{\gamma_2}\right]^{-1}=\\
	& = \frac{A}{\ln(10)L_*} \left[\left(\frac{L}{L_*}\right)^{\gamma_1'}+\left(\frac{L}{L_*}\right)^{\gamma_2'}\right]^{-1},
\end{split}
\end{equation}
where the $\gamma_1$ ($\gamma_1'$) and $\gamma_2$ ($\gamma_2'$) are respectively
the low-end and high-end luminosity power-law indices of the $dN/d\log L$ ($dN/dL$), and $A$ the normalization.

As for the evolutionary properties of the blazar population, usually three scenarios are proposed: a pure luminosity evolution (PLE, i.e.,  
sources are more/less luminous in the past, while their number density remains roughly constant), a pure density evolution (PDE, i.e., 
the number density of sources increases/decreases with redshift, but their typical luminosity remains constant) or a mixed luminosity-dependent density evolution (LDDE, i.e.,  sources density changes as a function of luminosity, which also
varies as redshift increases). In the three different scenarios the evolved luminosity function translates as follows:
\begin{enumerate}
\item PLE:
\begin{equation}
 \phi(L,V(z))=\phi(L/e(z),V(z=0)),
\end{equation}
\item PDE:
\begin{equation}
 \phi(L,V(z))=\phi(L,V(z=0))\times e(z),
\end{equation}
\item LDDE:
\begin{equation}
 \phi(L,V(z))=\phi(L,V(z=0))\times e'(z),
\end{equation}
\end{enumerate}
where the evolutionary factors are: 
\begin{equation}\label{eq:evol_fact}
e(z)=(1+z)^{k}{\rm e}^{z/\xi},
\end{equation}
with $z$ being the redshift, $k$ the redshift index and $\xi$ the evolutionary cut-off term; and 
\begin{equation}\label{eq:evol_fact_ldde}
e'(z)=\left[\left(\frac{1+z}{1+z_C(L)}\right)^{-p1}+\left(\frac{1+z}{1+z_C(L)}\right)^{-p2}\right]^{-1}
\end{equation}
where $z_C(L) = z_C(L/10^{48.6})^{\rho}$, $z_C$ being the characteristic redshift; $p1$, $p2$ and $\rho$ are the redshift indeces.

In this work, we test all three different evolution scenarios. For the PDE and PLE we evaluate both the simple power-law and smooth broken power-law (Equation~\ref{eq:pl_dn} and \ref{eq:bpl_dn}) shapes for the local luminosity function, $\phi(L,V(z=0))$. For clarity, the former is referred to as {\it simple} PDE/PLE (sPDE/sPLE), the latter as {\it modified} PDE/PLE (mPDE/mPLE) in the rest of the paper. For the LDDE case we only test the smooth broken power-law $\phi(L,V(z=0))$ (Equation~\ref{eq:bpl_dn}). 

\subsection{Maximum-likelihood fit}\label{sec:ml}

In order to determine the best fit X-ray luminosity function (XLF) we follow the maximum likelihood (ML) method originally put forward by \citet{Marshall_1983}.
The likelihood function is taken in its normalization-free form from \citet{Narumoto_Totani_2006} and can be written as
\begin{equation}
\mathcal{L} = \prod_{i=0}^{N_{\rm obs}} \frac{1}{N_{\rm exp}}\lambda(L_i,z_i).
\end{equation}
In the above, the product covers up to $N_{\rm obs}$, the total number of sources in the sample; $L_i$ and $z_i$ are the luminosity and redshift of the $i^{th}$ source and $\lambda(L,z)$ is defined as
\begin{equation}
\lambda(L,z)\equiv\phi(L,z)\omega(L, z), 
\end{equation}
where $\phi(L,z)$ is one of the luminosity function representations highlighted in the previous Section, and $\omega(L,z)$ is the sky coverage at a specific flux. Finally, $N_{\rm exp}$ is the expected number of observed sources for a particular $\phi(L,z)$ and is evaluated as:
\begin{equation}
	N_{\rm exp} = \int_{z_{\rm min}}^{z_{\rm max}}\int_{L_{\rm min}}^{L_{\rm max}} \lambda(L,z) \frac{dV}{dzd\Omega} dL dz,
\end{equation}
where the integrals limits are set to: $L_{\rm min}=10^{43}\,\rm erg~s^{-1}$, $L_{\rm max}=10^{50}\,\rm erg~s^{-1}$, $z_{\rm min}=0$, and $z_{\rm max}=6$. 
The minimum luminosity is chosen to be an order of magnitude lower than the minimum observed luminosity ($L_{\rm min, obs}\sim10^{44}\,\rm erg~s^{-1}$). The 
maximum values of redshift ($z_{\rm max}$) and luminosity ($L_{\rm max}$) do not influence the 
fit results, hence they are arbitrarily set to the ones reported above. 
The standard $C=-2\ln(\mathcal{L})$ is then calculated as:
\begin{equation}\label{eq:log-like}
C = -2\left[\left(\sum_{i=0}^{N_{\rm obs}}\ln\lambda(L_i, z_i)\right) - N_{\rm obs}\ln\left(N_{\rm exp}\right)\right].
\end{equation}
The free parameters in each representation of $\phi(L,z)$ are varied until the minimum value of $C$ is achieved, i.e., $\Delta C=1$ 
\citep[under the limit for which $\mathcal{L}\propto\exp(-\chi^2/2)$, with $\chi^2$ following a chi-square distribution, see e.g.,][]{Loredo_1989}. 
When the minimum is reached, the best-fit parameters and their associated $1\sigma$ errors are extracted (see Table~\ref{tab:res_ml}). For the task we use the \texttt{pyROOT} implementation of \texttt{Minuit}\footnote{\url{https://root.cern.ch/doc/master/classTMinuit.html}}. 
The Akaike Information Criteria (AIC, \citealp{akaike_1974}) is then employed to compare the different values of $C$ and determine the best-fit XLF model, ascribed to the lowest AIC value. 
The ML and AIC values are reported next to every tested model in Table~\ref{tab:res_ml} and we discuss the results in Section~\ref{sec:res}.

\begingroup
\renewcommand*{\arraystretch}{1.2}
\setlength{\tabcolsep}{4pt} %

\begin{sidewaystable*}[]
\vspace{+10cm}
\caption{Result of the maximum likelihood fit}
\scalebox{0.88}{
\centering
\hspace{-4cm}
\begin{tabular}{ l | l | c c c c c c c c | c c c c c}
	SAMPLE & LF & \multicolumn{8}{c}{Parameters} & $C$ & AIC &  KS$_{z}$ &  KS$_{L}$ & CXB \%\\
\hline
	Total &     &A$^{a}$ & & &  $\gamma$ & k & $\xi$ & & & & & & & \\
	& sPLE & $1.14\pm0.10$ & & & $1.32\pm0.08$ & $1.87\pm0.59$ & $-2.98\pm1.58$ & & & 1492.76 & 1500.76 & 0.10 & 0.31 & 2.89\% \\
	&     & A$^{a}$ & & & $\gamma$ & k & $\xi$  & & & & & & & \\ 
	& sPDE & $1.57\pm0.10$ & & &$1.31\pm0.08$ & $4.13\pm1.49$ & $ -1.28\pm0.74$ & & & 1492.76 & 1500.76 & 0.10 & 0.31 & 2.89\% \\
	&      & A$^{a}$ & $L_*^{b}$ & $\gamma_1$ & $\gamma_2$ & k & $\xi$ & & & & & & & \\ 
	& mPLE &$1.69\pm0.15$ &$1.47\pm0.93$ & $-0.51\pm1.15$ & $1.79\pm0.19$ & $2.79\pm0.53$ & $-2.23\pm0.81$ & & & 1465.43 &  1477.43 & 0.92 & 0.43 & 4.21\%\\ 
	&      & A$^{a}$ & $L_*^{b}$ & $\gamma_1$ & $\gamma_2$ & k & $\xi$ & & & & & & & \\
	& mPDE & $1.48\pm0.13$ &$ 1.50\pm0.82$ & $ -0.70\pm1.25$ & $1.76\pm0.17$ & $7.74\pm1.93$ & $-0.78\pm0.31$ & & & 1465.43 & 1477.85 & 0.96 & 0.44 & 19.58\%\\
	&      &A$^{a}$ & $L_*^{b}$  & $\gamma_1$ & $\gamma_2$ & $p_1$ & $p_2$ & $z_C$ & $\rho$ & & & & &\\ 
	& LDDE & $(1.92\pm0.17)\times10^{-9}$ & $ 997.7\pm733.8$ & $0.79\pm0.05$ & $1.14\pm0.19$ & $4.75\pm0.94$ & $-1.44\pm3.64$ & $6.00\pm3.82$ & $0.27\pm0.04$ & 1480.06 & 1496.06 & 0.73 & 0.39 & 0.08\%\\
\hline       
\hline
	BZQ &      & A$^{a}$ & &  & $\gamma$ & k & $\xi$ & & & & & & &\\ 
	& sPLE & $0.52\pm0.05$ & &  &$1.36\pm0.10$ & $2.70\pm0.61$ & $-1.76\pm0.62$ & & & 1293.84 & 1301.84 & 0.29 & 0.34 & 3.67\% \\
	&      & A$^{a}$ & &  & $\gamma$ & k & $\xi$ & & & & & & &\\ 
	& sPDE & $0.52\pm0.05$ & & &$1.35\pm0.10$ & $6.38\pm1.64$ & $-0.74\pm0.28$ & & & 1293.84 & 1301.84 & 0.29 & 0.34 & 3.65\% \\
	&      &A$^{a}$ & $L_*^{b}$ & $\gamma_1$ & $\gamma_2$ & k & $\xi$ & & & & & & &\\ 
	&mPLE  & $ 1.13\pm0.12$ &$1.01\pm0.66$ & $-1.00\pm2.07$ & $1.67\pm0.17$ & $3.23\pm0.57$ & ${-1.62\pm0.47}$ & & & 1281.87 & 1293.87 & 1.00 & 0.45& 2.55\% \\
	&      & A$^{a}$ & $L_*^{b}$ & $\gamma_1$ & $\gamma_2$ & k & $\xi$ & & & & & & &\\ 
	&mPDE  & $ 0.89\pm0.09$ &$1.15\pm0.83$ & $-0.88\pm2.02$ & $1.66\pm0.17$ & $8.60\pm2.00$ & $ -0.60\pm0.09$ &  & & 1281.94 & 1293.94 & 1.00 & 0.45& 10.35\%\\
	&      & A$^{a}$ & $L_*^{b}$  & $\gamma_1$ & $\gamma_2$ & $p_1$ & $p_2$ & $z_C$ & $\rho$ & & & & &\\
	&LDDE  & $(7.7\pm0.82)\times10^{-5}$ &$1.55\pm1.93$ & $-3.48\pm 1.98$ & $1.55\pm0.19$ & $ 4.54\pm0.99$ & $-9.43\pm11.03$ & $3.61\pm0.36$ & $(2.356.08\times10^{-8})\pm0.03$ &  1281.10 & 1297.10 & 0.99 & 0.44 & 4.38\% \\ 
\multicolumn{15}{l}{
  \begin{minipage}{\textwidth}
 \tablenotetext{a}{Normalization constant in units of $\rm Mpc^{-3}$}
 \tablenotetext{b}{Luminosity scale factor in units of $\rm 10^{44}\,\rm erg~s^{-1}$}
\end{minipage}
}

\end{tabular}
}\label{tab:res_ml}

\end{sidewaystable*}
\endgroup

\subsection{Consistency tests}\label{sec:test}
To test the consistency of our results, we further performs two checks:
\begin{enumerate}
	\item The Kolmogorov-Smirnov (KS) test. Given a empirical distribution function and a cumulative one (derived from the representative model), the test returns
		the probability that the data and the model are drawn from the same distribution. If the probability is too low (the threshold is here 
		set to KS$<30\%$), the model can be disregarded. This is applied to both the redshift (KS$_z$) and luminosity (KS$_L$) 
		distributions of our blazar sample. 
	\item The source count distribution ({\it logN-logS}). For any luminosity function, 
		the number of expected sources as a function of flux ($S$) can be computed as:
		\begin{equation}
			N(>S) = 4\pi\int_{0}^{z_{\rm max}}\int_{L(z,S)}^{\infty} \phi(L,z) \frac{dV}{dzd\Omega} dL dz.
			\label{eq:lognlogs}
		\end{equation}
		The prediction is then compared to the observed {\it logN-logS}
		\begin{equation}
			N(>S) = \sum_{i=0}^{N_{\rm obs}} \frac{1}{\omega(S_i)} \quad S_i>S,
			\label{eq:lognlogs_obs}
		\end{equation}
		where the above sum covers up to the the total number of observed 
		sources ($N_{\rm obs}$) and $\omega(S_i)$ is the sky coverage 
		evaluated at the flux of the $i^{th}$ source ($S_i$).
		Importantly, the bright-end slope of the {\it logN-logS} can also inform us about the evolution of the population.
		In a Euclidian space, if there was no evolution, a source class of a given luminosity would be distributed according to 
		$N(>S)\propto S^{-3/2}$. If instead the considered class of objects underwent a positive (negative) evolution,
		the slope of the {\it logN-logS} would be greater (lesser) than 1.5. 
\end{enumerate}

\begin{figure}
	\includegraphics[width=\columnwidth]{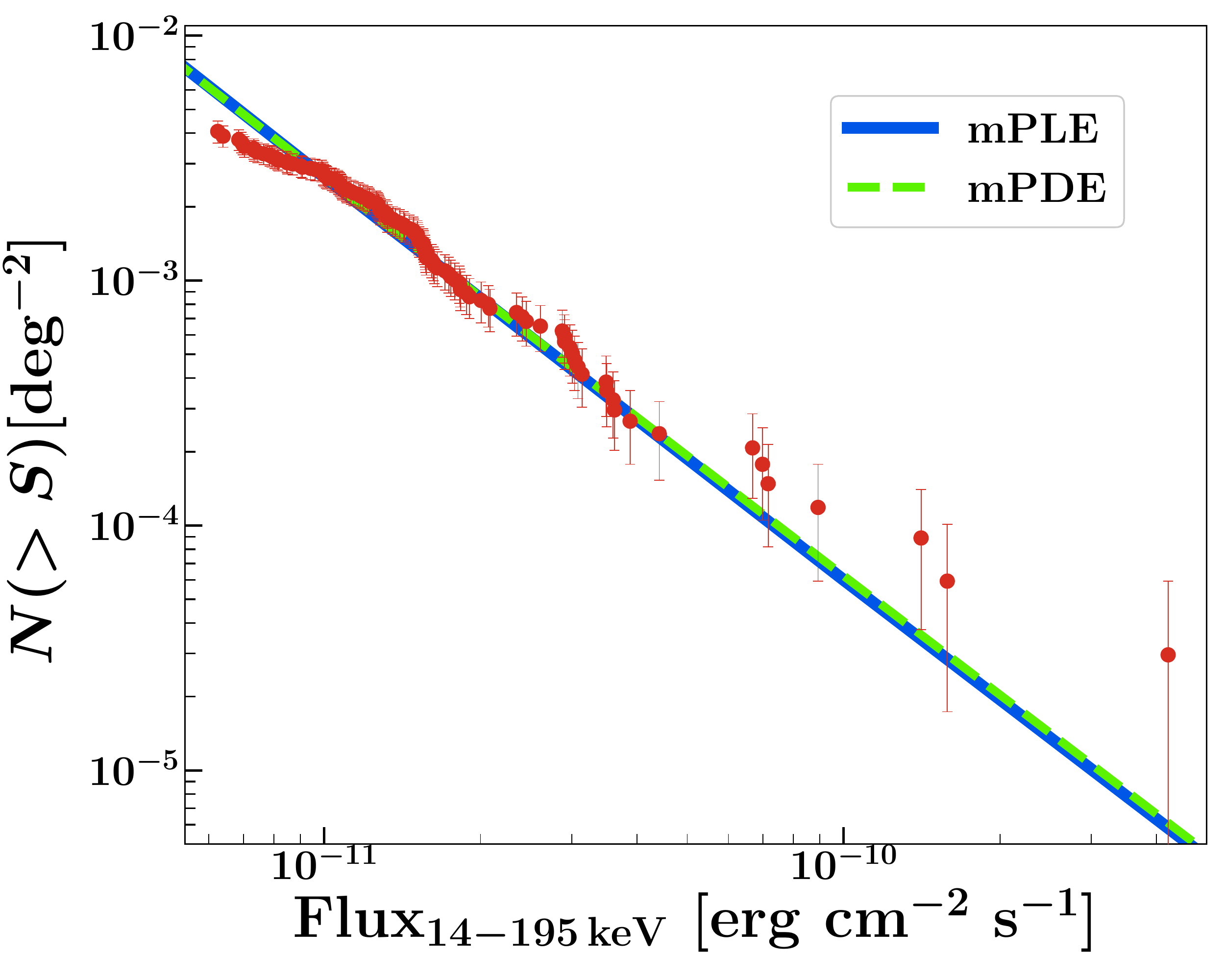}
    \caption{Cumulative BAT-blazars {\it logN-logS}, i.e., total number of sources above a certain flux corrected for the survey sky coverage $\omega(S)$, $N(>S)$, as a function flux   (Eq.~\ref{eq:lognlogs}-\ref{eq:lognlogs_obs}).
        The observed {\it logN-logS} is represented by the red data points, while the
        blue solid and green dashed lines show the prediction from the best-fit mPLE and mPDE models, respectively.
        We note that in this representation the data points are not independent from one another.
        As can be seen, both models explain reasonably well the observed distribution in the region $ 8\times10^{-12}\,\rm erg~cm^{-2}~s^{-1}$ to $10^{-10}\,\rm erg~cm^{-2}~s^{-1}$, where most source counts lie. As detailed in Appendix~\ref{appendix:A}, the slight discrepancy between the data and the model predictions below 
        $8\times10^{-12}\,\rm erg~cm^{-2}~s^{-1}$ does not influence the validity of our results. The derived spectral slope for the {\it logN-logS} is $1.62\pm0.05$, indicative of a positive evolution of this source class (see \S\,\ref{sec:test})}.\label{fig:logn}
\end{figure}

\begin{figure*} 
\centering
        \includegraphics[width=.49\textwidth]{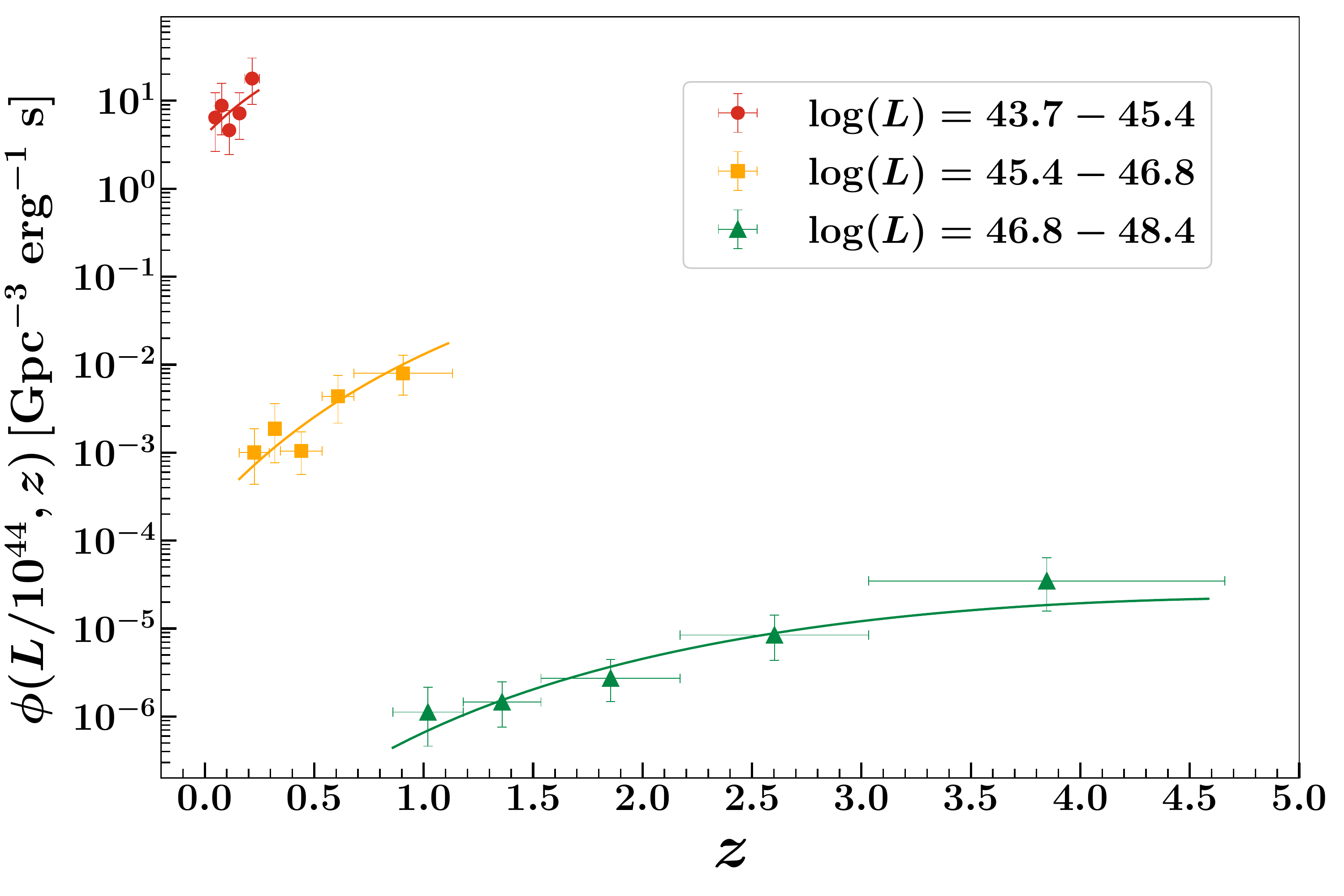}
        \includegraphics[width=.49\textwidth]{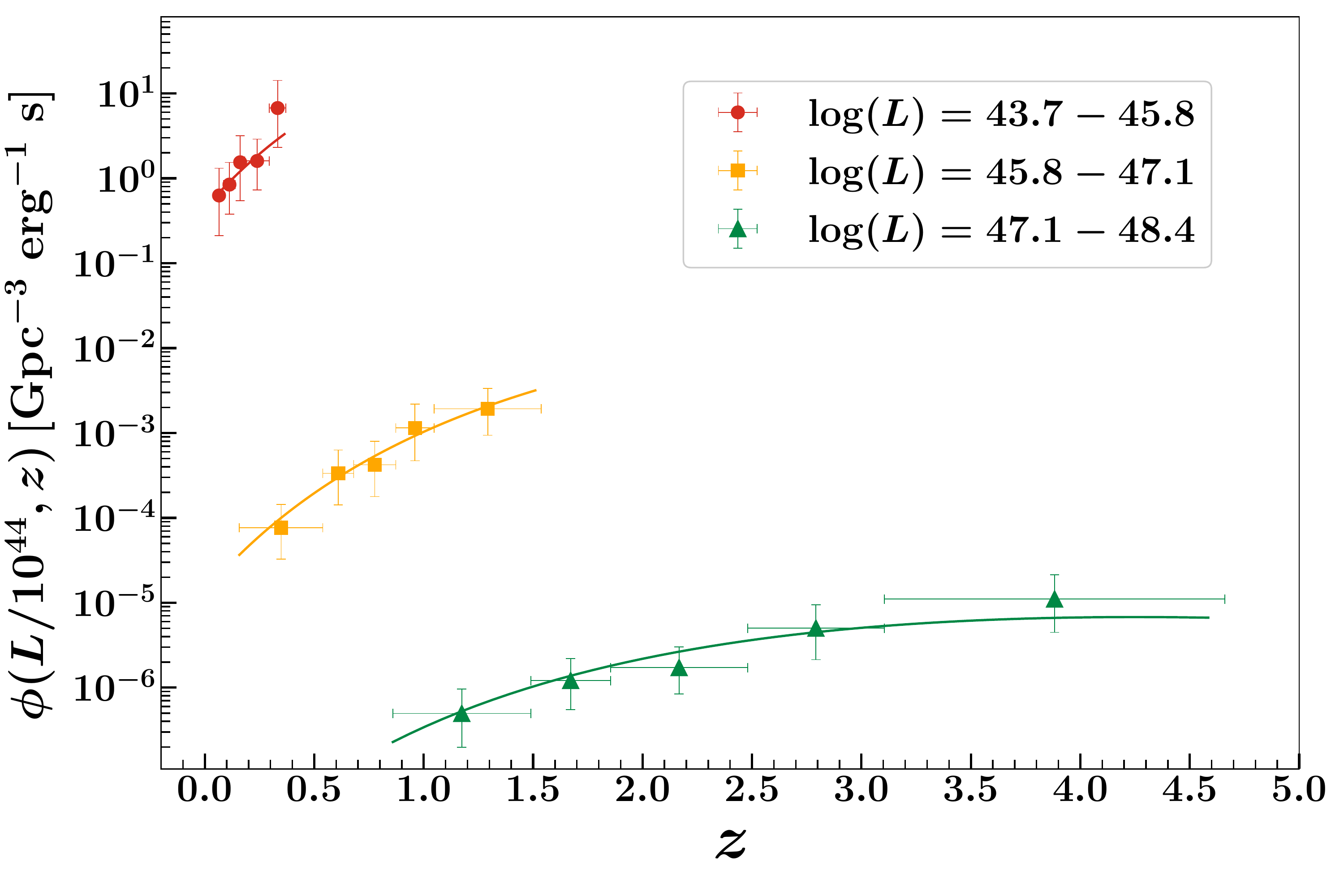}
        \includegraphics[width=.49\textwidth]{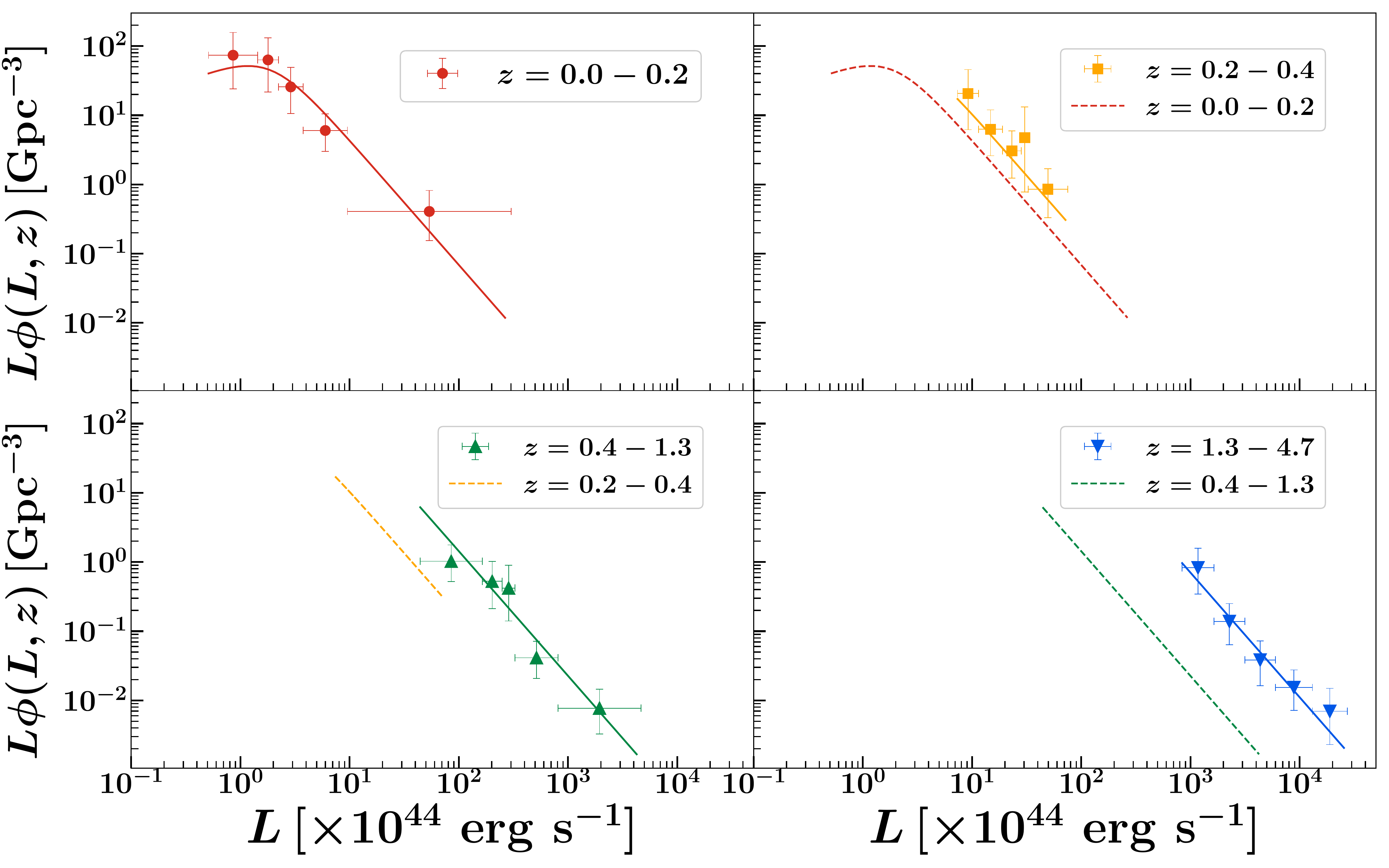}
        \includegraphics[width=.49\textwidth]{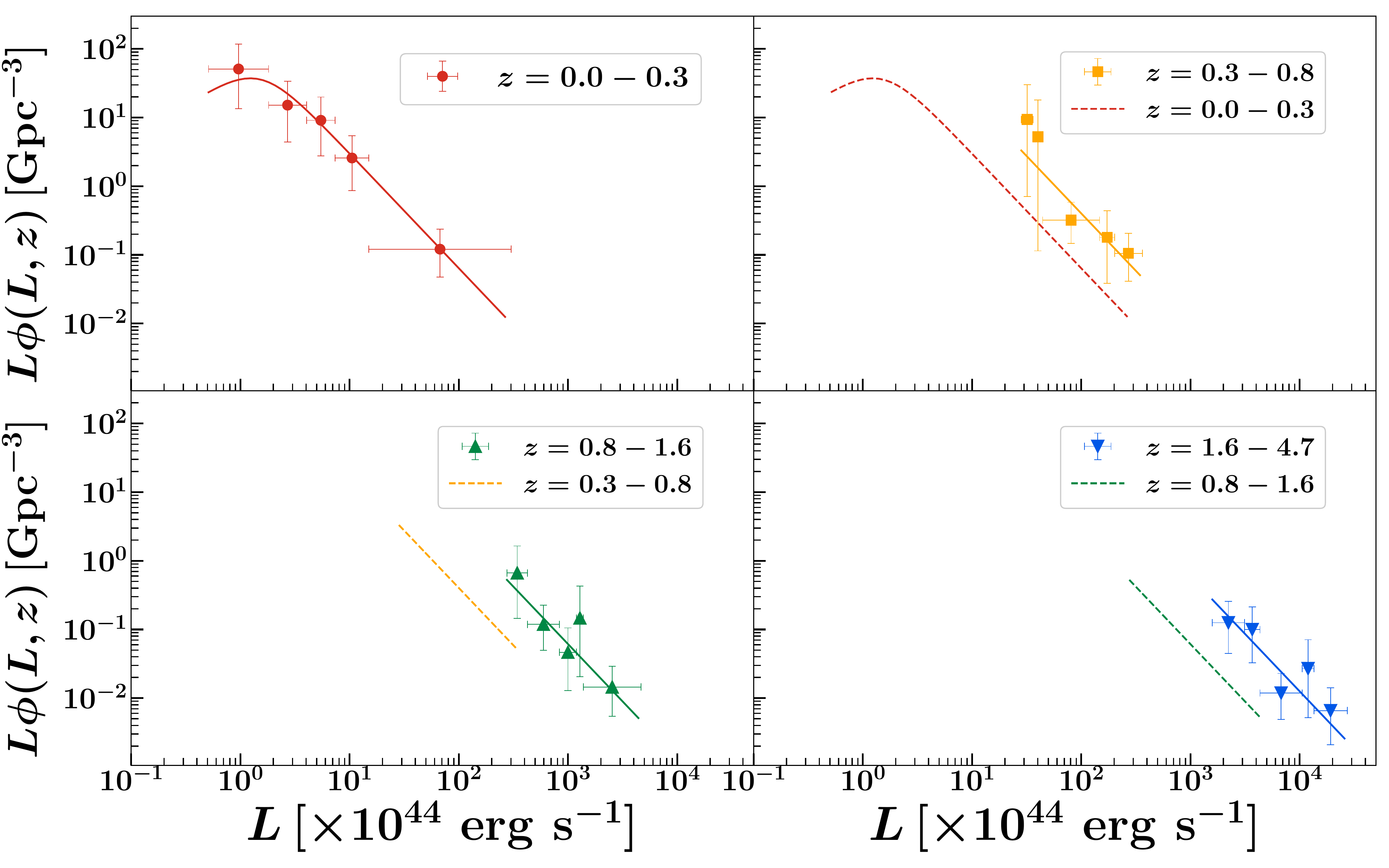}
	\caption{Total blazar (left) and FSRQs (right) X-ray luminosity function, $\phi(L,z)$ (where $L\equiv L_{\rm X}$). On the top 
	panels, it is shown as function of redshift in various luminosity bins; 
	in the bottom panel it is shown are function of luminosity at different redshift bins. The data points are the 
	one deconvolved via the $N^{\rm obs}/N^{\rm mdl}$ method described in Section~\ref{sec:lf_all}. The solid lines instead show 
	our best-fit mPLE model. In the bottom panels, to highlight the evolution at different luminosities, 
	the XLF from the previous redshift bin is overplotted with dashed lines.\label{fig:evol}}
\end{figure*}

\section{Best-fit Luminosity Function}\label{sec:res}
\subsection{All blazars}\label{sec:lf_all}
The results from the ML fit are three-fold.
Firstly, as can be noted in Table~\ref{tab:res_ml}, the use of a smooth broken power law to represent the local luminosity function greatly improves the fit results with respect to the simpler power-law case ($\Delta_{\rm AIC, sPLE-mPLE}\sim23$). The KS statistics values also show that the latter models are close or below the 30\% threshold set in Section~\ref{sec:test}, hence can be disregarded. This outcome was already noted in A09, where the authors emphasize how it is necessary to introduce a luminosity break to explain the observed redshift and luminosity distribution.

The second result is that it is not possible to %distinguish whether a luminosity or density %evolution is taking place in the BAT-blazar sample
distinguish between the luminosity evolution or the density evolution
scenarios.
This is hinted in the simple power-law scenario and confirmed by the broken power-law one. In fact, results on the ML fit and AIC values between sPDE/sPLE and mPDE/mPLE only slightly differ ($\Delta_{\rm AIC, mPDE-mPLE}\sim  0.4$), rendering the models comparable to each other. Moreover, the LDDE case does not improve the results upon either the mPDE or the mPLE. Indeed, both ML fit and AIC values are higher than the mPDE/mPLE cases ($\Delta_{\rm AIC, LDDE-mPLE}\sim 19$), and the best-fit redshift index is close to zero ($\rho= 0.27\pm0.04$), therefore removing the luminosity dependence from Equation~\ref{eq:evol_fact_ldde}.

Finally, the evolutionary parameters confirm a positive evolution of the blazars population ($k>2$) both for the mPDE and mPLE case. The evolutive parameter $\xi$ is coherently negative for both cases, indicating an exponential cut-off at high redshifts. Moreover, the slope derived from our best-fit luminosity function to the {\it logN-logS} (shown in Figure~\ref{fig:logn}) is $1.62\pm0.05$ for both the mPLE/mPDE results, in accordance with a positive evolution of the source population, a result already emphasized by A09. 

The best-fit index value of the distribution for the sPLE/sPDE is $\gamma=1.31\pm0.08$. 
This is significantly ($\sim3\sigma$) harder than the one reported A09 for the sPDE model ($\gamma_{\rm A09}=1.67\pm0.13$\footnote{For the simple power-law $\phi(L, V=(z=0))$ (Equation~\ref{eq:pl_dn}), 
we note that A09 lists $\gamma'=\gamma+1$ (i.e., the index of the $dN/dL$, not of the
$dN/d\log L$). Therefore, to provide a consistent comparison, 
we here report the $\gamma=\gamma'-1$ values of A09.}), though in agreement 
with their adopted sPLE one ($\gamma_{\rm A09}=1.26\pm0.07$, see Table 4 of A09).
For the mPDE/mPLE scenarios, the bright-end slopes of the distribution are softer with respect to the sPDE/sPLE ones, i.e., $\gamma_2=1.79\pm0.19$. 
It is important to note that these values are in very good agreement with the indices reported by A09 in the simple 
power-law scenario, and similar to the one found for the evolution of the unbeamed jetted AGN population \citep[$\gamma_{\rm FRII,  15 GHz}\sim1.65$,][see Section~\ref{sec:disc}]{Cara_Lister_2008}.
On the other hand, there is a difference at the $\sim 3\sigma$ level from the values found in A09 for the mPLE ($\gamma_{2,\rm A09}=2.73\pm0.37$) and mPDE ($\gamma_{2,  A09}=2.54\pm0.21$) cases . 
We note that this discrepancy can be accounted for the fact that our sample (1) reaches higher redshifts, 
(2) has three time the source number than the one used by A09, and (3) goes almost an order of magnitude deeper in flux.
It follows that our fit is able to more accurately constrain the shape of these distributions. 
The faint-end slope is flat ($\gamma_1<1$) for both mPLE/mPDE, although poorly constrained due to the absence of sources. 
A flattening of the luminosity function is expected at low luminosities as a result
of beaming \citep[][]{Urry_Shafer_1984}, with a predicted spectral index value 
between 1 to 1.5. Our derived $\gamma_1$ fit values are consistent with this slope, taking into account the errors.

In Figure~\ref{fig:evol} the LF prediction from the mPLE best-fit model are shown, as function of redshift in the top left and as function of $14-195\,\rm keV$ luminosity in the bottom left panel. The displayed data points are the deconvolved ones, obtained by calculating the scaling factor through the ${N^{\rm obs}}/{N^{\rm mdl}}$ technique. This has been shown to be an effective unbiased representation for any predicted LF given the real data \citep[see][]{LaFranca_1997,page_carrera_2000,Miyaji_2001}, which are unfolded as such:

\begin{equation}
\phi(L,z)^{\rm obs} = \phi(L,z)^{\rm mdl}\times \frac{N^{\rm obs}}{N^{\rm mdl}},
\end{equation}
where
\begin{equation}
\frac{N^{\rm obs}}{N^{\rm mdl}} = \frac{N^{\rm obs}(L_i,z_i)}{\int_{z_{\rm min}}^{z_{\rm max}}\int_{L_{\min}}^{L_{\rm max}} dL dz \lambda(L,z)},
\end{equation}
and $N^{\rm obs}$ is the observed number of blazars, while $N^{\rm mdl}$ is the number predicted to be observed given $\phi(L,z)$. We stress that the mPDE results look formally identical to the mPLE ones in this representation. In fact, it is clear that the XLF data as function of luminosity are distributed accordingly to a straight power law (in $\log-\log$ space) without any apparent turnover, which impedes the differentiation between a luminosity versus a density evolution scenario. This behavior, discussed further in Section~\ref{sec:jets}, is reflective of the fact that the BAT is sampling the high-luminosity tip of the blazar population, and it is still not able to detect
the break (except maybe at the lowest redshift, see below)
in the luminosity function expected as a result of beaming.
The mPLE or mPDE models are favored by the ML fit, indicating that indeed a break and a change in slope of the distribution are preferred. 
This not only is expected as a direct consequence of beaming, but also because if the distribution was a continuing straight power law, 
the hard X-ray sky would be dominated by too many low-luminosity blazars.
In Figure~\ref{fig:evol}, it can be noticed from the model prediction that the 
break in luminosity should start appearing in the lowest redshift bin ($z=[0.0, 0.2]$). Nevertheless, the statistical uncertainties on the
data do not allow us to see the break with high significance. 
In fact, mPLE and mPDE are still not differentiable in likelihood values as they are sampling a power-law distribution with a luminosity cut-off occurring at the minimum observed luminosity ($L_*\sim10^{44}\,\rm erg~s^{-1}$), and hence are formally identical to each other \citep{Bahcall_1977}. 

\subsection{FSRQs}\label{sec:lf_fsrq}
\begin{figure}
    \includegraphics[width=\columnwidth]{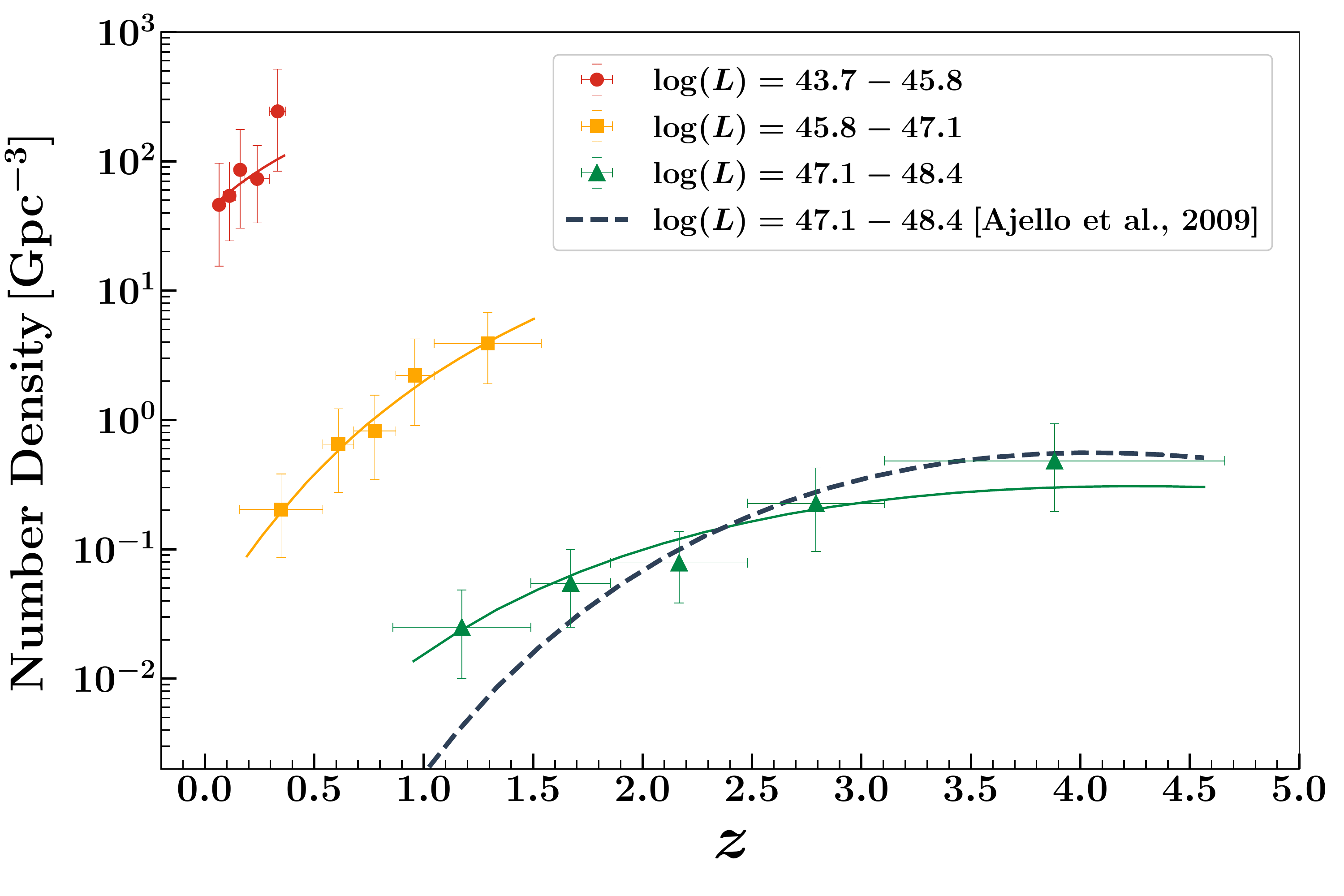}
    \caption{Number density plot of BAT FSQRs as function of redshift and luminosity bins. The data points are the one deconvolved via the $N^{\rm obs}/N^{\rm mdl}$ method and the solid lines show the
    best-fit mPLE model. For comparison, 
    number densities model prediction from A09 in the highest luminosity bin are shown ($\log L = 47.1-48.4\,\rm[erg~s^{-1}]$, black dotted line). 
    It can be seen how our best-fit XLF predicts a different XLF shape implying fewer sources at the highest redshifts, $z=[3-5]$.}
    \label{fig:num_dens}
\end{figure}

FSRQs outnumber BL Lacs in our sample ($75\%$ of the total, see Table~\ref{tab:sample}); they also are more luminous, and have better constrained redshifts which span a larger range (the farthest source being at
$z=4.65$, see Table~\ref{tab:sample}). To test their evolution, we fitted the same models as the ones used for the overall population. Results show that FSRQs drive the evolution of the whole BAT-blazar sample. Their XLF is similarly well described by a broken power law with a luminosity cut-off occurring at $L_*\sim 10^{44}\ergflux$. Both the mPDE/mPLE shapes give consistent fits, with the high-end slope being $\gamma_2=1.67\pm 0.17$. 
It is interesting to note that the results on the indices on the $\phi(L, V(z=0))$ are very similar to the ones 
reported on the LAT FSRQs studied in A12 (where $\gamma_{2, \rm 100\,MeV-100\, GeV}=1.60$, see Table 3 in their work).
The mPLE representation is shown on the right panels of Figure~\ref{fig:evol}, and Figure~\ref{fig:num_dens} displays the number density of these sources as a function of redshift. 

As previous works have found, this class of objects evolve positively in redshift ($k=5.05\pm0.79$, i.e., their number density or luminosity increases as function of redshift), although uncertainties on the evolutionary parameters remain quite high. This positive evolution is also confirmed by the slope of the {\it logN-logS} derived from the best-fit luminosity function which is $1.60\pm0.07$ for both the mPLE/mPDE models. The factor $e(z)$ (Equation~\ref{eq:evol_fact}) allows us to estimate the peak of the distribution, which occurs at $z_{\rm peak}=-1-k\xi$. Given the best-fit values of these parameters, the peak is located at $z_{\rm peak}\sim4.3$. Nevertheless, the errors associated with $k$ and $\xi$ allow for this value to range between $z_{\rm peak}\in[3.5,5.5]$, impeding the exact localization of the maximum
of this population of powerful jetted AGN.

We stress that for the first time our BAT-blazar sample contains 
one source that lies beyond $z=4$ (SWIFT 1430.6+4211, BAT index 1448). This enables us to set more solid constraints on the location of the peak, 
and to assess whether a turnover in the luminosity function of the most luminous sources is starting to appear in the data. 
As evident from Figure~\ref{fig:evol} (top right panel), even for the high-luminosity end of the population,
a turnover of the XLF remains undetected, placing the cut-off peak beyond $z=4$.

\begin{figure}
    \includegraphics[width=\columnwidth]{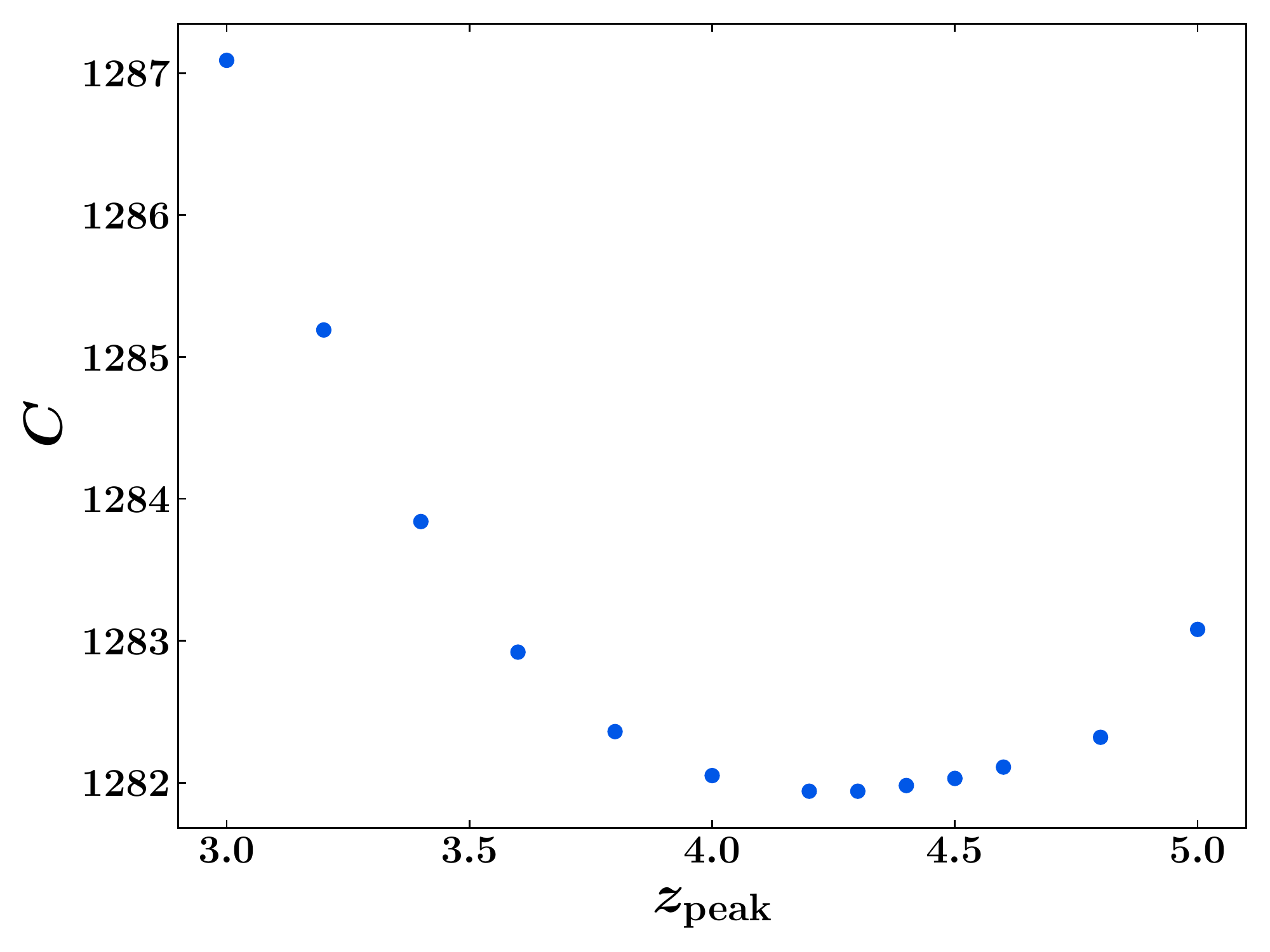}
    \caption{Log-likelihood results ($C$, see Equation~\ref{eq:log-like}) for the BAT FSRQs using the MPDE model fixing the position
	of the $z_{\rm peak}=[3,5]$. It can be noticed how the fit values reach a minimum 
	for a peak occurring at $z_{\rm peak}=4.3-4.4$, confirming that the luminous BAT FSRQ 
	population peaks at $z>4$.}\label{fig:log-like-zpeak}
\end{figure}

To further assess the likelihood that the peak lies at $z>4$ (or viceversa), we perform ML fits using the mPDE evolutionary 
model, forcing the peak to occur at a specific redshift in the range $z_{\rm peak}=[3,5]$. Figure~\ref{fig:log-like-zpeak} shows the 
results of ML values as a function of $z_{\rm peak}$. It can be seen that our fits support the scenario 
in which $z_{\rm peak}\sim4.3-4.4$ ($C$ is at a minimum) and could occur even at $4.6$, the maximum redshift of our sample. 
Instead, below $z<4$ results differ significantly from the minimum ML values, excluding the possibility of the
peak occurring at later cosmic times.
It is therefore with high confidence that we ascertain that the peak lies $z\geq4$, confirming
the results found by A09.
A deeper all-sky X-ray survey would most likely be necessary to pinpoint this peak.

Figure~\ref{fig:num_dens} shows the number density of FSRQs as function of redshift for different luminosity bins predicted by the mPLE best-fit. The data points are derived using the $N^{\rm obs}/N^{\rm mdl}$ technique. For comparison, the model prediction on FSRQ number densities derived from the MPLE best-fit reported in A09 in the highest luminosity bin are plotted ($ \log(L)=47.1-48.4$, black dotted line). It can be seen how our best-fit model predicts lower number densities than A09 (e.g., two times fewer sources per comoving volume at $z=4$), implying the existence of fewer highly luminous FSRQs than previously anticipated. 

\section{Average blazar SED}\label{sec:ave_sed}
The high-energy (hard X- to \gm-rays) blazars SED typically shows a double power-law shape with a peak located in the MeV to GeV range (depending on the source class and luminosity, \citealp[e.g.,][]{Ghisellini_1998,Ghisellini_2017}). Here we aim to phenomenologically characterize such high-energy SED in order to correctly account for their spectral shape in the whole $14\rm\, keV - 10\rm\, GeV$ range. This in turn allows us to extrapolate their contribution to the CXB as well as making prediction for the MeV background. We choose to consider only the sources belonging to the FSRQ class, 
as they are the dominating population (in luminosity) in our sample, and are expected to produce the higher contribution to the CXB (see A09).

\subsection{LAT detected BAT FSRQs}\label{sec:lat_det_bl}

At first, we
restrict ourselves to the sources which have both BAT and LAT data\footnote{Although the BAT and LAT surveys do not strictly cover the same time period, it is safe to assume that the data (given also the large uncertainties on the spectral parameters) are a good representation of the average source state both in hard X- and \gm-rays.}. This results in 46 objects (out of 88 BAT FSRQs). The chosen putative SED
spectral shape is the following:
\begin{equation}\label{eq:sed}
E^2\frac{dN}{dE} =  E^2\left\{K\left[\left(\frac{E}{E_b}\right)^{\eta_1}+\left(\frac{E}{E_b}\right)^{\eta_2}\right]^{-1}\right\},
\end{equation}
where K is the normalization constant, $E_b$ is the break position in the distribution, and $\eta_1$ and $\eta_2$ are the low-energy and high-energy spectral indices, respectively. In A12, the contribution from the extragalactic background light \citep[EBL, ][]{EBL_2018} was added to the same framework
as an exponential cut-off to Equation~\ref{eq:sed}. Since the contribution of the EBL is significant at $>10\,\rm GeV$, far from the background energies of interest to this work, we limit the fit to $<10\,\rm GeV$ and we avoid adding the EBL contribution.
The luminosity-dependent SEDs are obtained by multiplying both sides of Equation~\ref{eq:sed} by $4\pi D_L(z)^2$, with $D_L$ being the luminosity distance of the source \citep{Hogg_1999}, and all SEDs are shifted by $(1+z)$ in order to transform to the source rest-frame.
The hard X-ray and $\gamma$-ray spectra of the sources are obtained considering a power-law spectral shape in both bands, using the flux and index values from the BAT 105 catalog and the 4FGL, respectively.
The $1\sigma$ uncertainties in the spectra are accounted for by a bowtie spectrum using both flux and index error values provided by the two catalogs.
We fit Equation~\ref{eq:sed} simultaneously to both BAT and LAT data
and derive the best-fit parameters for every source using a standard minimizing $\chi^2$ technique.
In turn, this allows us to divide the sample in different luminosity bins, chosen such that they contain roughly the same number of sources.
\begin{figure*}
    \includegraphics[width=\columnwidth]{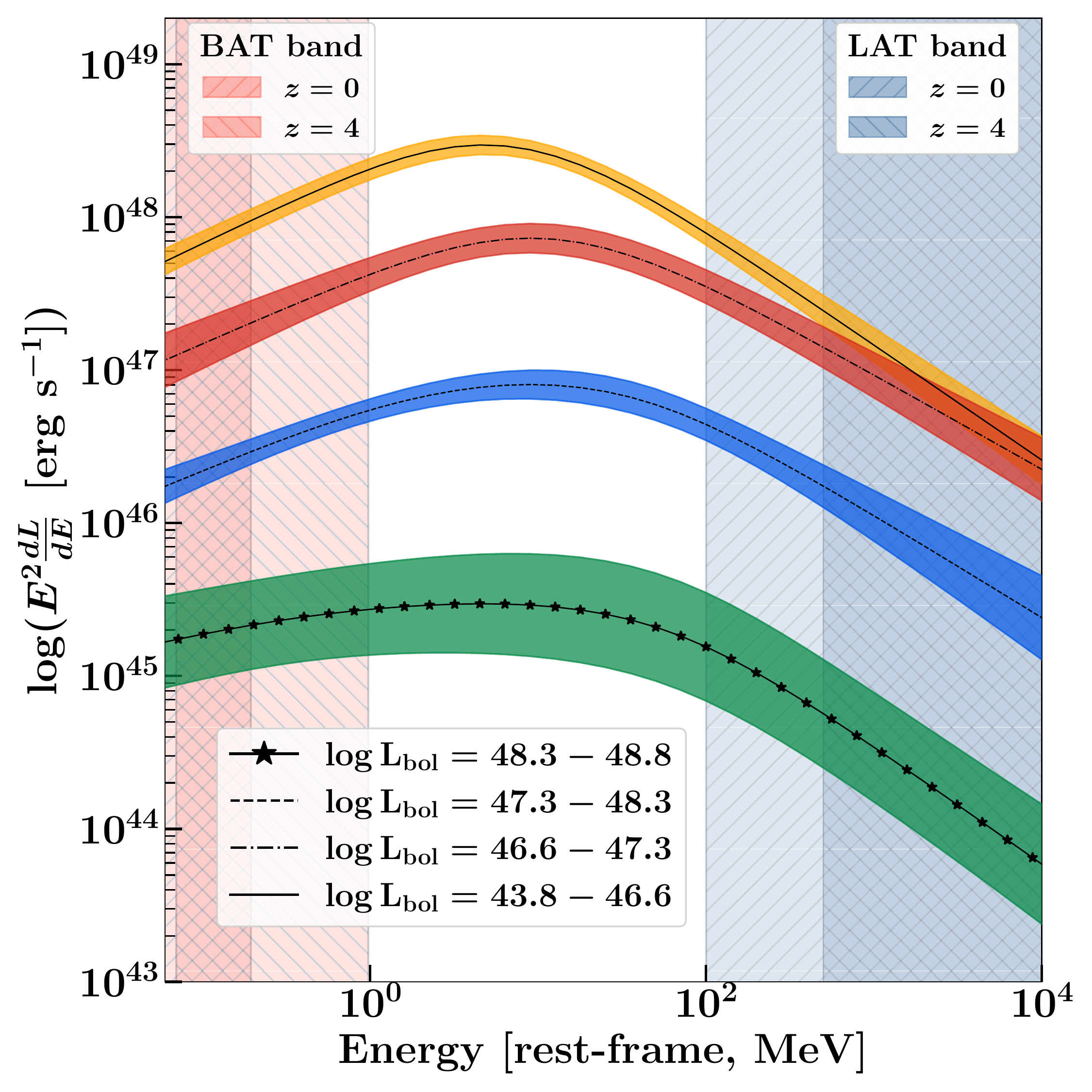} 
    \includegraphics[width=\columnwidth]{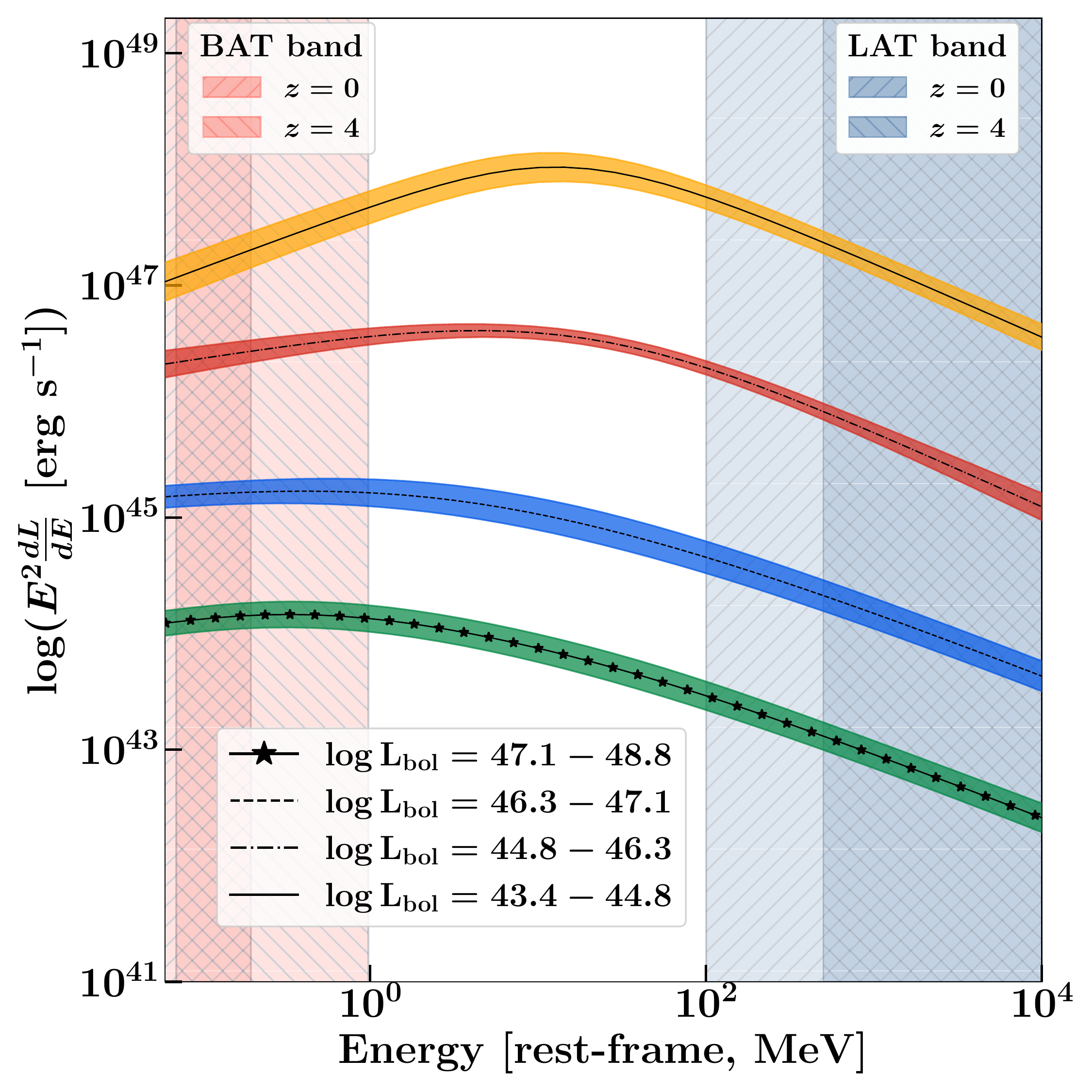} 
    \caption{Average high-energy SED of the FSRQs in our sample derived by fitting: (i) only sources which are BAT and LAT detected (46 sources, left panel) and (ii) the whole sample (88 sources, right panel). The details of the fits can be found in Section~\ref{sec:ave_sed}. The different lines represent the mean of the distribution at different luminosity bins (as labelled in the Figure). The errors (shown as colored shaded regions) are derived
    employing a Jackknife technique. The high-energy SED spectral form does not show strong dependence of the
    chosen luminosity bin. On the right panel it can be noticed that the lowest luminosity sources have on average an SED peaking at lower energies ($\sim 3\,\rm MeV$ versus $\sim17\,\rm MeV$ for the highest luminosity sources). This relation between the energy break position and the source luminosity has been calibrated (see Section~\ref{sec:lat_non_det_bl} and Figure~\ref{fig:index_vs_z}, right) in order to properly take into account the contribution to the MeV background. The shaded hatched pink and light blue regions represent the BAT and LAT energy bands, respectively,
    shifted for sources at $z=0$ (forward slash) and $z=4$ (backward slash).
    As can be seen, the MeV peak of the sources falls exactly in the region which still remains uncovered
    by both instruments and is critical to determine the entire SED.}\label{fig:ave_sed}   
\end{figure*} 

The binned luminosity SEDs are shown in Figure~\ref{fig:ave_sed} (left panel). The errors are computed using the Jackknife method \citep{Efron_1981}.
As can be seen, the shape of the FSRQs SED does not show a strong evolution in luminosity, i.e., both the break position and the spectral indices are very similar in every luminosity bin. This was already noted by previous works and is in contrast with the anti-correlation between the source luminosity and the low-energy synchrotron peak \citep[see A09,][]{Ghisellini_2010, Ghisellini_2017,Paliya_2019}. Importantly, it allows us to establish average SED parameters for these sources, which result to be:
$\eta_1=1.53\pm0.09$, $\eta_2=2.65\pm0.07$ and $E_{b,\rm 
rest-frame}=16.70\pm1.82\,\rm MeV$.
To further check for consistency, we extract the values of spectral indices and energy break from the average blazar SED models reported in \citet{Paliya_2019}. This results in $\eta_1\sim1.45-1.60$, $\eta_2\sim2.45-2.63$ and $E_{b, \rm rest-frame}\sim 9-12\,\rm MeV$, values that are in complete agreement with our fits. 
In Figure~\ref{fig:ave_sed}, the shaded hatched pink and light-blue regions display the BAT and LAT energy bands, respectively, shifted for sources at $z=0$ (forward slash) and $z=4$ (backward slash). As can be seen, the MeV peak of the sources falls exactly in the region which still remains uncovered by both instruments and would be critical to determine the full SED.

\subsection{LAT-undetected BAT FSRQs}\label{sec:lat_non_det_bl}

\begin{figure*}
    \includegraphics[width=0.45\textwidth]{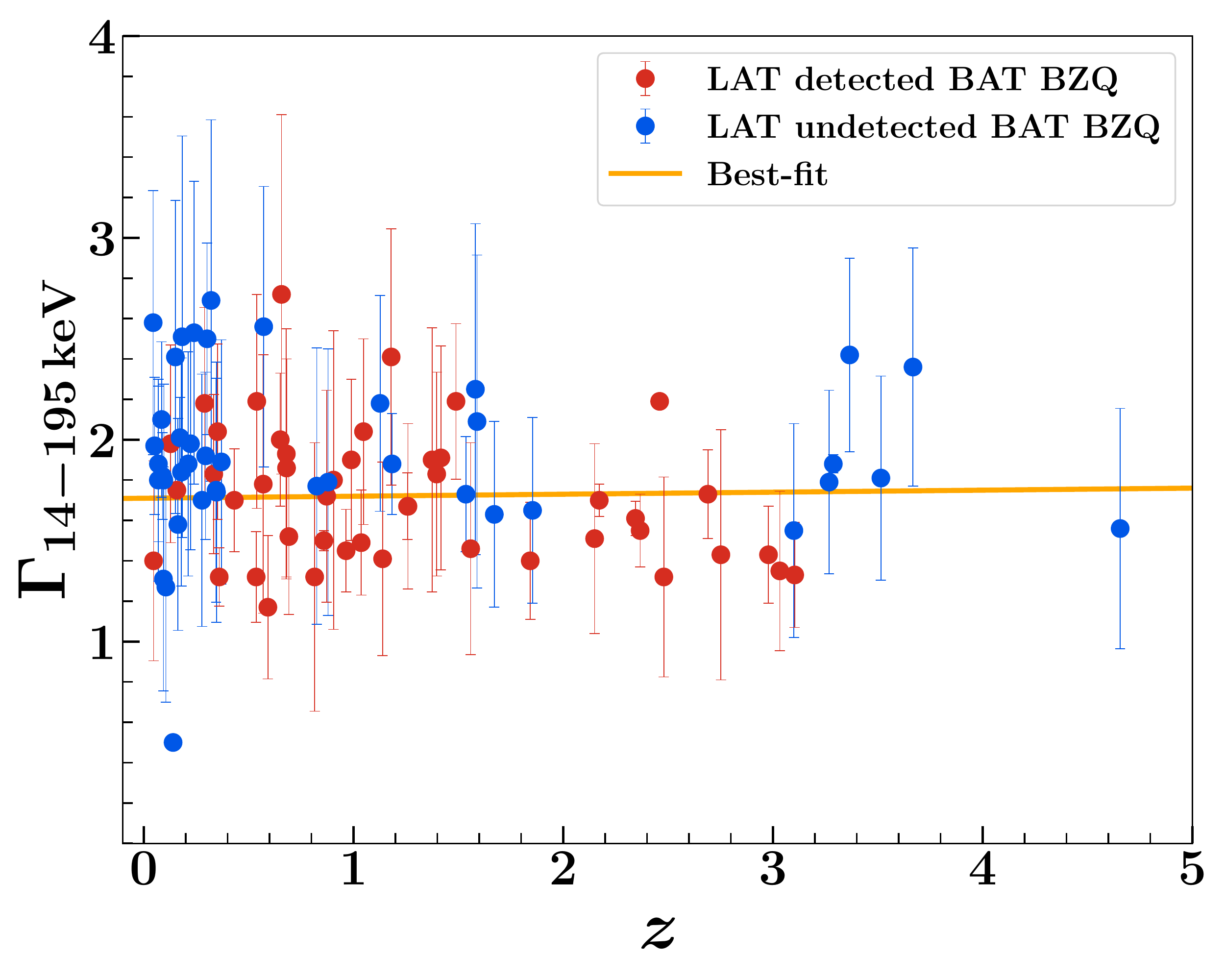}
    \includegraphics[width=0.45\textwidth]{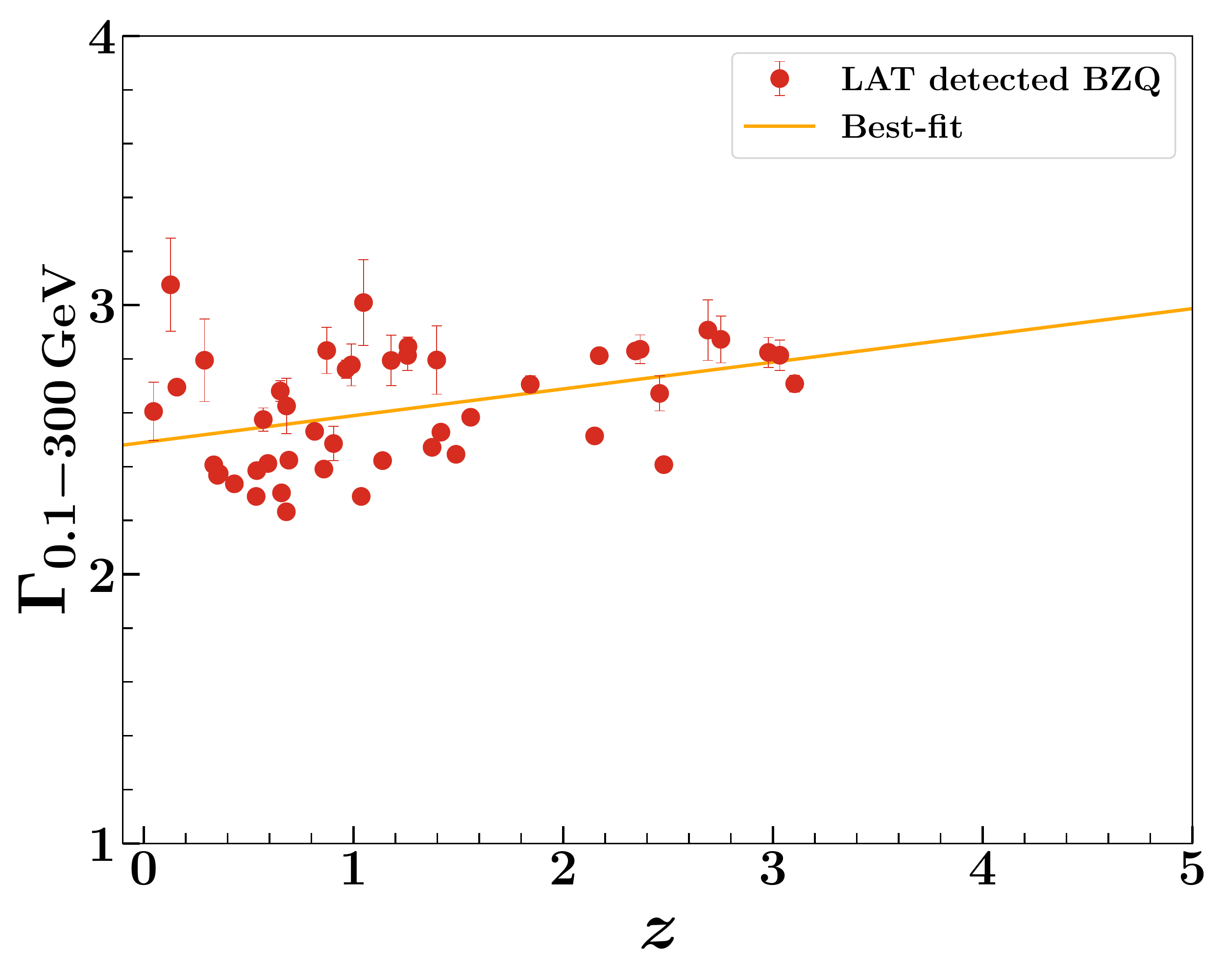}
    \caption{{\bf Left:} BAT ($14-195\,\rm keV$) photon index versus redshift of all BAT FSRQs in our sample. Red data points represent LAT-detected sources, while blue data points are the LAT-undetected ones. The yellow line is the best-fit linear relationship calculated for these data. The slope is very flat ($m_{\rm BAT}=0.01\pm0.01$) indicating that there is no evolution in redshift of the BAT photon index, and the intercept values gives us the average value of $c_{\rm BAT}=1.71\pm0.01$. {\bf Right:} LAT ($0.1-300\,\rm GeV$) photon index versus redshift of all LAT-detected BAT FSRQs in our sample (red data points). The best-fit line is shown in yellow. Similarly to the BAT regime, the slope is very flat ($m_{\rm LAT}=0.09\pm0.03$) and the intercept returns an average photon index $c_{\rm LAT}=2.48\pm0.05$. These empirical relations confirm that there is no dependence in redshift of the high-energy SED spectral slopes of BAT FSRQs.}
\label{fig:index_vs_z}   
\end{figure*}

In order to understand whether the LAT-undetected blazars would influence the above result, we tried establishing an average $14\,\rm keV$ to $10\,\rm GeV$ SED for these sources.
We note that there is neither a redshift nor a X-ray luminosity dependence between LAT-detected and -undetected blazars for our BAT-selected sample, i.e.~LAT-undetected sources lie in the same luminosity-redshift space as the LAT-detected ones.

As results in Section~\ref{sec:lat_det_bl} point to the non-evolution of the average blazar SED with luminosity, we wished to understand if this was consistent with the data. Therefore, we perform (i) a linear fit between the BAT ($14-195\,\rm keV$) photon index and redshift of all BAT FSRQs and (ii) a linear fit between the LAT ($0.1-300\,\rm GeV$) photon index and redshift of all BAT FSRQs detected by the LAT.
    In agreement with the results in Section~\ref{sec:lat_det_bl}, both fits return very flat slopes ($m_{\rm BAT}=0.01\pm0.01$ and $m_{\rm LAT}=0.09\pm0.03$) and there is no statistically significant difference between a linear or a constant behavior in redshift (i.e. $m=0$); therefore there is no dependence in redshift of either the BAT or LAT photon indices. In the hard X-ray regime, the best-fit value for the intercept is $c_{\rm BAT}=1.71\pm0.01$, and in the $\gamma$-ray regime $c_{\rm LAT}=2.48\pm0.05$. Figure~\ref{fig:index_vs_z} shows these results. The value of the BAT intercept is softer with respect to the best-fit average spectral index from the LAT-BAT SED ($c_{\rm BAT}=1.71$ vs.\ $\eta_1=1.53$). This behaviour is mostly influenced by the low-$z$ (low-luminosity) sources that are LAT-undetected (upper left corner of Figure~\ref{fig:index_vs_z}, left panel), and it is indicative of the fact that for these objects the BAT is sampling the SED closer to the high-energy peak, resulting in a softer photon index. 
With these relations in hand, we can assign a $\gamma$-ray spectral slope to all LAT-undetected FSRQs, knowing their redshift.
For each source that is undetected by the LAT, we assign a $\gamma$-ray flux that does not violate the 12-year upper limits computed at the position of that source.
The mock $14\,\rm keV$ to $10\,\rm GeV$ SED of the LAT-undetected sources is therefore constructed using BAT real data and LAT upper limits. 

We employ the same formalism as Section~\ref{sec:lat_det_bl} to derive the average SED using the total sample of 88 blazars. The result is shown in Figure~\ref{fig:ave_sed} (right panel). Overall the average SED shape is similar to what found in Section~\ref{sec:lat_det_bl}, and the best-fit average spectral slopes are $\eta_1=1.70\pm0.14$, $\eta_2=2.61\pm0.03$.
Interestingly, it can be seen that as the luminosity decreases, the peak position shifts slightly towards lower energies ($E_{\rm b, rest-frame}\sim 3\,\rm MeV$ for the lowest luminosity bin versus $E_{\rm b, rest-frame}\sim 17\,\rm MeV$ for the highest luminosity bin). This finding is closely related to the softer photon index detected by the BAT for the low-$z$ (low-luminosity) sources and it can be explained by the fact that the lower luminosity FSRQs have on average lower Doppler factors, making them less luminous and more low-energy peaked \citep[see also Section~\ref{sec:beam_res}, e.g.][]{Ghisellini_2015, Sbarrato_2015}.  
We take into account the shift of the SED to lower energies at lower luminosities (as in Figure~\ref{fig:ave_sed}, right)  using $\log{E_{b,\rm 
rest-frame}[\rm MeV]}=0.35\times\log{L_{\rm BAT}-15.44}$ to estimate the contribution of blazars to the MeV background.

\section{Contribution to the high-energy Backgrounds}\label{sec:bkg}
Resolving the CXB in its different components has so far been a challenge. Although above 10 keV low-luminosity, unbeamed AGNs are expected to contribute the most to the CXB \citep[e.g.,][]{Ajello_2012_AGN,Ueda_2014,Aird_2015,Ananna_2019}, blazars have been predicted to account up to $\sim 10-20\%$ of it in the hard X-ray band covered by BAT ($14-195\rm\,keV$, e.g., A09). 
With the best-fit blazar X-ray luminosity function in hand, it is possible to infer the contribution of the blazar population, and in particular of FSRQ one, to the total CXB. We calculate this contribution as follows:
\begin{equation}
\begin{split}
	&F_{\rm CXB_{\rm 14-195\,\rm keV }} = \int_{V_{\rm min}}^{V_{\rm max}}\int_{L_{\rm min}}^{L_{\rm max}} dV' dL' F(L', z')\phi(L', z') \\
&= \int_{z_{\rm min}}^{z_{\rm max}}\int_{L_{\rm min}}^{L_{\rm max}} dz' dL' \frac{dV}{dz'd\Omega } F(L, z')\phi(L', z')
\end{split}
	\label{eq:back}
\end{equation}
where $F(L, z)$ is the flux of a source with luminosity $L$ at redshift $z$ and $\phi(L, z)$ is the best-fit XLF. The limits on the integral are: $z_{\rm min}=0$ , $z_{\rm max}=6$, $L_{\rm min}=10^{43}\,\rm erg~s^{-1}$ and $L_{\rm max}=10^{50}\,\rm erg~s^{-1}$.

The results for the various background contribution are as follows. For the mPLE model: $F_{\rm CXB_{\rm 14-195\,\rm keV}, ALL} = 6.5\times10^{-9}\,\rm erg~cm^{-2}~s^{-1}~sr^{-1}$, $F_{\rm CXB_{\rm 14-195\,\rm keV}, FSRQ} = 3.5\times10^{-9}\,\rm erg~cm^{-2}~s^{-1}~sr^{-1}$. For the mPDE model: $F_{\rm CXB_{\rm 14-195\,\rm keV}, ALL} = 2.78\times10^{-8}\,\rm erg~cm^{-2}~s^{-1}~sr^{-1}$, $F_{\rm CXB_{\rm 14-195\,\rm keV}, FSRQ} =  1.5\times10^{-8}\,\rm erg~cm^{-2}~s^{-1}~sr^{-1}$. The intensity on the CXB in the $14-195\,\rm keV$ as measured by \citet{Ajello_2008} is $I_{\rm CXB, 14-195\,\rm keV}=1.5\times10^{-7}\,\rm erg~cm^{-2}~s^{-1}~sr^{-1}$. It follows that, employing the mPLE model, the BAT blazars contribute $\sim 4\%$ to this background, while the mPDE predicts a $\sim19\%$ contribution. If we consider FSRQs alone, the values are, respectively, $\sim2\%$ and $\sim10\%$. 
It is obvious that, although their contribution is non zero, BAT blazars are not the dominant source class that can account for the whole CXB in the $14-195\,\rm keV$ regime, confirming A09 results. 

\begin{figure*} 
\centering
        \includegraphics[width=.69\textwidth]{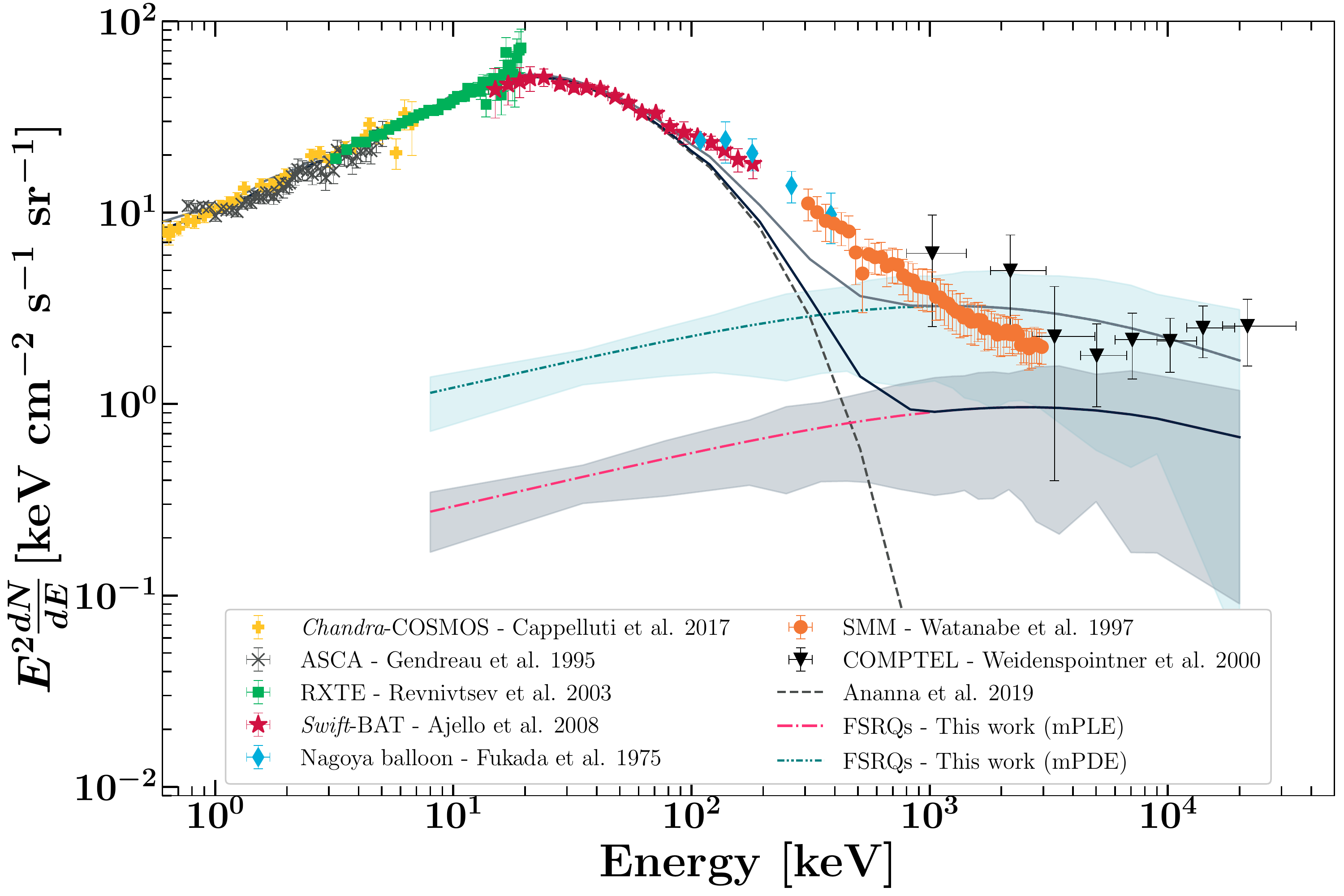}
        \caption{Intensity spectrum of the CXB and derived contribution of the FSRQs using the best fit mPLE model (pink dash-dotted line 
	 and gray shaded region) and the mPDE one (light blue dash-dotted line and cyan
	 shaded region).
         The data points represent the different measurements of the cosmic diffuse background, and have been extracted from the 
	 works indicated in the label \citep{Fukada_1975, Gendreau_1995, Watanabe_1997, COMPTEL_2000, Revnivtsev_2003,
	 Ajello_2008, Capelluti_2017}. 
         The dashed black line is the model prediction of the total contribution of AGNs to the CXB as derived by \citet{Ananna_2019}. 
	 The solid black and gray lines represent the sum of the total contribution from AGNs and FSRQs using the mPLE and mPDE models, respectively. The spectrum of 
	 FSRQs has been taken from the best-fit SED derived in Section~\ref{sec:ave_sed}. The shaded areas are derived from 
	 Monte Carlo simulations employing the best-fit parameter ranges. \label{fig:cxb}}
\end{figure*}

In Figure~\ref{fig:cxb} we show the intensity spectrum of the cosmic high-energy background (from $\sim0.5\,\rm keV$ to $\sim30\,\rm MeV$) and the predicted contribution of FSRQs\footnote{BL Lacs sources
have been found to be subdominant in this regime (see A09), hence are excluded from this calculation.} from $14\,\rm keV$ to $30\,\rm MeV$, employing the best-fit mPLE and mPDE models.
We adopt Equation~\ref{eq:back} and the spectral shape derived from 
Section~\ref{sec:lat_non_det_bl} to extrapolate the blazar contribution to the MeV regime, where $F(L, z)$ becomes $F(L, z, E)$, hence introducing 
the dependence on the blazar spectrum. The position of the blazar SED energy break ($E_{b, \rm rest-frame}$, Eq.~\ref{eq:sed}) is allowed to vary depending on the luminosity, following the results of Section~\ref{sec:lat_non_det_bl}. The $1\sigma$ uncertainty range (shaded gray/light blue areas in Figure~\ref{fig:cxb}) is calculated through a Monte Carlo approach, producing realization of the luminosity function $\phi(L,z)$ that take into account the errors associated with the best-fit parameters.
It can be seen how, although in the BAT regime FSRQs are subdominant, conversely in the $>500\,\rm keV$ regime their contribution is sufficient to explain the entire MeV background ($0.5-30\rm\,MeV$). It is important to note how the mPLE model can explain $\sim50-70\%$ of the MeV background above $1\,\rm MeV$. On the other hand, the mPDE model slightly over-predicts the MeV background above few $\rm MeV$, although the $1\sigma$ uncertainty band is quite large in this energy range.
It needs to be pointed out that other source classes have been anticipated to non negligibly contribute in this band. For example,
supernovae could contribute $\sim10-30\%$ to the MeV background \citep[e.g.,][]{Iwabuchi_2001,Ruiz-Lapuente_2016}. This hints to the fact that
the mPLE model is preferred with respect to the mPDE one. As highlighted by Figure~\ref{fig:cxb},
the mPLE extrapolation to MeV energies allows for contribution from other source classes.

Furthermore, to check whether our prediction of the blazar contribution well fits within population synthesis studies of other AGN classes, we also consider the contribution of AGNs to the CXB recently derived by \citet{Ananna_2019}. 
The black and gray solid lines in Figure~\ref{fig:cxb} represent the total model contribution of the 
two classes of sources. It can be noted that between $100\,\rm keV < E < 500\,\rm keV$, added contributions of AGN and blazars undergoing a mPLE evolution
fall short of completely explaining
the background. Conversely, if we employed the mPDE model, we could easily recover this gap though over predicting the MeV regime. 
Nevertheless, we caution that uncertainties in the $100-1000\,\rm keV$ range ascribed to (i) the measurements of the background,
(ii) models of non-jetted AGN components that could contribute in this regime (e.g., shape of the X-ray corona, \citealp{Inoue_2008,Fabian_2015}), and 
(iii) the extrapolation of the blazar model distributions, could naturally account for the discrepancy. 

Recent work from \citet{Toda_2020} predicts that the maximum contribution from FSRQs to the MeV 
background is $\sim3\%$. 
In their work, the authors perform the study with 53 BAT FSRQ (vs.~88 in this work) and they
do not recalculate the sky coverage (which could lead to biases in the results, see Section~\ref{sec:eff}) but use the all-sky one from \citet{BAT_105}.
Moreover, they work under the assumption that this source class follows a LDDE evolutionary paradigm.
As derived from our fits (Section~\ref{sec:res}), 
this (more complex) modelling is not significantly required from our data, as indicated by the fact that the luminosity dependence on the redshift evolution is compatible with zero. Moreover, we note that the spectral indices of the XLF derived for the LDDE scenario ($\gamma_{1}= -3.41\pm1.98$ and $\gamma_{2}=1.55\pm0.19$, Table~\ref{tab:res_ml}) are compatible with the mPDE/mPLE ones (within the statistical uncertainty), further emphasizing how these models are yet to be disentangled for this population. Finally, complexities of the sky coverage and sensitivity may lead to divergent
results from their work to ours.

\section{Distribution of jet properties}\label{sec:jets}
Significant selection effects have to be considered when dealing with relativistically beamed source emission. Indeed, Doppler effects can both enhance or de-boost the intrinsic jet emission by hundreds if not thousand of times \citep[e.g.,][]{Kellermann_2003, Ghisellini_2017, Yuan_2018, Lister_2019}.
On the other hand, it is possible to obtain information about the parent population taking into account the same beaming effects. 
The observed luminosity $L$ from a highly beamed relativistic source is related to its intrinsic (unbeamed) luminosity $\mathscr{L}$ by
\begin{equation}
L = \delta^p \mathscr{L},
\end{equation}
where $\delta$ is the kinematic Doppler factor
\begin{equation}
\delta = \frac{1}{\Gamma(1-\beta\cos\theta)}. 
\end{equation}
In the above, $\Gamma$ is the Lorentz factor of the jet, i.e., how fast are the electron moving along the structure (i.e. what is they bulk flow motion); it is related to the velocity of the emitting plasma $\beta = v/c$ by $\Gamma = 1/\sqrt{(1-\beta^2)}$, where $c$ is the speed of light. 
The power $p$ depends both on the radiation emission processes at the considered frequencies and on the jet configuration.
In the hard X-ray regime covered by BAT, blazar SEDs are in general dominated by the rising non-thermal power law, a result of inverse Compton scattering experienced by the electrons in the jet. The relativistic particles could be interacting with the same photons they produced once accelerated via synchrotron process, hence boosting them in a synchrotron self Compton scenario \citep[SSC, e.g.,][]{Ghisellini_1989}; it is also expected that photon fields external to the jet are enhanced by these same electrons via external Compton process \citep[EC, e.g.,][]{Sikora_1994}. In \citet{Dermer_1995} relationships between observed flux density and powers of the kinematic Doppler factor are provided for the EC and SSC cases; for the EC $p=4+2\alpha$ and for the SSC $p=3+\alpha$, where $\alpha$ is the source spectral index.

Another factor that has to be taken into account is how beaming alters the shape of the intrinsic luminosity function. 
\citet{Urry_Shafer_1984} solved this problem for the case where the intrinsic luminosity function, $\phi(\mathscr{L})$, starts off as a 
simple power law of the form:
\begin{equation}
\phi(\mathscr{L}) = k_1\mathscr{L}^{-B}, 
\end{equation} 
where $B$ is the index of the distribution, and $k_1$ the normalization.
Under the assumption that the jets orientation are distributed uniformly, $P_{\theta}(\theta)\propto\sin\theta$, and that the plasma moves along the jet with a single $\Gamma$ value, $P_{\Gamma}(\Gamma)\propto\delta(\Gamma-\Gamma_0)$, \citet{Urry_Shafer_1984} showed that the observed luminosity function, $\phi(L)$, just by the act of beaming, becomes a broken power law.
The break coincides with the value $L_{\rm b} = \delta_{\max}^p\mathscr{L}_{\rm b}$. The low luminosity population is concentrated before the break (mostly with $\mathscr{L}\sim\mathscr{L}_{\rm b}$), and follows a distribution with index $(p+1)/p$ which, for reasonable values of $p$ (i.e., $p=2-10$), is within 1 and 1.5. Above the break, the luminosity function maintains the same spectral shape as the parent population (i.e., $\propto L^{-B}$), although of course the normalization differs.
It is important to note for the following discussion that, as pointed out by \citet{Urry_1991} and \citet{Lister_2003}, the resulting boosted luminosity function under more complex, and physically relevant, assumptions (e.g., $\phi(\mathscr{L})$ is a broken power law, or the distribution of $\Gamma$ values is not a $\delta$ function) shows the same sort of broken power-law behavior as \citet{Urry_Shafer_1984}.

Due to the lack of significant detection of a break in our XLF, we limit our analysis to the simple power-law $\phi(\mathscr{L})$.
It is sensible to assume that $\Gamma$ values observe a power-law distribution (of index $\mu$):
\begin{equation}
	P_\Gamma(\Gamma)\propto\Gamma^{\mu}.
\label{eq:pgamma}
\end{equation}
The analytical solutions for this more complex scenario are reported in \citet{Lister_2003} and \citet{Cara_Lister_2008}, and we follow their 
formulation
of the problem.

Adopting the modification on the $\Gamma$ distribution (Equation~\ref{eq:pgamma}), the probability density of $\delta$ is
\begin{equation}
P_{\delta}(\delta) = \frac{1}{\delta^2}\int_{f(\delta)}^{\Gamma_2}\frac{P_{\Gamma}(\Gamma)}{\sqrt{(\Gamma^2-1})} d\Gamma,
\end{equation}
where the lower limit $f(\delta)$ is reported in the Appendix of \citet{Lister_2003}. 
Finally all the ingredients are present to determine the observed luminosity function:
\begin{equation}
\phi(L) = k_1 L^{-B}\int_{\delta_1(L)}^{\delta_2(L)}P_{\delta}(\delta)\delta^{p(B-1)} d\delta,
\label{eq:obsphi}
\end{equation} 
where the limits of integration ($\delta_1(L)$ and $\delta_2(L)$) are taken from A12.
The left hand side of Equation~\ref{eq:obsphi}, $\phi(L)$, can be derived by de-evolving the best-fit luminosity function, $\phi(L,z)$, to redshift $z=0$ using the
weighted $1/V_{\rm max}$ method \citep[see][A09, A12]{Schmidt_1968, Della_Ceca_2008}. 
Once obtained, we can then perform a multivariate fit to derive the best-fit parameters that describe the distribution of jetted sources. 

On account of the fact that the above formulation requires numerous input factors (many also interdependent),
here we provide a list of the most relevant constraints for the parameters employed 
to the scope of our fit:
\begin{enumerate}
\item The limits for $\Gamma$ are taken from average properties of radio-loud blazars, $\Gamma_1=5$ and $\Gamma_2=40$ \citep[e.g.,][]{Lister_2009, Saikia_2016, Paliya_2019, Lister_2019}. 
\item The possible range of values for the $p$ parameter is derived by taking into account the fact that the average spectral index of BAT FSRQs is $1.78$ (see Table~\ref{tab:sample}), with the minimum being $0.8$ and the maximum $2.8$. Thus, following \citet{Dermer_1995}, $p$ is allowed to span a range $p\sim[4,6]$ in the SSC case, and $p\sim[5,9]$ in the EC one. Therefore, we 
test several $p$ values (kept frozen during the fit) in the range $p\in[4,9]$ with an increment of $+1$ at every fit. 
\item The lower limit on the intrinsic luminosity, $\mathscr{L}_1$, is dictated by the relation $\mathscr{L}_1=L_{\rm min}/\delta_{\rm max}^p$, where $L_{\rm min}$ is the minimum observed luminosity of $\phi(L)$ and $\delta_{\rm max} = \Gamma_2+\sqrt{(\Gamma_2^2-1)}$. The upper limit on the intrinsic luminosity does not influence the results of the fit and is arbitrarily chosen to be $\mathscr{L}_2=10^6\mathscr{L}_1$.
\end{enumerate}
The only free parameters left are the normalization $k_1$, the intrinsic luminosity index $B$ and the $\Gamma$ distribution
index value $\mu$. 
The fit is performed by employing a standard $\chi^2$ minimizing technique implemented via \texttt{Minuit}.

\subsection{Beaming results}\label{sec:beam_res}

\begingroup
\renewcommand*{\arraystretch}{1.2}
\begin{table}[]
\caption{Beaming fit values for $p=5$ and $p=7$}\label{tab:fit_distr}
\centering
\scalebox{1}{
%\hspace{-4cm}
\begin{tabular}{ c | c c}
 &   $p=5$  & $p=7$ \\
 \hline
	$\log(k_1)$ & $-17.0\pm0.8$ & $-23.9\pm 0.3$ \\ 
	$\mu$ & $-3.33\pm1.30$ & $-1.95\pm1.53$\\
	$B$ & $2.72\pm0.05$ & $2.71\pm0.04$\\
	$<\theta>$ & $2.77^{\circ}$ & $1.79^{\circ}$ \\
	$<\Gamma>$ & $8.3^{+3.5}_{-1.4}$ & $12.1^{+8.1}_{-4.0}$\\
\hline
	$\chi^2$/d.o.f. & $1.40$  &  $1.40$\\

\end{tabular}}
\end{table}
\endgroup

\begin{figure}
    \includegraphics[width=\columnwidth]{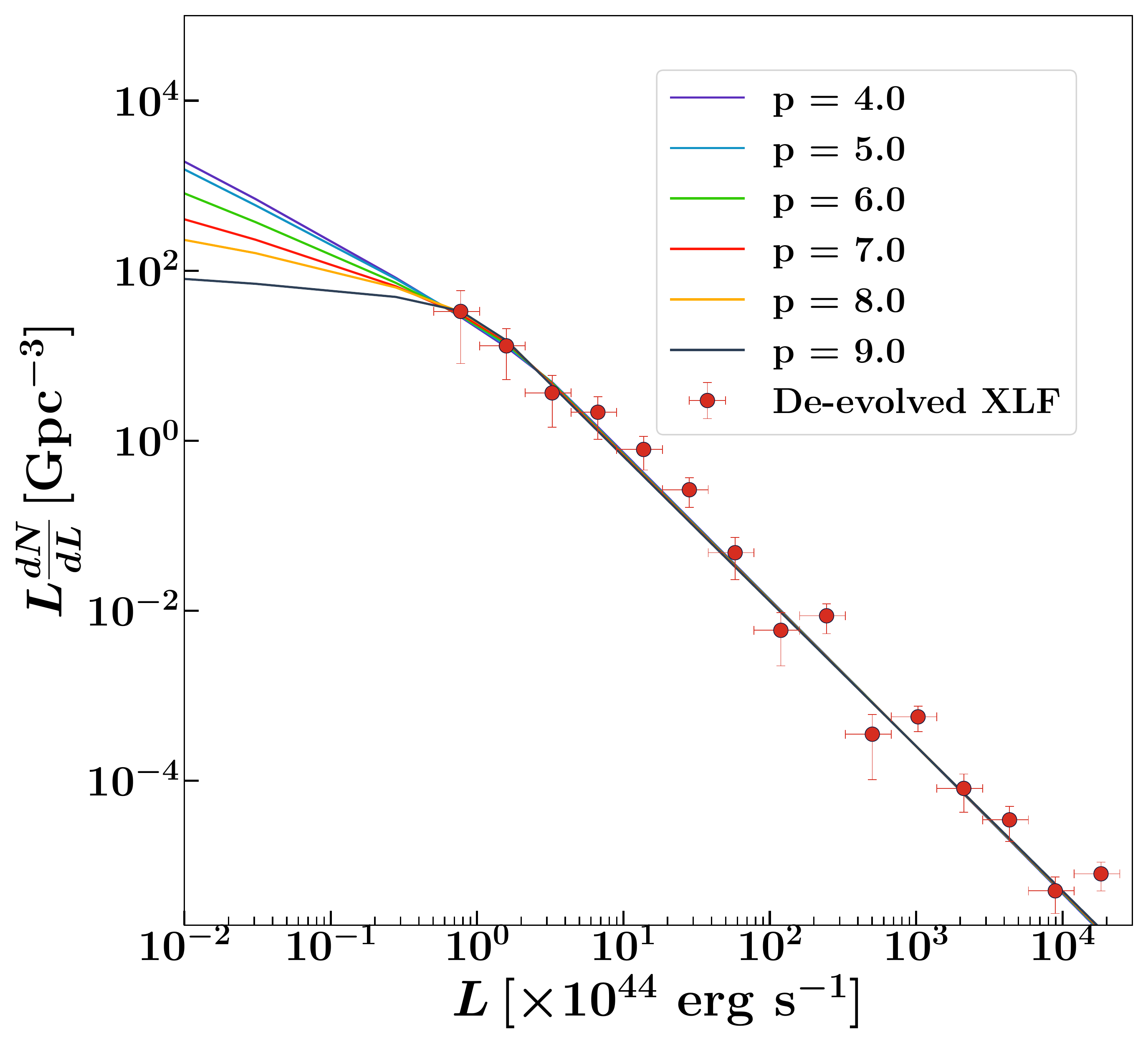}
    \caption{Best-fit to the de-evolved XLF ($\phi(L,0)$, red data points) using various values of the kinematic Doppler factor distribution index ($p$ parameter). Higher values of $p$ ($p>5$) are more likely attributed to EC process in the jet, while lower values are associated with the SSC process. As can be seen, for $p>7$ the XLF is predicted to show a break around the lowest luminosity bin of our sample ($L\sim10^{44}\,\rm erg~s^{-1}$) while for lower values of $p$ this break would happen at $L<10^{42}\,\rm erg~s^{-1}$. From our best fits we are not able to discern which values are better representing the population of these jets, though the range $p=[5,7]$ seems to be the more favorable overall (see details in Section~\ref{sec:beam_res}). }\label{fig:p_val}
\end{figure}

\begin{figure*}
    \includegraphics[width=\columnwidth]{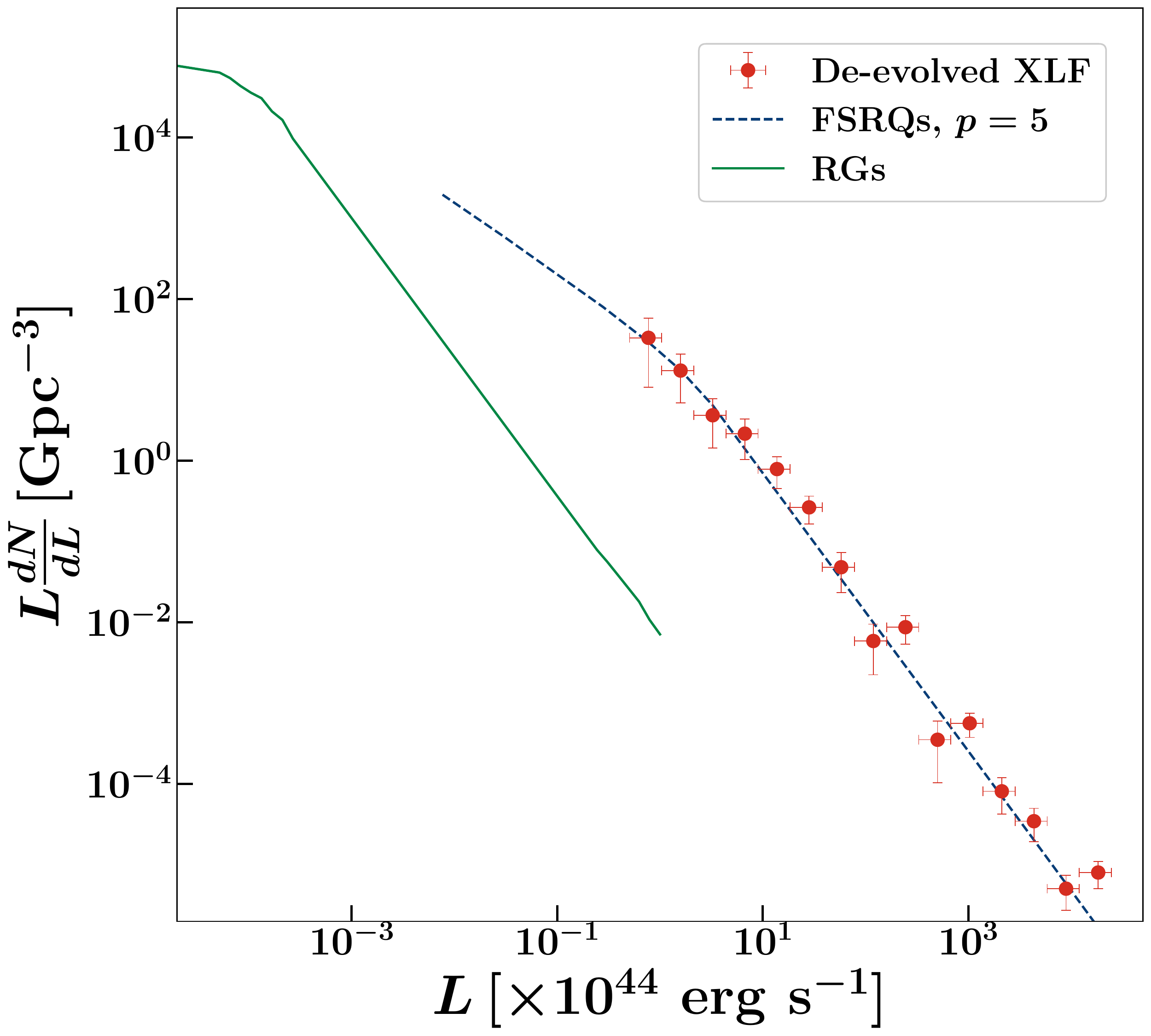}
    \includegraphics[width=\columnwidth]{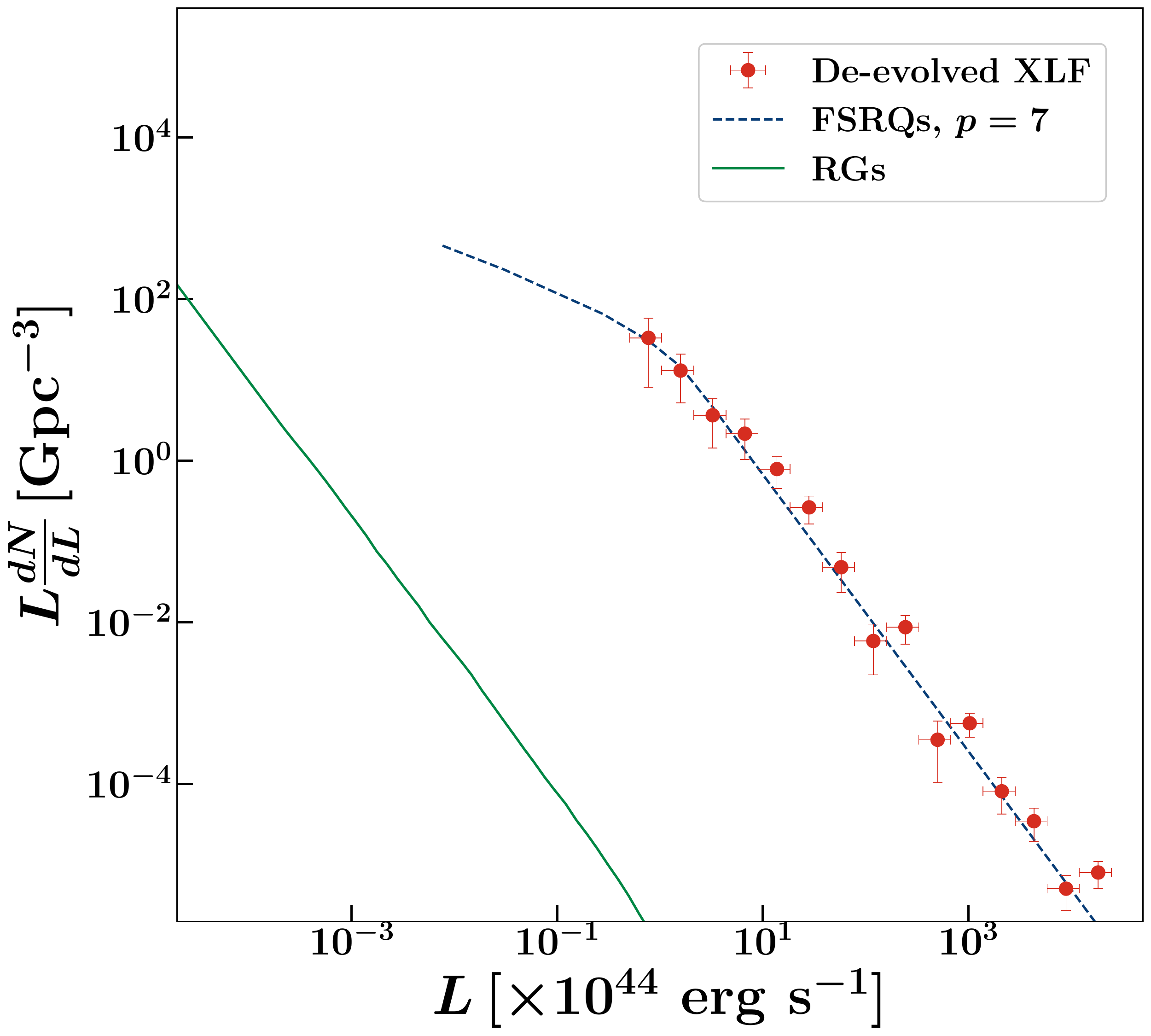}
 \caption{Results of the beaming fit to the BAT FSRQ population for $p=5$ (left) and $p=7$ (right). The red data points are the de-evolved luminosity function,
	$\phi(L,0)$, the blue dotted line is the fit derived for the FSRQ population, and the green line shows
	the distribution of the parent population (labeled here as RGs). As can be noted, different values of $p$ predict different 
	number densities of the parent population at different luminosity bins. For $p=5$, RGs would show a break in the distribution at $L\sim10^{41}\,\rm erg~s^{-1}$, luminosity at which they would be found with number densities of $10^4\,\rm Gpc^{-3}$. For $p=7$ instead this break would appear at even lower luminosities and their number densities at  $L\sim10^{41}\,\rm erg~s^{-1}$ would be two order of magnitudes lower with respect to $p=5$. 
	\label{fig:fit_distr}}
\end{figure*}

\begin{figure*}
    \includegraphics[width=\textwidth]{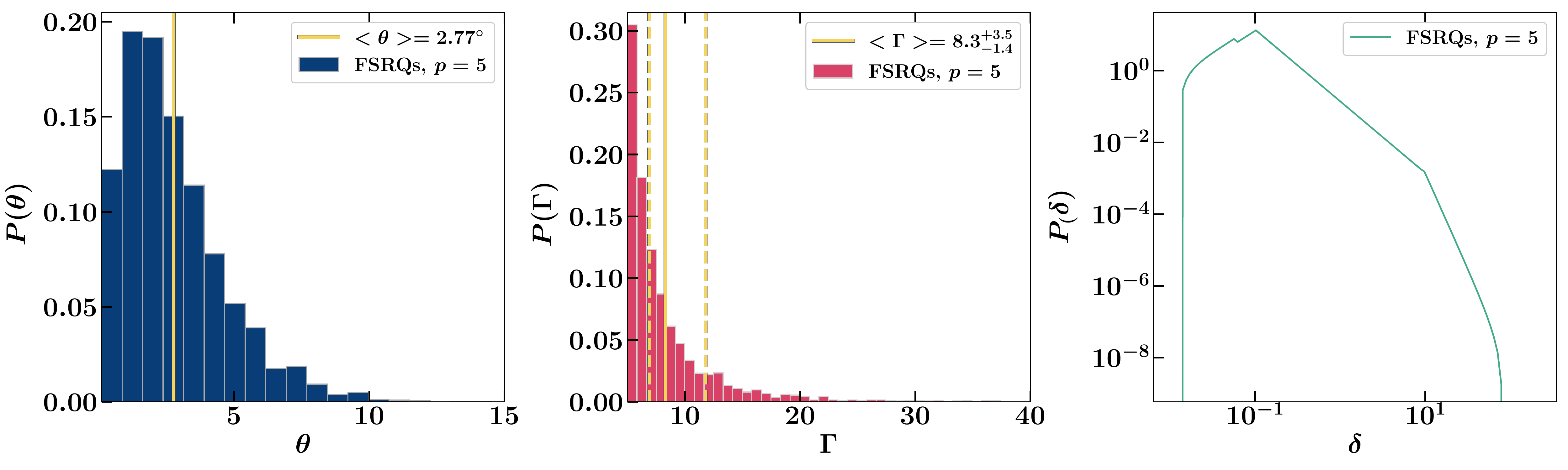}
    \includegraphics[width=\textwidth]{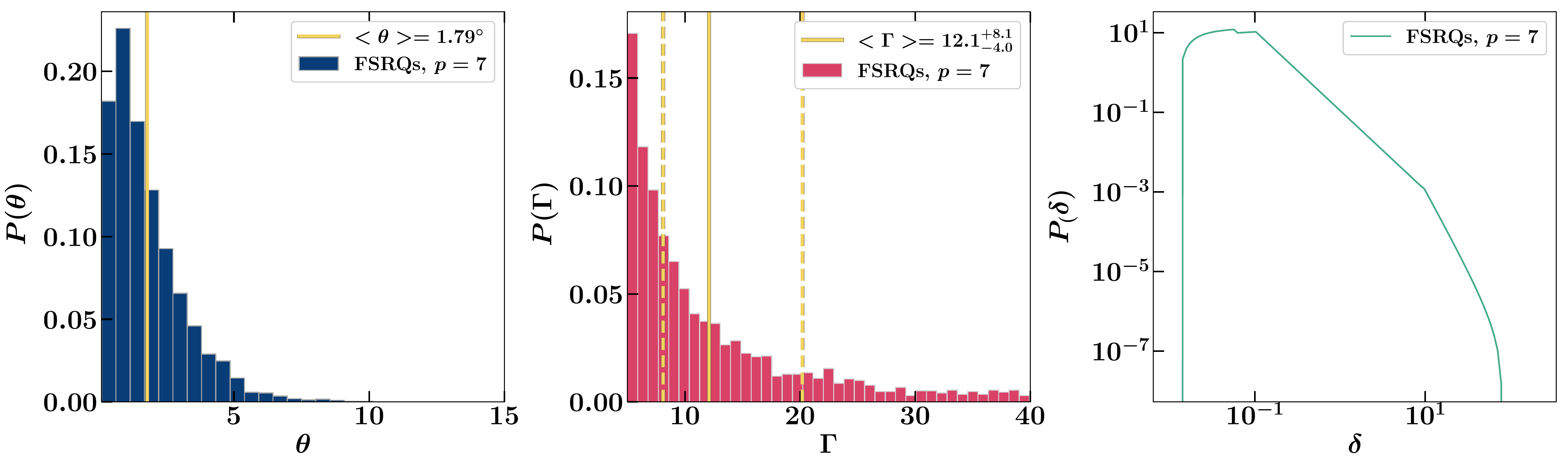}
	\caption{Distribution of $\theta$ ($P(\theta)$, normalized to 1), $\Gamma$ ($P(\Gamma)$, normalized to 1) 
	and $\delta$ ($P(\delta)$) for the beaming fit to the BAT FSRQ population
	for $p=5$ (top) and $p=7$ (bottom).
	The average for the $\theta$ and $\Gamma$ distribution are shown by the yellow solid line. For the latter, the associated errors are represented by the yellow dashed lines. We note how a lower value of $p$ predicts lower average $\Gamma$, as well as a narrower range of predicted $\Gamma$ values, with respect to a higher value of $p$. On the other hand, the viewing angle distribution is broader for lower value of $p$, though they remain distributed 
	between $1^{\circ}$ and $10^{\circ}$ and have an average $\theta\sim2-3\degree$, quite independently of $p$. 
	\label{fig:fit_distr_angle}}
\end{figure*}

Results of the beaming fit with different values of the $p$ parameter
are shown in Figure~\ref{fig:p_val}.
The reduced $\chi^2$ estimate for all fits are very similar
($\chi^2_{\nu}\sim1.4$), impeding us to disentangle
which value of the $p$ parameter better represents the distribution of these jets.
However, errors on the fit parameters get progressively larger at $p>7$. In the case $p=9$ the value of $\mu$ is completely unconstrained by the fit ($\mu\sim9\pm14$), rendering these higher values of $p$ less likely. Moreover, from Figure~\ref{fig:p_val}, it can be seen that 
higher values for $p$ (e.g., $p>7$, more likely attributed to EC process) predict a turnover of BAT FSRQ luminosity function at the level of the faintest observed luminosity bin of our sample ($L\sim10^{44}\,\rm~erg~s^{-1}$).
Lower values of $p$ instead place this peak at lower luminosities. 
It would indeed be necessary to detect this turnover to draw firm conclusions.
The best-fit parameter values for $p=5$ and $p=7$ are reported in Table~\ref{tab:fit_distr}. For the same two cases, Figure~\ref{fig:fit_distr}-\ref{fig:fit_distr_angle} show both the beamed and unbeamed luminosity function of the jets ($\phi(L)$ and $\phi(\mathscr{L})$, blue dashed and green solid line) as well as the normalized distributions of $\theta$, $\Gamma$ and $\delta$ factors.

\begin{figure*}
    \centering
    \includegraphics[width=0.8\textwidth]{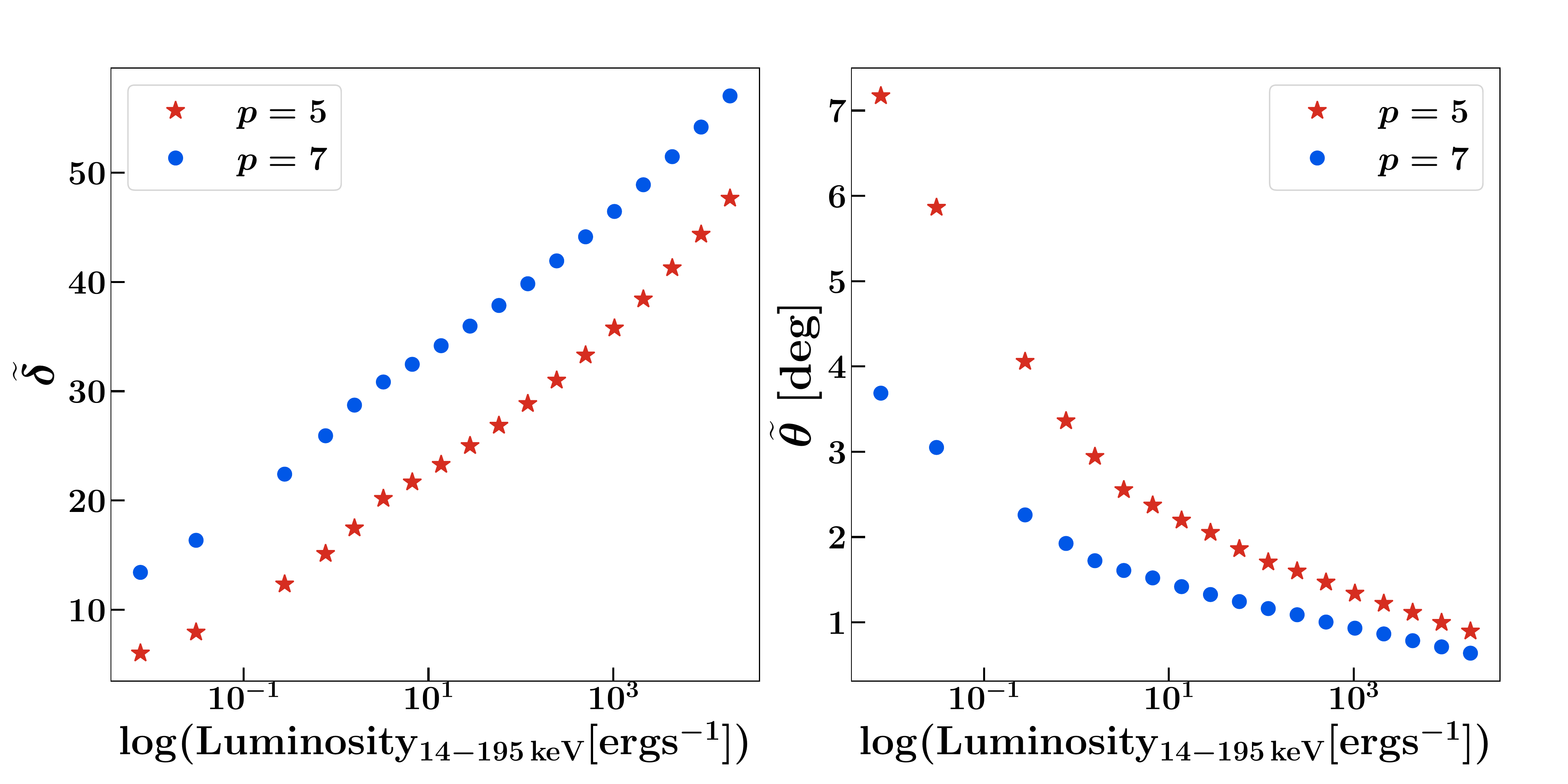}
    \caption{Median Doppler factors ($\widetilde{\delta}$, left) and median viewing angles ($\widetilde{\theta}$, right) as function of $14-195\,\rm keV$ luminosity derived from the beaming fit to the BAT FSRQ population
for $p = 5$ (red stars) and $p=7$ (blue dots). As the luminosity of the sources decreases, the $\delta$ factors become smaller and viewing angles get larger (corresponding to lower $\Gamma$ factors). }\label{fig:median_delta_theta}
\end{figure*}

The index of the jet Lorentz factors distribution is $\mu=-3.30\pm1.29$ for $p=5$ and $\mu=-1.95\pm1.53$ for $p=7$. 
The corresponding average $\Gamma$ factor are $<\Gamma>=8.3^{+3.5}_{-1.4}$ ($p=5$) and
$<\Gamma>=12.1^{+8.1}_{-4.0}$ ($p=7$). It can be noted how for higher values of $p$ the distribution of jets has on average higher bulk Lorentz factors, and their distribution is broader in the chosen $\Gamma$ range of values. 
As expected, the distribution of jet viewing angles derived through the fit 
is mostly confined in the range $\theta\in[0^{\circ},10^{\circ}]$ with an average value, $<\theta>\sim2-3^{\circ}$, and for higher values of $p$ this distribution is narrower than for lower ones. 
The beaming fit also enables us to derive how Doppler factors and viewing angles (therefore $\Gamma$) change as function of hard X-ray luminosity. This is shown in Figure~\ref{fig:median_delta_theta}. It can be seen that the lowest luminosity sources have lower $\delta$ and can be detected at larger viewing angles (hence have smaller $\Gamma$ factors); high-luminosity sources instead have higher $\delta$ but can only be detected at very narrow viewing angles (and higher $\Gamma$).
Lastly, the shape derived for the intrinsic LF recovers the $B=2.73\pm0.05$, quite independently from the adopted value for $p$. 

The best-fit value for the normalization ($k_1$) of the intrinsic luminosity function decreases as $p$ increases. This has implication on the predicted number density of the jetted parent sources as function of luminosity.
As shown in Figure~\ref{fig:fit_distr}, for $p=7$ our fit predicts $\sim10^{2}\,\rm Gpc^{-3}$
misaligned jets at $L\sim 10^{39}\,\rm erg~s^{-1}$, and the break of the distribution remains undetected in the plotted luminosity range, indicating its location to be at even fainter luminosities. Instead, for $p=5$
the fit predicts $\sim5\times10^{4}\,\rm Gpc^{-3}$, with a break of the population occurring
at $L\sim 5\times10^{39}\,\rm erg~s^{-1}$.
Thanks to this fit, we can derive the percentage of FSRQs to the total number density of their parent population. For $p=5$, this fraction results to be $\sim0.1\%$, while for $p=7$ the fraction becomes $\sim0.001\%$.

Considering the derived number densities of blazars and the distribution of $\Gamma$, we can estimate 
the number densities of parent population applying the $2\Gamma^2$ correction\footnote{For every jet 
found pointed close to our line of sight,
one can estimate the number of sources at the same redshift, with the same black hole mass,
but with jets pointed away from us. 
This estimate can be obtained by geometrical arguments assuming (1) the jets to be on both sides of the AGN, and (2) both jets have an opening angle ($\theta$) of $1/\Gamma$ (where $\Gamma$ is the bulk Lorentz factor
of the jet).
The number of misaligned jetted sources therefore
can be estimated as follows:
\begin{equation}
        N_{\rm misaligned}=\frac{A_{\rm sphere}}{A_{\rm jets}}= \frac{4\pi}{2\pi}\times \Gamma^2=2\Gamma^2. 
\end{equation}
}.
We note that recently \citet{Lister_2019} derived the properties of the parent population of radio jets and 
pointed out how the $2\Gamma^2$ correction is invalid. The authors find that for $\Gamma>15$ there is
a shallow increase in the predicted number of parent jets for each jet found with a particular Lorentz factor ($\sim15\times\Gamma$ instead
of $2\Gamma^2$); for $\Gamma<15$ instead the parents are distributed according to $(2\Gamma)^{p}$.
The number densities of FSRQs in our work (see Figure~\ref{fig:num_dens}) range between $[50,200]\,\rm Gpc^{-3}$ for the luminosity bin $\log (L)=[43.7,45.8]\,\rm~erg~s^{-1}$,
$[0.2,4]\,\rm Gpc^{-3}$ for the luminosity bin $\log (L)=[45.8,47.1]\,\rm~erg~s^{-1}$, and $[0.02,0.4]\,\rm Gpc^{-3}$ for the luminosity bin $\log(L)=[47.1, 48.4]\,\rm~erg~s^{-1}$.
The parent population densities for the case $p=5$ and $p=7$, using both the standard $2\Gamma^2$ correction and the modified formulation from \citet{Lister_2019} are listed in Table~\ref{tab:number_dens_parents}. The results show that, depending on $p$ and on the luminosity bin, the number of parents range between $2\,\rm Gpc^{-3}$ and $10^4\,\rm Gpc^{-3}$.

\begingroup
\renewcommand*{\arraystretch}{1.2}
\begin{table*}[]
\centering
\caption{Number densities of the jet's parent population derived from the beaming for $p=5$ and $p=7$ using the simple $2\Gamma^2$ correction and the modification from \citet{Lister_2019}.}\label{tab:number_dens_parents}
\begin{tabular}{ c | c | c | c}
$\log(L)[\,\rm erg~s^{-1}]$ & $[43.7,45.8]$ & $[45.8,47.1]$ & $[47.1, 48.4]$ \\
\hline
\hline
$p=5$ & & & \\
\hline
$2\Gamma^2$ & $[9.0\times10^3,3.6\times10^4]\,\rm Gpc^{-3}$ &  $[36,723]\,\rm Gpc^{-3}$  &  $[3.5,70]\,\rm Gpc^{-3}$ \\
\citet{Lister_2019} &  $[6.6\times10^3,2.6\times10^4]\,\rm Gpc^{-3}$  & $[26,530]\,\rm Gpc^{-3}$ & $[2,53]\,\rm Gpc^{-3}$\\
\hline
\hline
$p=7$ & & & \\
\hline
$2\Gamma^2$ & $[2\times10^4,8.2\times10^4]\,\rm Gpc^{-3}$ &  $[82, 1.6\times10^{3}]\,\rm Gpc^{-3}$  &  $[8,164]\,\rm Gpc^{-3}$ \\
\citet{Lister_2019} & $[9.6\times10^3,3.8\times10^4]\,\rm Gpc^{-3}$ & $[38,772]\,\rm Gpc^{-3}$ &  $[3,77]\,\rm Gpc^{-3}$  \\
 \hline
\end{tabular}
\end{table*}
\endgroup

\section{Discussion}\label{sec:disc}

In this work, we derive the most up-to-date BAT blazar luminosity function. 
We rely on a clean, significance limited, sample of 118 blazars (88 belonging to the FSRQ class) 
detected by the BAT at $>5\sigma$ above Galactic latitudes $|b|>10^{\circ}$ in 105 months of survey. 
An important thing to keep in mind for our work is that the FSRQ population dominates the inferences about our entire blazar sample.
Being more numerous and detected to higher redshifts, these sources 
set the more stringent constraints for the derived luminosity function.
BL Lac sources are mostly concentrated at low redshifts (only 2 sources have $z>0.36$), and 70\% of them reside in the lowest luminosity bin of the derived XLF ($\log(L)=43.7-45.4\,\rm[erg~s^{-1}]$, see Table~\ref{tab:sample} and Figure~\ref{fig:evol}). These objects have been found by previous work to show zero to mildly negative evolution (e.g., A14) and to contribute negligibly to the $14-195\,\rm keV$ background (see A09). In fact, looking at the consistent results between the evolution of FSRQs (Section~\ref{sec:lf_fsrq})
and the whole population (Section~\ref{sec:lf_all}), we have deemed unnecessary to 
further investigate the BL Lacs evolution in the BAT energy range. It follows that, throughout this discussion, the word blazar is used as a synonym of FSRQ 
(and vice versa). 
Our main findings are listed and discussed below. 

\subsection{The blazar X-ray luminosity function}\label{sec:XLF_res}

The blazar X-ray luminosity function derived in this work highlights several results. First, as discussed in Section~\ref{sec:res}, it is yet not possible to 
discern which kind of evolution takes place in this source class. Both density and luminosity evolution give compatible fits, 
of maximum likelihood values comparable to each other.
This result is easily comprehended when plotting the luminosity function both in terms of redshift and 
luminosity (Figure~\ref{fig:evol}). The lack of any significant break in the distribution is evident, 
which translates into the fact that PDE and PLE models are essentially indistinguishable
from each other \citep{Bahcall_1977}.
This leads us to conclude that the BAT survey is still sampling only the
high-luminosity blazar population at every redshift, while missing the bulk of the low-luminosity one. A slight 
hint of the occurrence of a break is present in the XLF model prediction for the lowest redshift
bin (bottom plots of Figure~\ref{fig:evol}). However, 
the statistical uncertainties related to its position remain large and its evolution with redshift and luminosity undetermined.

Another result is the fact that introducing a double power law to describe the local luminosity function, $\phi(L, V(z=0))$, significantly 
improves the fit. This feature is an expected consequence of the beaming effect and was already noticed by A09. 
The errors on the low-luminosity end slope are still quite large due to the lack of detection of many
low-luminosity sources, but its value is compatible with the anticipated 1 to 1.5 from beaming predictions.
Another expected and confirmed result is that the 
high-luminosity end of the BAT blazar distribution displays an index of $\gamma_2'\sim2.7$.
In A09, this slope was found in the simple power-law luminosity function scenario but was not recovered 
in the double power-law case ($\gamma_{2',\rm A09}\sim3.7$), possibly due to the limited sample size.
Therefore, our results using a larger and deeper sample confirm that the bright-end of the blazar luminosity function 
is consistent with the slope estimated for the unbeamed jetted AGN population 
(FRIIs luminosity function has been found to display a bright-end slope of $\gamma'_{\rm FRII}\sim2.65$, \citealp[see][]{Cara_Lister_2008}).

In agreement with previous results, the derived evolutionary parameters, as well as the {\it logN-logS} trend, point to 
a positive evolution of the BAT blazar class (i.e., more/more luminous
sources at earlier times). 
The values of $k$ are congruent between our work and A09, and they are consistently $>2$.
For our best-fit mPLE, $k_{\rm mPLE}=3.23\pm0.57$ (see Table~\ref{tab:res_ml}) and in A09 $k_{\rm mPLE}=3.67\pm0.48$.
Moreover, we confirm that the peak of the high-luminosity BAT-FSRQ population is located at $z>4$.
Differently than constrained only by upper limits in A09,
our dataset extending up to $z=4.65$ allows us to measure the redshift peak directly.
The peak position predicted by our best fit is $z_{\rm peak}=4.3\pm1.0$, which is in good agreement with the $z_{\rm peak}=4.3\pm0.5$ 
reported A09. Though our uncertainties on the position are larger than in A09, 
ML fits performed forcing the peak to be at a specific redshift confirm that it is more likely to occur between $z=4.3-4.4$ (see Section~\ref{sec:lf_fsrq} for detailed description). 
Even at $z_{\rm peak}=4.6$ the fit results return maximum likelihood and parameter values consistent with our best-fit scenario (see Figure~\ref{fig:log-like-zpeak}).
This behaviour has strong implication in the number density of blazars expected at high redshift. As shown 
in Figure~\ref{fig:num_dens}, number density values predicted at $z\sim4$ are two times lower than those derived in A09.

This early evolutionary peak still remains very puzzling. In fact, LAT blazars (see Section~\ref{sec:mev_gev}), non-jetted AGN, galaxies, 
and star formation history of the universe are all found to coherently peak at $z\sim2$. As pointed out in A09, the only other class of objects 
that resembles in evolution the most luminous BAT blazars is the massive elliptical galaxies one \citep{DeLucia_2006}. 
Giant elliptical galaxies are understood to be a by-product of major merger events, which have also been shown to be a quick 
and viable mean to fast black hole accretion. In the chaotic high-redshift universe merger fractions have been found to be
larger for highly massive galaxies even at $z>3$ \citep[e.g.,][]{Bluck_2009,Whitney_2021}. 
Therefore, one can picture a scenario in which there is a strong link between enhanced merging activity (leading to higher mass black holes and 
elliptical host galaxy morphology) and a jetted phase of the AGN in the early universe. 
The first evidence of a nascent jet possibly triggered by a merger was 
reported by \citet{Paliya_2020_a}, further strengthening this scenario. Moreover the more luminous blazars
are powered by the most massive black holes (see later), pointing to a further connection between 
jet activity and supermassive black hole growth.
This envisaged paradigm also fits in the picture of cosmic ``downsizing" of AGN activity, in which 
most massive and active black holes form quite early on, following the hierarchical 
build up of structures in the universe, and less powerful AGNs exist at later times \citep[e.g.,][]{Trakhtenbrot_2012,Shen_2020}.
Finally, as invoked in A09, if jets are powered through the \citet{B_Z_1977} mechanism, then they may be tapping into an extra 
reservoir of energy provided by the black hole spin. Positive evidence that jets are indeed more powerful than 
their accretion disks has been found even for sources at $z>3$ \citep[e.g.,][]{Ghisellini_2014, Paliya_2017}. Calculations for major merger events also show how these can produce maximally spinning black holes \citep[see][]{Berti_Volonteri_2008, Volonteri_2010}
and that high-redshift black holes spin faster than their lower redshift counterpart \citep[e.g.,][]{Volonteri_2016}. 
Therefore not only these powerful blazars help us trace the evolution of massive black holes in the early universe, but also 
the history of black hole spins.

\subsection{Number Density of Blazars and Parent Population} 
The Doppler boosting affecting the jet emission allows us to derive the properties of the parent population 
of randomly oriented jets (Section~\ref{sec:jets}). A few caveats have to be taken into consideration while examining these results. Firstly, an expected
outcome of beaming on the local luminosity function is a flattening of the low-end of $\phi(L, V(z=0))$, which is not yet clearly 
seen in our sample. 
Therefore, the fit parameters have large uncertainties. Secondly, the possible explored values of $p$ (the index of the 
kinematic Doppler factor distribution) strongly relies on the jet configuration as well as emission processes. The finding that the 
high-energy SED peak location does not strongly depend on the source luminosity (Section~\ref{sec:ave_sed}), 
is suggestive of the fact that EC process is 
dominant in the considered sample. Therefore, higher values of $p$ (e.g., $p=5-7$) should be considered as more relevant for the population. 

The fit enables us to derive the power-law distribution bulk Lorentz factors, which is shown in its normalized version in the middle panels of Figure~\ref{fig:fit_distr_angle} for two values of $p$. Our result show that higher values of $p$ predict
a distribution of jets with on average higher bulk Lorentz factors and has a larger spread (between the chosen $\Gamma=[5,40]$). 
If we compare with the results of A12 on LAT blazars, for $p=5$ 
the $\Gamma$ distribution is less steep ($\mu_{\rm LAT}=-2.43\pm0.11$ vs.~$\mu_{\rm BAT}=-3.33\pm1.30$ ) and the average $\Gamma$ is higher ($<\Gamma_{\rm LAT}>=10.2_{-2.4}^{+4.8}$ vs.~$<\Gamma_{\rm BAT}>=8.3_{-1.4}^{+3.5}$). 
This is in very good agreement with the outcome from radio monitoring of LAT blazars by the VLBA \citep[e.g.,][]{Lister_2009}, 
which shows that LAT detected blazar jets have, on
average, higher velocities that non-LAT detected ones. 
It is interesting to notice that recent single source studies on high-redshift blazars with 
combined hard X-ray and $\gamma$-ray detection \citep[e.g.,][]{Ajello_2016,Marcotulli_2017,Ackermann_2017, Marcotulli_2020_a} 
find that the jet powers could be lower than previously thought for these sources (i.e., average $\Gamma\sim9-10$, instead of
the canonical $15$, \citealp{Sbarrato_2015}).
Lower values of $\Gamma$ for these high-redshift luminous blazars \cite[cf.][]{Paliya_2020_b} have strong implication
on supermassive black holes space densities (Section~\ref{sec:smbh_space_dens}). 

The distribution of jet viewing angles derived through the fit is very narrow (left panels of Figure~\ref{fig:fit_distr_angle}).
This can be understood in view of the fact that the most luminous and highest redshift sources have to be extremely aligned (and have high $\Gamma$ factors) for them to be detected in the BAT band. On the other hand, the lower luminosity sources, residing at lower redshifts (see Figure~\ref{fig:blazar_sample}) can be found a larger viewing angles ($\theta_{\rm max}\sim10\degree$) and with lower $\Gamma$ factors. The distributions of $\delta$ factors and viewing angles as function of BAT luminosity (Figure~\ref{fig:median_delta_theta}) highlight these strong geometrical selection effects, which are also visible in the $14\,\rm keV$ to $10\,\rm GeV$ average SED of these sources (see right panel of Figure~\ref{fig:ave_sed} and Section~\ref{sec:lat_non_det_bl}). As the source luminosity decreases, the peak position slightly shifts towards lower energies ($\Delta E_{b, \rm rest-frame} \sim10\,\rm MeV$), indicative of lower $\delta$ of the jets \citep[e.g.][]{Ghisellini_2015,Sbarrato_2015, Lister_2019, Paliya_2019}.
Finally, the shape derived for the intrinsic luminosity function ($B=2.73\pm0.05$) is in agreement the spectral slope of the blazar XLF ($\gamma_2'\sim2.7$) as well as the one of the unbeamed jetted AGN.

Number densities of the parent population are strongly affected by the value of $p$, and can change by orders of magnitude depending on the resulting fit (see Figure~\ref{fig:fit_distr}). 
The derived percentage of FSRQs to the total number density of their parent population is $\sim0.1\%$ for the case $p=5$ (i.e., one jet out of a hundred would be detected in the blazar orientation), while higher values of $p$ make these estimate decrease by order of magnitudes (i.e.~for $p=7$, one jet in a thousand would be detected as blazar).
Considering the derived number densities of blazars and the distribution of $\Gamma$,
we could estimate the number densities of parent population applying the $2\Gamma^2$ correction as well as its modified formulation from \citet{Lister_2019}.
The results are quite similar in values with the two approaches, implying the existence from a few to more than 10 thousands jets in the universe (depending on the chosen luminosity bin, see Table~\ref{tab:number_dens_parents}).
The low-mid luminosity estimates are consistent with the estimated number density of FRII \citep{Cara_Lister_2008}, which is $\sim1.6\times10^{3}\,\rm Gpc^{-3}$ (for $L_{\rm 15 GHz}\geq1.3\times10^{25}\,\rm W~Hz^{-1}$), though the 
total number of parents is higher than the one so far predicted for FRIIs.
Overall, values of the $p$ parameter in the range 
$5-7$ provide a better description of our jet distribution. 

\begin{figure}[t!]
    \includegraphics[width=\columnwidth]{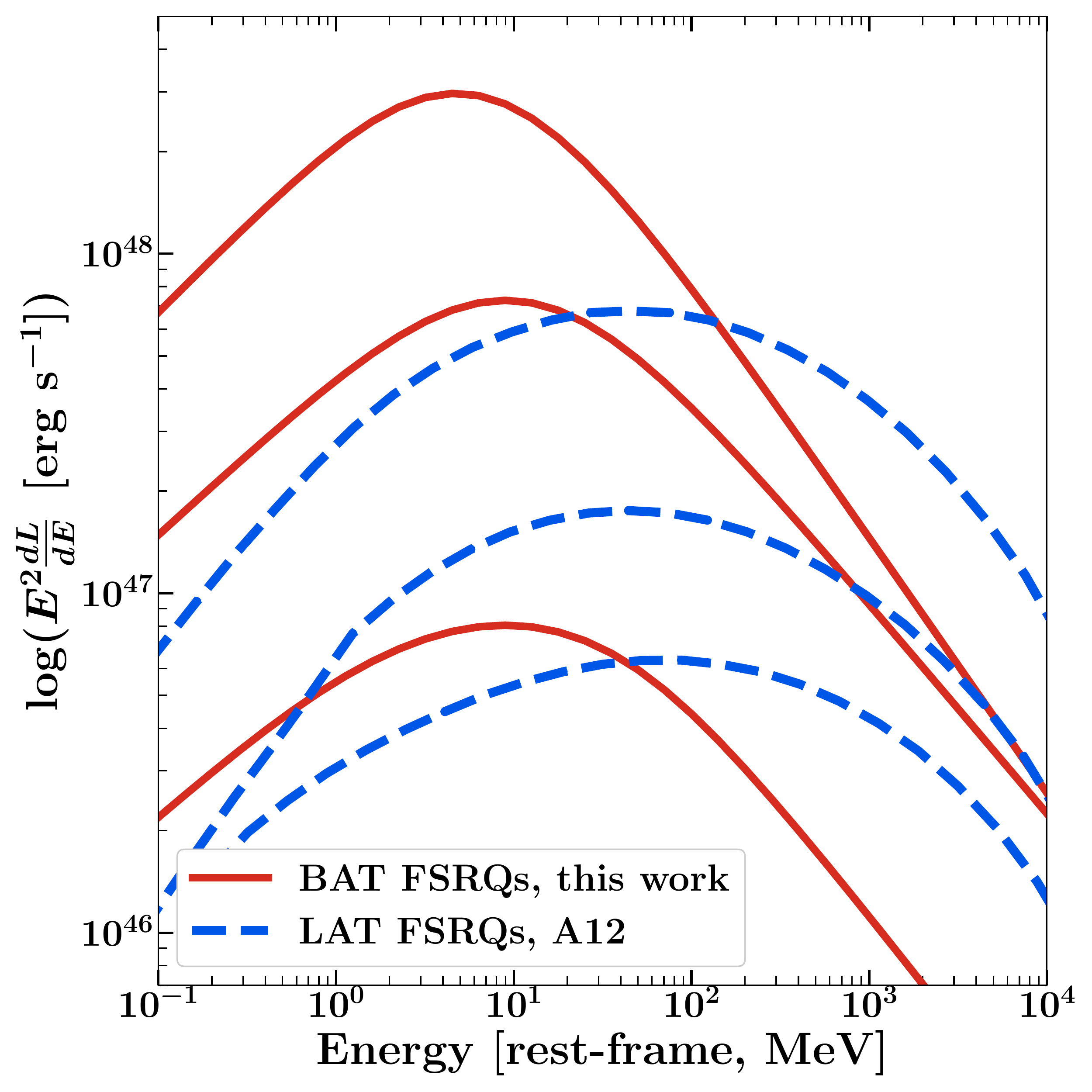}
	\caption{Comparison between average BAT-blazar and LAT-blazar SEDs. BAT blazar SEDs are shown by the magenta solid lines and are the ones derived from this work in the X-ray luminosity bins ($\log L_{\rm BAT} = [46.6, 47.3], [47.3,48.3], [48.3, 48.8]\,\rm erg~s^{-1}$, see also Figure~\ref{fig:ave_sed}). The LAT SEDs are the ones reported in A12 for the bolometric luminosity (derived in their work from the $\gamma$-ray luminosity) bins $\log L_{\rm bol} = [47.6, 47.9], [47.9,48.5], [48.5, 49.3]\,\rm erg~s^{-1}$ and are shown by the blue dashed lines. 
	This Figure highlights that BAT blazars have their high-energy peak located at lower energies than the LAT sources. Moreover, both classes of objects do not show strong evolution of the SED shape and/or peak depending on the chosen luminosity bin.}\label{fig:lat_bat_sed}
\end{figure} 

\subsection{MeV versus GeV blazars}\label{sec:mev_gev}
Hard X-ray and $\gamma$-ray blazars have been surmised to belong to the same source
population, but carrying slightly different characteristics.
The most luminous FSRQs detected by 
BAT belong to the class referred to as `MeV blazars'. Their high-energy 
SED, as derived in Section~\ref{sec:ave_sed}, peaks in the MeV band ($E_{b, \rm rest}\sim 3-17\,\rm MeV$), 
and they are extremely luminous, capable of reaching $L_{\rm bol}>10^{49}\,\rm erg~s^{-1}$. 
Single source studies of these MeV blazars usually find that their high-energy emission
is dominated by the EC process\footnote{For this interpretation, we assume a pure leptonic emission
model, and we do not consider the more complex, but equally valid, hadronic emission scenarios.} rather than SSC.
Further evidence confirming this scenario lies in our derived average $ 14\,\rm KeV- 10\,\rm GeV$ SED (Section~\ref{sec:ave_sed} and Figure~\ref{fig:ave_sed}), which shows similar slopes as well as 
peak positions, independently of the luminosity bin considered.
This behaviour is in contrast with the anti-correlation found instead for the low-energy synchrotron peak component
\citep[e.g.,][]{Padovani_1998, Ghisellini_2017} and in the high-energy SED of high-synchrotron peaked BL Lacs (which instead agree with the SSC paradigm). 
The lack of strong luminosity dependence on the peak position of the SED makes the jet parameters independent on X-ray luminosity or 
redshift. In turn, this translates into the fact that the BAT FSRQs belong to a population with homogeneous properties.
In Figure~\ref{fig:lat_bat_sed} we overlay our average BAT-blazar SED and the LAT-blazar SED
from A12. As can be seen, the high-energy SED of LAT FSRQs shows a peak between $E_{b, \rm rest-frame}\sim50-100\,\rm MeV$, 
and they are on average less luminous than the BAT ones.
Their emission, similarly to BAT FSRQs, is attributed to EC emission from the jet and their spectral shape 
does not show a strong evolution as function of luminosity.
Interestingly, the shape of the XLF at $z=0$ is very similar to the $\gamma$-ray luminosity function (GLF) in terms of 
spectral indices (while the normalization and typical break luminosity differs). 

Slight differences between the two classes appear in the derived jet properties of the parent population.
The BAT-blazar jets are found to be on average slower than the LAT detected ones (for $p=5$, $\Gamma_{\rm BAT} = 8.3$ 
vs.~$\Gamma_{\rm LAT} = 10.2$).
Predicted number densities for LAT FSRQs are $\sim1500\,\rm Gpc^{-3}$ while for BAT FSRQs these may be as high as 
$10^4\,\rm Gpc^{-3}$. 
Major differences between the two classes are (1) their typical average luminosities, and (2) their evolutionary properties. 
In fact, in terms of evolution, it had been surmised that since the high-energy emission is ascribed to the same radiative
process for both classes, then their evolution should be similar, if not the same. However, discrepancies have been 
noticed. LAT FSRQs have a peak in the evolution at $z_{\rm peak}\sim1.6$ after which their space density decreases quickly.
This peak occurs significantly later than the one derived here for the most luminous BAT FSRQs ($z_{\rm peak}\sim4.3$). Recent works 
have also highlighted that a PDE evolution may be taking place in the LAT blazar sample \citep{Marcotulli_2020_b} while in this work 
we find that a luminosity evolution seems to be favored. 
Furthermore, BAT FSRQs are powered on average by more massive black holes than the LAT FSRQs \cite[see,][]{Ghisellini_2017,Paliya_2019, Paliya_2021}.

These properties suggest that hard X-ray and $\gamma$-ray blazars belong to the same family of sources, with the same homogeneous properties, which has undergone a two-phase evolutionary sequence.
In line with the discussion in Section~\ref{sec:XLF_res}, this could mean that these jets
and their black holes are following the cosmic downsizing: the most extreme and luminous jets powered by the most massive black holes (BAT blazars) 
peak earlier in cosmic history, to then become less powerful in the late universe (LAT blazars),
tracing more closely the evolution of the non jetted AGNs \citep[e.g., A12,][]{Ueda_2014,Aird_2015,Shen_2020}.

Differences in evolutionary paths of BAT and LAT blazars have been already reported in previous works \citep[A09, A12, A14,][]{Toda_2020}, and possibly a combined LAT-BAT-detected blazar luminosity function study is due. 
Alternatively, an all-sky MeV instrument (e.g., COSI, \citealp{Tomsick_2019}; AMEGO-X, \citealp{Caputo_2022}, see Section~\ref{sec:mev_pred}) would enable us to 
produce new measurements of the background in the MeV energy range, 
to pinpoint the peak of the SED of these powerful blazars, and to 
understand what type of evolutionary scenario is taking place in this source class.

\begin{figure*}
\centering
\includegraphics[width=0.40\textwidth]{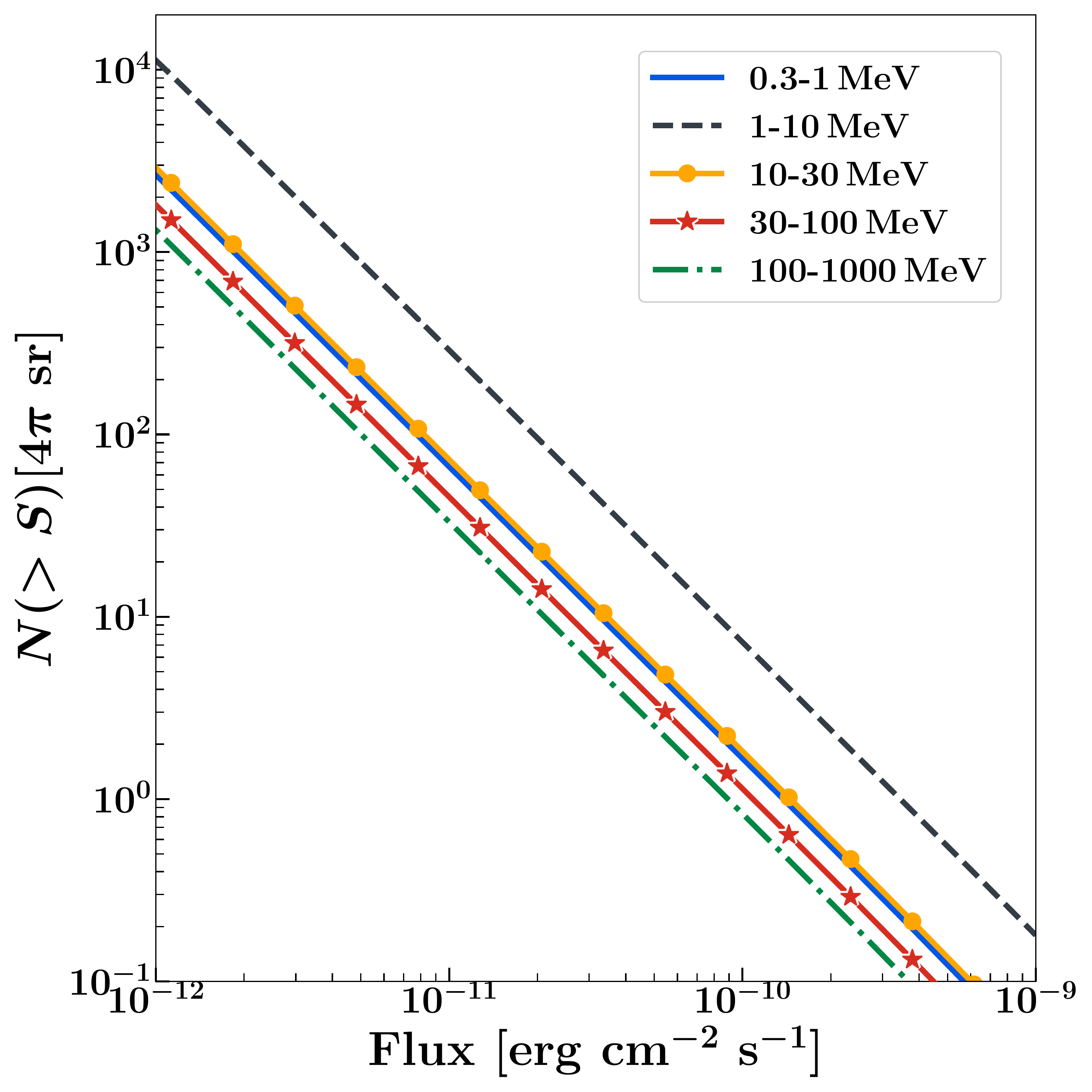}
\includegraphics[width=0.51\textwidth]{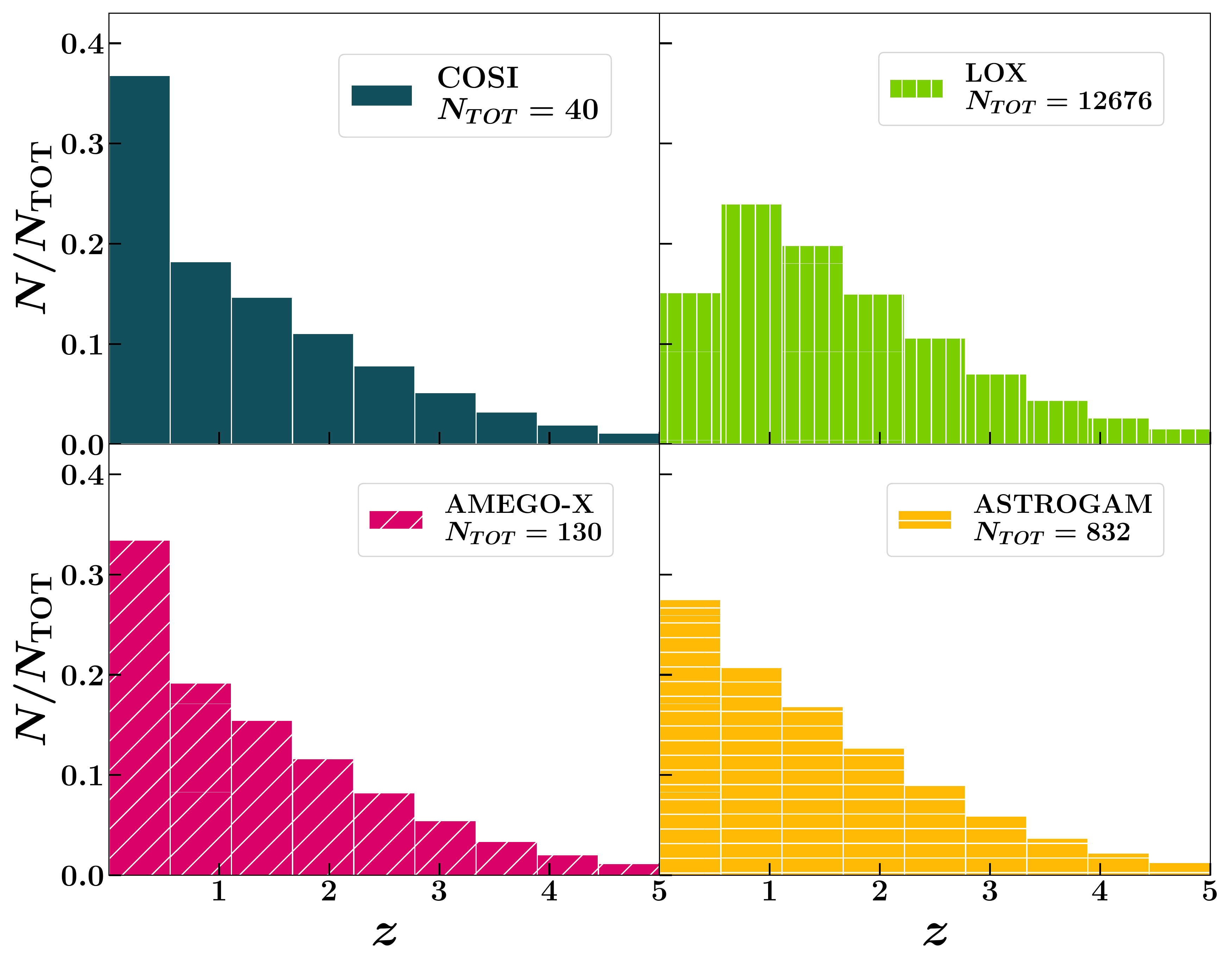}
    \caption{{\bf Left}: predicted {\it logN-logS} from our 
	best-fit XLF and SED models in different MeV bands. It can be seen that an instrument with a uniform sensitivity of $S=10^{-12}\,\rm erg~cm^{-2}~s^{-1}$ would be able to detect between $10^3$ and $10^5$ sources (depending on the energy range). {\bf Right:} histogram of fraction of sources as function of redshift expected to be detected by proposed MeV mission (see Section~\ref{sec:mev_pred}).\label{fig:mev_pred}}
\end{figure*}

\subsection{Supermassive black hole space density}\label{sec:smbh_space_dens}

The work of \citet{Sbarrato_2015} showed how for radio-loud (i.e., jetted) quasars
the supermassive black hole (SMBH, $M_{\rm BH}>10^9\Msun$) space density peaks earlier than radio-quiet (i.e., no jets) AGNs.
The authors considered the evolution of BAT blazars derived by A09 and employed the $2\Gamma^2$ correction to infer these densities.
However, two key assumptions had taken place: first that all $L>10^{47}\,\rm erg s^{-1}$ host $M_{\rm BH}>10^9\Msun$,
and second that the average $\Gamma$ factor is $\sim15$. 

Firm spectroscopic mass measurements are confirming 
that the most powerful blazars indeed host black holes with $M_{\rm BH}>10^9\Msun$. 
On the other hand, in our work we find that on average the value of $\Gamma$ factors
derived for these jets could be lower than previously assumed. This has implication on SMBHs
space densities of jetted sources. In fact, the authors derive the number density of SMBH traced by radio-loud
sources to be $\sim60\,\rm Gpc^{-3}$ at $z_{\rm peak}\sim4$. A Lorentz factor such as $6-8$ could lower this estimate by almost an order of magnitude.
Finally, we note that according to our derived evolutionary function, the number density of 
luminous jets is lower at $z>4$ than previously derived, and recall that the $2\Gamma^2$ may be an incorrect 
approximation when assessing the size of the parent population. Possibly a combination of better constrained LF and 
derived jet properties can help up shed a light on the population of supermassive black holes in the early universe.
This conundrum could also be helped by a deeper X-ray survey combined with multi-wavelength single source studies.

\subsection{Prospects for the MeV range}\label{sec:mev_pred}

The contribution of blazars to the CXB background, according to our best fit model, can range between $5-20\%$ in the $14-195\,\rm keV$ band. 
Therefore they are subdominant in this energy range. 
It is worth nothing that these limits will not significantly improve with longer BAT surveys, owing to the fact that the significance gain goes as $\sqrt{t}$ ($t$ being the time covered by the survey).
On the other hand, this source population 
has been found to contribute $>50\%$ of the EGB (depending on the 
energy range even up to $100\%$, see e.g. \citealp{Ackermann_2015, DiMauro_2018, Marcotulli_2020_b}). The most powerful of these jets have also been surmised to account for most of the MeV background. 
Indeed, as derived in 
Section~\ref{sec:ave_sed}, the average SED of FSRQs peaks at $\sim3-17\,\rm MeV$, i.e., the most of their energy output falls in the MeV band. 

In this work, we have extrapolated the contribution of powerful BAT FSQRs to the MeV background. From Figure~\ref{fig:cxb},
it can be seen how indeed the MeV background could be entirely produced by blazars alone. It is known that other source populations 
could contribute as well, up to few \%, to the MeV background. Our best-fit mPLE model convolved with the average SED
does have large uncertainties in the MeV band, hence allowing for contribution from other source classes. 
Only an MeV mission would provide the opportunity to constrain the background level and shape to a higher 
significance (latest reported by COMPTEL, \citealp{COMPTEL_2000}), as well as unveil the bulk of these blazar sources. 

With the best fit XLF and the blazar SED in hand, we can make prediction on the expected {\it logN-logS} from these powerful 
BAT FSRQs in the MeV band. 
The cumulative source count distribution, $N (> S)$,
is given by Equation~\ref{eq:lognlogs}. To extrapolate this function to a different energy range, 
it is necessary to allow the lower limit of the integral to become energy dependent, $L_{\rm min}(S_{\rm E_1, E_2}, z)$.
Therefore, the luminosity relies on the SED shape, and $S_{\rm E_1, E_2}$ is the flux limit in the requested 
energy range. It follows that the lower limit in luminosity depends on the flux sensitivity of the mission, at specific energy 
ranges. For the purpose of this derivation, however, we consider an arbitrary minimum flux of $10^{-13}\,\rm erg~cm^{-2}~s^{-1}$. 
The results are shown in the left panel of Figure~\ref{fig:mev_pred} (right panel) for the energy bins $[0.3,1,10,30,100,1000]\,\rm MeV$, 
similarly to the ones chosen by \citet{Inoue_2015}. It can be seen how the number of blazars detectable by an instrument 
with sensitivity $S>10^{-12}\,\rm erg~cm^{-2}~s^{-1}$ will be of the order of $10^3$ to $\sim\times10^4$.

\begingroup
\renewcommand*{\arraystretch}{1.2}
\setlength{\tabcolsep}{4pt} %
\begin{table*}[]
\centering
\caption{Prediction of total number of sources
	per $4\pi$ sterad and number of sources as a function of redshift for future MeV mission (and eROSITA).}\label{tab:mev_pred_z}
\resizebox{\textwidth}{!}{
\hspace{-2cm}
\begin{tabular}{ c | c | c | c | c c c c c}
	\label{tab:mev_pred}
	& Band & Sensitivity & Total & $0\leq z<1$ & $1\leq z<2$ & $2\leq z<3$ & $3\leq z<4$ & $4\leq z<5$ \\
	&  & $[\rm erg~cm^{-2}~s^{-1}]$ & $0\leq z<5$& & & & & \\
 \hline
	COSI  & $0.2-5\,\rm MeV$ &  $S>4\times10^{-11}$ (2 yrs, \citealp{Tomsick_2019})$^{\rm a}$  & 40    &  20 & 10 & 6 & 3 & 1 \\
	AMEGO-X & $1-10\,\rm MeV$ &  $S>1.6\times10^{-11}$ (3 yrs, \citealp{Caputo_2022})  & 130  & 63 & 34 & 20 & 9 & 4 \\
	 ASTROGAM  & $1-10\,\rm MeV$  &  $S>5\times10^{-12}$ (1 yr)$^{\rm b}$ & 832 & 367 & 241 & 137 & 63 & 24 \\
	LOX  & $0.1-3\,\rm MeV$  &  $S>10^{-12}$ (1 yr, \citealp{Miller_2019})  & 12676 & 4279 & 4339 & 2472 & 1135 & 451 \\
	eROSITA & $0.2-2\,\rm keV$ & $S>10^{-14}$ (1 yr, \citealp{Predehl_2021}) & 230023 & 23681 & 86786 & 70403 & 34981 & 14172\\
\hline
\multicolumn{9}{p{\textwidth}}{
\tablenotetext{a}{Point source sensitivities obtained via private communication.}}
\end{tabular}}
\end{table*}
\endgroup

Finally, with the obtained XLF we can make predictions on how many sources per redshift bin would be 
detectable by an MeV experiment with a certain sensitivity. 
We consider estimated sensitivities of proposed and accepted MeV missions: 
COSI \citep[point source sensitivities obtained via private communication, see][]{Tomsick_2019};
AMEGO-X \citep{Caputo_2022};
ASTROGAM (point source sensitivities reported in the mission design proposed for the M7 mission call of ESA, obtained via private communication);
LOX \citep{Miller_2019}.
Table~\ref{tab:mev_pred} reports the predicted sensitivities and number of sources that will be detected by such missions in various redshift bins, 
and the normalized histogram showing the fraction of sources detected per redshift bin by these missions is 
shown in the right panel of Figure~\ref{fig:mev_pred}. For comparison we also list the prediction for 
eROSITA \citep{Predehl_2021}.
If we use the mPLE evolution model, it can be seen how we expect that tens of blazars would be detected even 
beyond $z>4$, and several up to a $z\sim5-6$. 
Recently, \citet{Wolf_2021} has pointed out how the XLF of QSOs may predict more sources than expected at higher redshifts.
In particular, the source they studied with eROSITA data is located at $z=5.81$ and may harbor a nascent jet. This could be a progenitor
to radio-loud quasars considered in this work, and might imply the existence of many such 
powerful sources earlier than $z>4$. Some of these high-$z$ blazars, or nascent jets may have already been detected
at $z>4$ \citep[see][]{Zhu_2019}. Only an instrument with a deeper sensitivity either in hard X-rays or in the MeV 
band could unveil the bulk of these high-$z$ powerful jets.

\subsection{Neutrino Predictions}\label{sec:neutrinos}
Blazars in general, and MeV blazars in particular, are thought to be possible sources of extragalactic neutrinos \citep[e.g.,][]{Murase_2014, Murase_2016,Kadler_2016,Aartsen_2017,Krauss_2018,IceCube_2018,Buson_2022}. Taking advantage of the derived best-fit high-energy SED and of the up-to-date XLF, we  calculate the number of neutrinos expected to be found in coincidence with MeV blazars detected by a forthcoming MeV mission \citep[][]{astrogam_2018,Tomsick_2019, Miller_2019, Caputo_2022}. We follow the methodology described in \citet[][and references therein]{Krauss_2018}, whose main assumptions are: (i) the integrated neutrino flux between $30\,\rm TeV$ and $10\,\rm PeV$ is equivalent to the high-energy flux emitted by MeV blazars integrated between $0.1\,\rm keV$ and $1\,\rm TeV$ ($\phi_{\nu} = \phi_{\gamma}$, see also \citealp[][]{Krauss_2014}); (ii) the neutrino spectrum follows a power law ($dN_{\nu}/dE_{\nu}\propto E^{-\Gamma_{\nu}}$) of index $\Gamma_{\nu}=2.58$ \citep{Icecube_2015}. 
The total number of neutrinos expected from one single MeV blazar is therefore: 
\begin{equation}
\begin{split}
    N_{\nu}=\sum_{i=0}^{N-1}&\int^{E'_{i+1}}_{E'_i}\frac{dN_{\nu}}{dE_{\nu}}dE_{\nu}= \\
    =&\frac{\phi_{\gamma}(2-\Gamma_{\nu})}{E_{2,\nu}^{2-\Gamma_{\nu}}-E_{1,\nu}^{2-\Gamma_{\nu}}(1-\Gamma_{\nu})}T_{\rm eff}\times\\
    \times&\sum_{i=0}^NA_{\rm eff}(E_{\nu}^\prime)(E_{i+1}^{\prime 1-\Gamma_{\nu}}-E_{i}^{\prime 1-\Gamma_{\nu}}).\\
\end{split}
\label{eq:neutr_numb_tot}
\end{equation}

In the above, $A_{\rm eff}$ is the IceCube effective area evaluated in the energy bin $[E^{\prime}_{i}, E^{\prime}_{i+1}]$; $E_{\nu}$ is the mean energy of the $i^{th}$ bin; the sum runs from $E^{\prime}_{0} =E_{1,\nu}=30\rm\,TeV$ to $E^{\prime}_{N}=E_{2,\nu}=10\rm\,PeV$; $\phi_{\gamma}$ is calculated from the best-fit SED of Section~\ref{sec:lat_non_det_bl}. For simplicity, we take the available IceCube neutrino effective area from \citet[][Figure 1]{Krauss_2018} calculated using 4 years of data ($T_{\rm eff}=1347$ days).
Furthermore, to obtain a more realistic number of detectable neutrinos, $N_{\nu}$ has to be corrected by: (1) an empirical factor ($f=0.009$, \citealp{Kadler_2016}) that takes into account physically motivated blazar spectra; (2) the ratio between the 4\,yr IceCube exposure and the exposure of the considered MeV instrument ($t=T_{\rm MeV}/T_{\rm eff}$). This reduces the expected number of neutrinos from one source to:
\begin{equation}
    N_{\nu,det}=N_{\nu}\times ft. 
\end{equation}

Considering an MeV mission, with its sensitivity limit ($S_{\rm lim}$) over a certain time period ($T_{\rm MeV}$) in a specific energy band, it is possible to estimate the expected number of detectable sources and their flux distribution (see Section~\ref{sec:mev_pred}). For every detectable source, we randomly extract its MeV flux from the relevant {\it logN-logS} and calculate $N_{\nu,det}$. Importantly, we  estimate the number of coincident detections  as the number of blazars that are able to produce at least one neutrino, which could be detected by IceCube, within the observing window of the MeV mission ($T_{\rm MeV}$). %factor (i.e.~is the source bright to produce at least one neutrino), set equal to 1 in case $N_{\nu,det}>1$, 0 otherwise.
\begingroup
\renewcommand*{\arraystretch}{1.2}
\begin{table}[]
\centering
%\scriptsize
	\caption{Predictions on number of coincidence detections of neutrinos from MeV blazars by IceCube,  considering future MeV missions (Section~\ref{sec:neutrinos}).}\label{tab:neutrino_pred}
\begin{tabular}{ c | c }
	 &  Number of Coincident Detections$^{\rm a}$ \\
 \hline
	COSI     &  $2^{+1}_{-1}$  \\
	AMEGO-X  &  $4^{+1}_{-1}$ \\
	ASTROGAM &  $5_{-1}^{+1}$  \\
	LOX      &  $11_{-2}^{+1}$ \\
\hline
\multicolumn{2}{l}{%
  \begin{minipage}{\columnwidth}
 \tablenotetext{a}{Median values of the  number of blazars able to emit at least one $>$100\,TeV neutrino and  corresponding $1\sigma$ CL.}
\end{minipage}%
}
\end{tabular}

\end{table}
\endgroup
We perform this calculation 10000 times and extract the median number of coincident detections for all MeV missions listed in Table~\ref{tab:mev_pred_z}.
The results are reported in Table~\ref{tab:neutrino_pred}. The  median number of coincident detections ranges from $2-12$. 
This implies that, for example, COSI will detect (in its 2\,year survey) 2$^{+1}_{-1}$ blazars with coincident detection by IceCube.
We note that these numbers are quite conservative as we do not take into account the possibility of flares. Blazars are known to be extremely variable at all wavelengths, and in particular at $\gamma$-ray energies where the emission is dominated by the higher energy particles \citep[e.g.,][]{Abdo_2010,IceCube_2018,Nesci_2021, Malik_2022}. A significant increase in neutrino flux  is therefore expected when the sources are detected in their flaring state \citep[see e.g.,][]{Murase_2018}. Indeed, even a factor of 5 flux increase would imply a number of coincident detections of $9_{-2}^{+1}$ in the COSI $2\,\rm yrs$ survey. Finally, a longer exposure time when IceCube and an MeV mission are both operative would ensure a larger number of coincident detections.

\section{Summary \& Conclusions}
In this work we have derived the most up-to-date BAT blazar X-ray luminosity function in the $14-195\rm\,keV$ range. The main results are summarized as follows:
\begin{enumerate}
    \item BAT blazars evolve positively in redshift, indicating the presence of more (or more luminous) sources at earlier cosmic times. The peak in space density of this population is located at $z_{\rm peak}\simeq4.3\pm1.0$. In terms of number densities, these sources are predicted to be less numerous at higher redshifts compared with previous works.
    \item The blazar XLF at every redshift bin is distributed (in $\log-\log$ space) according to a straight power law of index $\gamma_2=1.67\pm0.19$. Lack of any spectral break in the XLF impedes us to confirm whether the evolution of these source class happens in luminosity or density. Nonetheless, fit results confirm with high significance that a break in the local XLF at the level of the last observed luminosity ($L_{\rm min}=10^{44}\,\rm erg~s^{-1}$) is favored, and the index of the low-luminosity end flattens as expected from beaming.
    \item We derive the average SED of BAT FSRQs (with and without LAT detection) in the range $14\,\rm keV$-$10\,\rm GeV$ in order to understand their contribution to the cosmic high-energy background. In the $14-195\,\rm keV$ blazars can contribute at most $\sim20\%$. On the other hand, these powerful sources are found to be able to potentially contribute $\sim70-100\%$ of the MeV background in the $0.5-30\,\rm MeV$ range (depending on the evolutionary model). Predictions from the density evolution model are shown to slightly overestimate the MeV background level, implying that a luminosity evolution is more likely favored by these sources as it allows for contribution from other known MeV emitting sources.
    \item Properties of the jets parent population are derived in this work. The intrinsic luminosity function is confirmed to be distributed similarly to the unbeamed jetted AGNs ($B=2.7$). For the selected blazar sample, the average viewing angles are very narrow ($\theta=2-3\degree$) and the average speeds of the jet plasma is lower than previously expected ($<\Gamma>=8-12$). This has implication on the power of these sources, which could be lower than previously assumed, as well as on their number densities which could decrease up to an order of magnitude (considering the $2\Gamma^2$ correction, see Section~\ref{sec:smbh_space_dens}).
    \item BAT blazars (mostly comprising FSRQ-type sources) are found to be more luminous than LAT detected sources and host more massive black holes. 
    Lack of strong evolution in the SED and XLF properties suggest that these two source classes belong to the same parent population. The difference in evolution (i.e., BAT blazar peak at earlier cosmic times and most likely follow a luminosity evolution while LAT blazars likely undergo a density evolution and peak at $z\sim2$) indicate that blazars may be following the paradigm of AGN cosmic downsizing. 
    \item Finally, prediction for number counts of sources in the MeV range are derived, implying detection on the order of hundreds or thousands of sources up to redshift $5-6$, making the prospects for MeV blazar science very promising 
    in light of an upcoming MeV mission. 
    Furthermore, the estimated number of high-energy neutrino detection by IceCube in coincidence with MeV blazars within the observing window of forthcoming MeV missions is derived. The expectation is that IceCube should detect 2 coincident neutrinos in a 2 year MeV survey like COSI.
\end{enumerate}

\acknowledgments
We thank the anonymous referee for their very insightful comments on the manuscript.

The authors acknowledge funding under NASA contract: 80NSSC20K0044.

Support for this work was provided by NASA through the NASA Hubble Fellowship grant \#HST-HF2-51486.001-A 
awarded by the Space Telescope Science Institute, which is operated by the Association of Universities 
for Research in Astronomy, Inc., for NASA, under contract NAS5-26555.
 
MB acknowledges support from the YCAA Prize Postdoctoral Fellowship.
BT acknowledges support from the Israel Science Foundation (grant number 1849/19) and from the European Research Council (ERC) under the European Union's Horizon 2020 research and innovation program (grant agreement number 950533).
CR acknowledges support from the Fondecyt Iniciacion grant 11190831 and ANID BASAL project FB210003. KO acknowledges support from the Korea Astronomy and Space Science Institute under the R\&D program(Project No. 2022-1-868-04) supervised by the Ministry of Science and ICT and from the National Research Foundation of Korea (NRF-2020R1C1C1005462).

We acknowledge support from NASA through ADAP award NNH16CT03C.

The \textit{Fermi} LAT Collaboration acknowledges generous ongoing support
from a number of agencies and institutes that have supported both the
development and the operation of the LAT as well as scientific data analysis.
These include the National Aeronautics and Space Administration and the
Department of Energy in the United States, the Commissariat \`a l'Energie Atomique
and the Centre National de la Recherche Scientifique / Institut National de Physique
Nucl\'eaire et de Physique des Particules in France, the Agenzia Spaziale Italiana
and the Istituto Nazionale di Fisica Nucleare in Italy, the Ministry of Education,
Culture, Sports, Science and Technology (MEXT), High Energy Accelerator Research
Organization (KEK) and Japan Aerospace Exploration Agency (JAXA) in Japan, and
the K.~A.~Wallenberg Foundation, the Swedish Research Council and the
Swedish National Space Board in Sweden.

\appendix
\section{Source Count Distribution}\label{appendix:A}
In order to assess whether the discrepancy between the derived {\it logN-logS} and the model predictions (seen in Figure~\ref{fig:logn}) at fluxes $F_{\rm 14-195\, keV}<8\times 10^{-12}\,\rm erg~cm^{-2}~s^{-1}$ affects our results, we perform an XLF fit to the total blazar sample removing sources with $F_{\rm 14-195\, keV}<8\times 10^{-12}\,\rm erg~cm^{-2}~s^{-1}$ (12/118 sources, 3 of which have $z>1$). For consistency, we also use the efficiency above this threshold flux. We employ the same maximum-likelihood fitting technique described in Section~\ref{sec:ml} and test both the mPDE and mPLE parametrizations.
The obtained fits show that both the mPDE and the mPLE results are consistent with the ones derived for the whole sample (cfr. Table~\ref{tab:res_ml} and Table~\ref{tab:res_ml_cut}). 

\begin{figure} 
\centering
        \includegraphics[width=0.5\columnwidth]{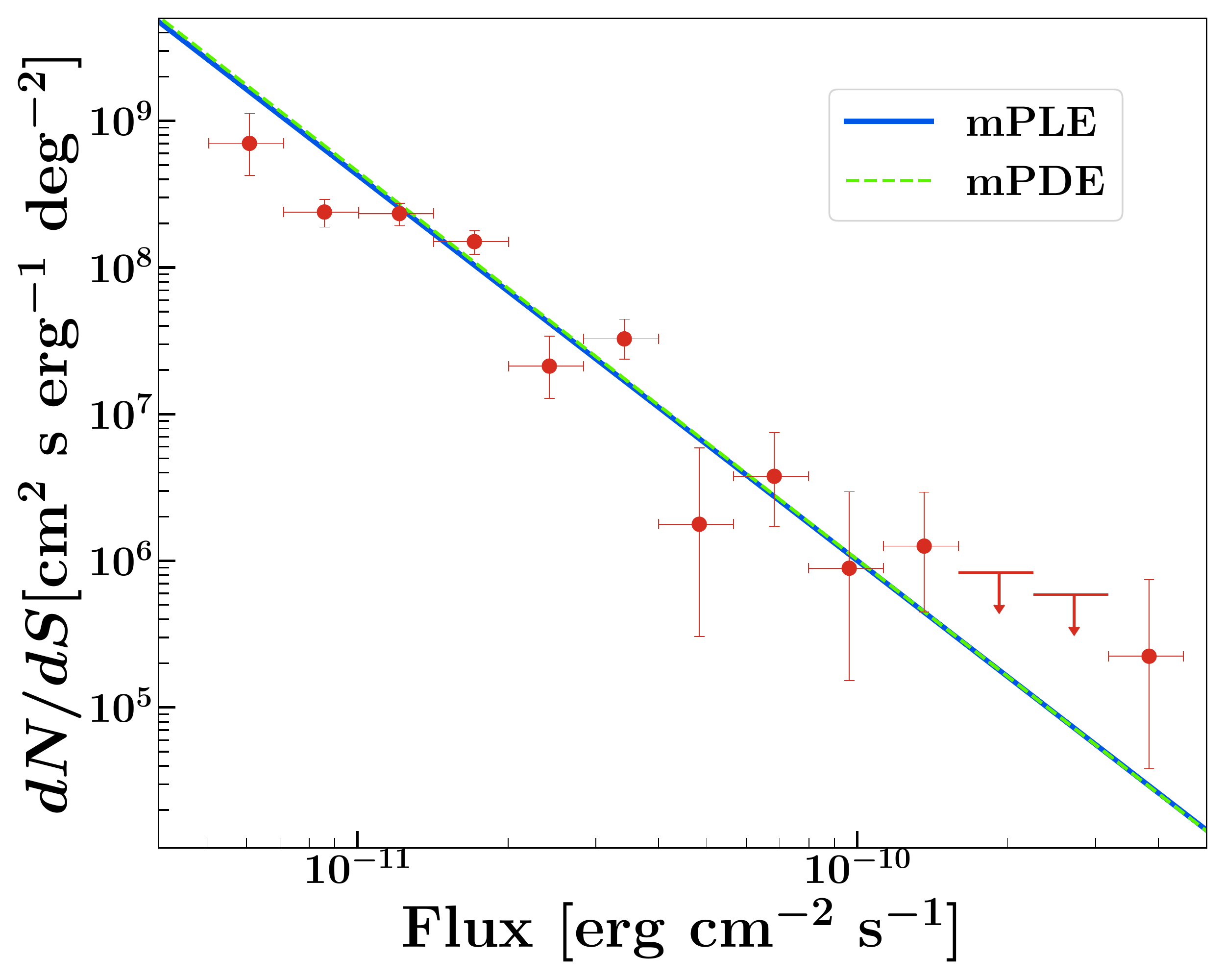}
        \caption{Differential BAT-blazars {\it logN-logS}, i.e., number of sources per flux bin corrected for the survey sky coverage, $dN/dS$, as a function flux.
        The observed differential {\it logN-logS} is represented by the red data points, while the
        blue solid and green dashed lines show the prediction from the best-fit mPLE and mPDE models, respectively. The error bars are calculated using Poisson statistics.
         \label{fig:diff_lognlogs}}
\end{figure}

The differential {\it logN-logS} for the total 118 beamed AGN sample is also shown in Figure~\ref{fig:diff_lognlogs}. As can be seen, within the statistical uncertainties the best-fit model predictions are consistent with the data.
This allows us to conclude that the full sample of 118 sources is representative of the BAT-blazar population and that possible discrepancies at low fluxes (possibly arising from sky coverage calculations or sources missed as blazars at low fluxes) do not influence our results. 
\begingroup
\renewcommand*{\arraystretch}{1.2}
\begin{table*}[h!]
\centering
\scriptsize
\caption{Result of the maximum likelihood fit only including sources $F_{\rm 14-195\,keV}>8\times10^{-12}\,\rm erg~cm^{-2}~s^{-1}$ (see Appendix~\ref{appendix:A}).}
\begin{tabular}{ l | l | c c c c c c | c c}% c c c c}
	SAMPLE & LF & \multicolumn{6}{c}{Parameters} & $C$ &  CXB\\ 
\hline
	Total &     &A$^{a}$ & $L_*^{b}$ & $\gamma_1$ & $\gamma_2$ & k & $\xi$ & &\\ %& & & & \\
	& mPLE &$1.70\pm0.16$ &$1.60\pm1.05$ & $-0.38\pm1.19$ & $2.03\pm0.23$ & $3.59\pm0.53$ & $-1.57\pm0.40$ & 1308.42 & 10.23\% \\ 
	&      & A$^{a}$ & $L_*^{b}$ & $\gamma_1$ & $\gamma_2$ & k & $\xi$ & &\\% & & & & \\
	& mPDE & $1.59\pm0.15$ &$1.54\pm0.84$ & $-0.65\pm1.32$ & $1.97\pm0.20$ & $10.52\pm2.23$ & $-0.53\pm0.16$ & 1308.62 & 115.14\% \\
\hline       
\multicolumn{10}{l}{%
  \begin{minipage}{\columnwidth}
 \tablenotetext{a}{Normalization constant in units of $\rm Mpc^{-3}$}
 \tablenotetext{b}{Luminosity scale factor in units of $\rm 10^{44}\,\rm erg~s^{-1}$}
\end{minipage}%
}
\end{tabular}\label{tab:res_ml_cut}
\end{table*}
\endgroup
\clearpage

\clearpage
\setlength\LTleft{0pt}
\setlength\LTright{0pt}
\begingroup
\renewcommand*{\arraystretch}{1.2}
\setlength{\tabcolsep}{3pt} 
\begin{longtable} {@{\extracolsep{\fill}}c|c|c|c|c|c|c|c|c|c@{}}
\caption{Table of clean sample used in the analysis. }\\

      No.$^{a}$ & Name$^{b}$ &  RA$^{c}$  & DEC$^{c}$ & Assoc. Ctpt.$^{d}$ & $\sigma^{e}$ & Type$^{f}$ &  $F_{\rm 14-195\rm keV}^{g}$ &  $\Gamma_{\rm 14-195\rm keV}^{h}$ &  $z^{i}$ \\
	& [SWIFT] & $[\rm ^{\circ}, J2000]$ & $[\rm ^{\circ}, J2000]$ & & & & $[10^{-12}\,\rm~erg~cm^{-2}~s^{-1}]$ & & \\
	\hline
\endfirsthead
\multicolumn{10}{c}%
{\tablename\ \thetable\ -- \textit{Continued from previous page}} \\
     No.$^{a}$ & Name$^{b}$ &  RA$^{c}$  & DEC$^{c}$ & Assoc. Ctpt.$^{d}$ & $\sigma^{e}$ & Type$^{f}$ &  $F_{\rm 14-195\rm keV}^{g}$ &  $\Gamma_{\rm 14-195\rm keV}^{h}$ &  $z^{i}$ \\
	& [SWIFT] & $[\rm ^{\circ}, J2000]$ & $[\rm ^{\circ}, J2000]$ & & & & $[10^{-12}\,\rm~erg~cm^{-2}~s^{-1}]$ & & \\
	\hline
\endhead
\hline \multicolumn{10}{c}{\textit{Continued on next page}} \\
\endfoot
\hline
\endlastfoot
\hline
  8&     J0010.5+1057  &   2.61 &  10.96  & Mrk 1501&                  14.02 & BZQ & $30.34_{-2.80}^{+3.06}$ & $1.82\pm0.21$          & 0.089 \\               
  9&     J0017.1+8134  &   4.48 &  81.56  & [HB89] 0014+813&           8.57 & BZQ & $11.39_{-1.86}^{+2.08}$ & $2.42_{-0.43}^{+0.53}$  & 3.366 \\ 
  30&    J0042.9+3016B &  10.70 &  30.28  & 2MASX J00423991+3017515&   11.4 & BZQ & $26.08_{-9.51}^{+12.67}$ & $0.5_{-...}^{+1.36}$   & 0.140 \\  
  48&    J0103.8-6437  &  15.94 & -64.63  & PKS 0101-649&              7.19 & BZQ & $14.3_{-1.48}^{+2.48}$ & $1.58_{-0.51}^{+0.54}$   & 0.163 \\ 
  59&    J0113.8+2515  &  18.41 &  25.29  & WISE J011322.69+251853.2&  5.52 & BZQ & $8.70_{-2.43}^{+2.98}$ & $2.09_{-0.72}^{+0.93}$   & 1.589 \\
  71&    J0122.9+3420  &  20.77 &  34.37  & SHBL J012308.7+342049&     9.23 & BZB & $11.27_{-1.62}^{+2.06}$ & $2.94_{-0.48}^{+0.61}$  & 0.272 \\
  1228&  J0131.5-1007  &  22.88 & -10.13  & SDSS J013126.71-100931.1& 6.48 & BZQ & $12.84_{-2.84}^{+3.04}$ & $1.81_{-0.48}^{+0.53}$  & 3.515 \\ 
  1235&  J0156.5-5303  &  29.13 & -53.03  & RBS 259&                  5.49 & BZB & $7.32_{-2.46}^{+1.56}$ & $2.31_{-0.51}^{+0.60}$   & 0.3043   \\  
  1237&  J0201.0+0329  &  30.29 &   3.57  & [HB89] 0158+031&          6.93 & BZQ & $10.71_{-2.87}^{+2.46}$ & $2.25_{-0.65}^{+0.99}$  & 1.581 \\
  120&   J0218.0+7348  &  34.38 &  73.80  & [HB89] 0212+735&           15.5 & BZQ & $34.98_{-2.43}^{+2.31}$ & $1.55\pm0.18$           & 2.367 \\          
  125&   J0225.0+1847  &  36.25 &  18.80  & RBS 0315 &                  14.15 & BZQ & $31.37_{-2.91}^{+3.47}$ & $1.73\pm0.22$          & 2.690 \\         
  132&   J0232.8+2020  &  38.18 &  20.28  & QSO B0229+200&             13.85 & BZG & $23.46_{-2.27}^{+2.62}$ & $2.28_{-0.28}^{+0.31}$ & 0.139 \\
  1243&  J0233.8+0243  &  38.43 &   2.43  & IGR J02341+0228&          5.13 & BZQ & $7.02_{-1.26}^{+2.72}$ & $2.69_{-0.71}^{+1.08}$   & 0.321 \\  
  1244&  J0244.8-5829  &  41.18 & -58.29  & BZB J0244-5819&           8.39 & BZB & $10.13_{-1.13}^{+2.46}$ & $2.43_{-0.42}^{+0.52}$  & 0.264 \\
  148&   J0245.2+1047  &  41.32 &  10.77  & 4C+10.08&                 8.05 & BZG & $18.3_{-2.44}^{+2.71}$ & $1.70_{-0.43}^{+0.48}$   & 0.070 \\   
  1246&  J0250.8-3626  &  42.69 & -36.26  & 6dF J0250552-361636&      5.23 & BZQ & $10.01_{-1.53}^{+3.36}$ & $1.73_{-0.60}^{+0.03}$  & 1.536 \\
  1254&  J0304.6+3348  &  46.18 &  33.80  & 4C +33.06&                5.25 & BZQ & $11.89_{-3.29}^{+2.51}$ & $1.86_{-0.51}^{+0.57}$  & 0.681 \\  
  1257&  J0310.7+3917  &  47.74 &  39.23  & WISE J031024.53+391057.9& 6.1 & BZQ & $9.90_{-3.19}^{+2.42}$ & $1.89_{-0.55}^{+0.66}$    & 0.370 \\   
  168&   J0311.8-7653  &  47.75 & -76.86  & [HB89] 0312-770&           6.89 & BZQ & $10.47_{-2.06}^{+2.24}$ & $1.98_{-0.48}^{+0.57}$  & 0.223 \\
  178&   J0326.0-5633  &  51.47 & -56.52  & 2MASX J03252346-5635443&   6.62 & BZG & $8.49_{-2.21}^{+2.35}$ & $2.06_{-0.54}^{+0.66}$   & 0.060 \\  
  188&   J0336.6+3217  &  54.12 &  32.29  & 4C+32.14&                 16.9 & BZQ & $44.17_{-3.05}^{+3.16}$ & $1.67_{-0.17}^{+0.16}$  & 1.258 \\
  192&   J0349.2-1159  &  57.36 & -11.98  & QSO B0347-121&             9.75 & BZG & $15.68_{-2.25}^{+3.31}$ & $2.20_{-0.36}^{+0.42}$  & 0.187 \\
  195&   J0353.4-6830  &  58.28 & -68.53  & PKS 0352-686&              10.03 & BZB & $12.24_{-1.44}^{+1.67}$ & $2.52_{-0.36}^{+0.43}$ & 0.087 \\
  206&   J0404.0-3604  &  60.96 & -36.07  & PKS 0402-362&              6.6 & BZQ & $10.65_{-1.87}^{+2.48}$ & $1.91_{-0.50}^{+0.61}$   & 1.417 \\
  208&   J0405.5-1307  &  61.35 & -13.14  & [HB89] 0403-132&           5.75 & BZQ & $11.03_{-3.44}^{+2.37}$ & $1.78_{-0.59}^{+0.69}$  & 0.570 \\ 
  1271&  J0407.9-1219  &  61.98 & -12.19  & [HB89] 0405-123&          5.89 & BZQ & $8.14_{-2.34}^{+1.89}$ & $2.56_{-0.60}^{+0.79}$   & 0.572 \\
  210&   J0413.3+1659  &  63.34 &  16.96  & MG1 J041325+1659&          6.51 & BZQ & $15.17_{-4.00}^{+2.77}$ & $1.88_{-0.51}^{+0.60}$  & 0.212 \\
  1278&  J0506.6-1937  &  76.69 & -19.62  & 2MASX J05064796-1936507&  5.03 & BZQ & $13.56_{-3.38}^{+3.16}$ & $1.31_{-0.55}^{+0.56}$  & 0.094 \\ 
  259&   J0507.7+6732  &  76.91 &  67.53  & 87GB 050246.4+673341&      6.7 & BZB & $9.03_{-2.28}^{+2.18}$ & $2.50_{-0.57}^{+0.81}$    & 0.314 \\ 
  276&   J0525.1-2339  &  81.27 & -23.66  & PMN J0525-2338&            5.78 & BZQ & $13.11_{-3.59}^{+2.80}$ & $1.55_{-0.51}^{+0.55}$  & 3.100 \\ 
  1282&  J0538.5-4411  &  84.63 & -44.11  & [HB89] 0537-441&          5.12 & BZB & $14.86_{-2.77}^{+2.58}$ & $0.88_{-...}^{+0.69}$   & 0.894 \\ 
  296&   J0539.9-2839  &  84.96 & -28.69  & [HB89] 0537-286&           11.67 & BZQ & $29.01_{-3.50}^{+2.67}$ & $1.33\pm0.26$          & 3.104 \\       
  306&   J0550.7-3212A &  87.68 & -32.27  & PKS 0548-322&              26.92 & BZB & $18.21_{-5.29}^{+6.16}$ & $3.23_{-1.23}^{+2.34}$ & 0.069 \\
  1289&  J0608.9-5507  &  92.20 & -55.07  & PKS 0607-549&             5.14 & BZQ & $6.24_{-3.72}^{+3.68}$ & $2.19_{-...}^{+5.84}$    & 2.460 \\  
  323&   J0612.2-4645  &  93.06 & -46.74  & PMN J0612-4647&           5.82 & BZQ & $6.85_{-1.41}^{+1.78}$ & $2.47_{-0.62}^{+0.88}$   & 0.317 \\
  327&   J0623.3-6438  &  95.85 & -64.61  & 2MASX J06230765-6436211&   7.31 & BZQ & $11.64_{-1.62}^{+2.09}$ & $1.98_{-0.45}^{+0.53}$  & 0.128 \\
  340&   J0635.8-7514  &  99.01 & -75.27  & PKS 0637-752&              9.57 & BZQ & $16.52_{-1.83}^{+2.78}$ & $2.00_{-0.30}^{+0.35}$  & 0.651 \\
  361&   J0710.3+5908  & 107.63 &  59.14  & 2MASX J07103005+5908202&   15.38 & BZB & $24.06_{-2.29}^{+2.77}$ & $2.28_{-0.24}^{+0.26}$ & 0.125 \\
  1308&  J0721.0+7133  & 110.49 &  71.33  & [HB89] 0716+714&          6.39 & BZB & $19.04_{-2.48}^{+2.68}$ & $1.15_{-0.43}^{+0.39}$  & 0.300 \\ 
  377&   J0733.9+5156  & 113.40 &  51.93  & 2MASX J07332681+5153560&   5.38 & BZG & $8.17_{-2.17}^{+2.27}$ & $2.32_{-0.71}^{+0.93}$   & 0.065 \\
  387&   J0746.3+2548  & 116.58 &  25.80  & B20743+25&                12.69 & BZQ & $36.01_{-3.50}^{+3.16}$ & $1.43_{-0.24}^{+0.24}$ & 2.979 \\
  407&   J0805.2+6145  & 121.31 &  61.75  & CGRaBS J0805+6144&         6.83 & BZQ & $17.53_{-3.79}^{+2.50}$ & $1.35\pm0.39$           & 3.033 \\ 
  1329&  J0836.6-2025  & 129.16 & -20.25  & [HB89] 0834-201&          5.48 & BZQ & $14.23_{-3.38}^{+3.05}$ & $1.43\pm0.62$           & 2.752 \\ 
  428&   J0841.4+7052  & 130.34 &  70.88  & [HB89] 0836+710&           38.17 & BZQ & $69.81_{-2.49}^{+2.34}$ & $1.70\pm0.08$          & 2.172 \\
  1336&  J0842.0+4021  & 130.46 &  40.30  & 2MASSi J0842037+401831&   4.96 & BZQ & $6.93_{-1.98}^{+2.18}$ & $2.41_{-0.65}^{+0.89}$   & 0.151 \\
  445&   J0909.0+0358  & 137.26 &   3.94  & 1RXS J090915.6+035453&     8.04 & BZQ & $15.65_{-2.76}^{+2.92}$ & $1.88_{-0.37}^{+0.27}$  & 3.288 \\
  1352&  J0923.2+3850  & 140.82 &  38.77  & B2 0920+39&               5.78 & BZG  & $12.22_{-2.73}^{+3.52}$ & $1.44_{-0.55}^{+0.57}$ & 1.137 \\
  454&   J0923.6-2136  & 140.91 & -21.61  & PKS 0921-213&              9.54 & BZQ & $18.79_{-2.62}^{+2.80}$ & $1.97_{-0.32}^{+0.36}$  & 0.052 \\
  1354&  J0930.1+4987  & 142.52 &  49.87  & 2MASS J09303759+4950256&  5.78 & BZB & $7.44_{-1.71}^{+2.21}$ & $2.59_{-0.50}^{+0.64}$   & 0.186 \\
  1355&  J0934.0-1721  & 143.58 & -17.33  & 2MASS J09343014-1721215&  5.64 & BZB & $7.93_{-1.97}^{+2.55}$ & $2.73_{-0.72}^{+1.16}$   & 0.249 \\
  500&   J1031.5+5051  & 157.85 &  50.90  & 2MASX J10311847+5053358&   7.71 & BZB & $7.85_{-1.31}^{+1.68}$ & $2.85_{-0.59}^{+0.83}$   & 0.360 \\
  1375&  J1044.8+8091  & 161.18 &  80.91  & [HB89] 1039+811&          6.09 & BZQ & $11.73_{-2.89}^{+2.05}$ & $1.67_{-0.39}^{+0.43}$  & 1.260 \\
  526&   J1103.5-2329  & 165.86 & -23.47  & 2MASX J11033765-2329307&   7.03 & BZB & $10.80_{-3.01}^{+1.66}$ & $2.53_{-0.51}^{+0.67}$  & 0.186 \\
  527&   J1104.4+3812  & 166.10 &  38.21  & Mrk 421&                   129.26 & BZB & $141.00_{-2.00}^{+1.11}$ & $2.76_{-0.03}^{+0.02}$ & 0.030\\
  1382&  J1105.4+0200  & 166.34 &   2.00  & ICRF J110538.9+020257&    5.43 & BZQ & $15.64_{-3.39}^{+3.15}$ & $1.27_{-0.55}^{+0.59}$  & 0.105 \\
  545&   J1130.1-1447  & 172.53 & -14.79  & PKS 1127-14&               13.62 & BZQ & $28.74_{-3.92}^{+2.13}$ & $1.88_{-0.24}^{+0.26}$ & 1.184 \\
  551&   J1136.7+6738  & 174.10 &  67.64  & 2MASX J11363009+6737042&   10.19 & BZG & $12.73_{-1.53}^{+2.24}$ & $2.33_{-0.33}^{+0.39}$ & 0.134 \\
  1396&  J1153.0+3311  & 178.22 &  33.08  & 7C 1150+3324&             6.32 & BZQ & $10.08_{-2.12}^{+2.84}$ & $1.83_{-0.47}^{+0.54}$  & 1.397 \\
  578&   J1153.6+4931  & 178.34 &  49.49  & 4C+49.22&                 8.18 & BZQ & $12.78_{-1.73}^{+2.39}$ & $1.83_{-0.37}^{+0.42}$  & 0.334 \\
  1397&  J1153.9+5848  & 178.47 &  58.48  & [HB89] 1217+023&          6.27 & BZQ & $8.46_{-1.77}^{+1.82}$ & $2.18_{-0.48}^{+0.56}$   & 0.202 \\
  1402&  J1220.2+0202  & 185.04 &   1.99  & FBQS J1221+3010&           5.87 & BZQ & $8.03_{-1.69}^{+1.98}$ & $2.53_{-0.62}^{+0.88}$   & 0.240 \\
  610&   J1221.3+3012  & 185.34 &  30.15  & 4C+04.42&                 10.04 & BZB & $10.62_{-1.68}^{+1.29}$ & $2.94_{-0.48}^{+0.63}$ & 0.183 \\
  612&   J1222.4+0414  & 185.58 &   4.21  & SDSS J122358.97+404409.3& 14.37 & BZQ & $36.22_{-3.56}^{+3.29}$ & $1.45\pm0.20$          & 0.965 \\
  614&   J1224.9+2122  & 186.22 &  21.40  & PG 1222+216&               12.6 & BZQ & $24.50_{-2.87}^{+2.65}$ & $1.70_{-0.25}^{+0.26}$  & 0.432 \\
  619&   J1229.1+0202  & 187.27 &   2.04  & 3C 273&                    197.62 & BZQ & $421.57\pm3.08$ & $1.75\pm0.02$                 & 0.158 \\
  1410&  J1238.4+5349  & 189.59 &  53.49  & SDSS J123807.76+532555.9& 5.33 & BZQ & $8.47_{-1.91}^{+2.38}$ & $1.74_{-0.62}^{+0.67}$   & 0.347 \\
  1412&  J1254.9+1165  & 193.72 &  11.65  & QSO B1252+119&            6.51 & BZQ & $13.24_{-2.43}^{+3.00}$ & $1.72_{-0.49}^{+0.55}$  & 0.872 \\
  645&   J1256.2-0551  & 194.05 &  -5.79  & 3C 279&                    12.29 & BZQ & $38.82_{-4.16}^{+3.67}$ & $1.32_{-0.23}^{+0.21}$ & 0.536 \\
  656&   J1305.4-1034  & 196.38 & -10.56  & PKS 1302-102&              5.4 & BZQ & $13.72_{-2.99}^{+3.80}$ & $1.70_{-0.58}^{+0.66}$   & 0.278 \\
  1417&  J1306.4-7603  & 196.61 & -76.04  & 2MASX J13071558-7602451&  5.07 & BZQ & $7.33_{-1.98}^{+1.77}$ & $2.51_{-0.76}^{+1.23}$   & 0.183 \\
  675&   J1331.6-0504  & 202.98 &  -5.15  & PKS 1329-049&              5.83 & BZQ & $15.50_{-3.46}^{+3.33}$ & $1.51_{-0.46}^{+0.48}$  & 2.150 \\
  681&   J1337.7-1253  & 204.41 & -12.94  & [HB89] 1334-127&           6.67 & BZQ & $13.21_{2.65}^{+2.75}$ & $2.19_{-0.48}^{+0.57}$   & 0.539 \\
  690&   J1347.1+7325  & 206.61 &  73.37  & 2MASSi J1346085+732053&    7.59 & BZQ & $10.34_{-2.30}^{+2.05}$ & $2.18_{-0.43}^{+0.52}$  & 0.290 \\
  727&   J1428.7+4234  & 217.14 &  42.65  & 1ES 1426+428&              17.63 & BZB & $20.85_{-1.04}^{+1.50}$ & $2.56_{-0.22}^{+0.23}$ & 0.129 \\
  1448&  J1430.6+4211  & 217.64 &  42.11  & B3 1428+422&              4.98 & BZQ & $9.90_{-2.58}^{+2.46}$ & $1.56_{-0.57}^{+0.62}$   & 4.655 \\
  752&   J1458.9+7143  & 224.47 &  71.71  & 3C 309.1&                  5.15 & BZQ & $7.65_{-1.77}^{+1.94}$ & $1.80_{-0.69}^{+0.79}$   & 0.905 \\
  763&   J1512.8-0906  & 228.21 &  -9.08  & PKS 1510-08&               18.62 & BZQ & $66.8_{-3.40}^{+3.16}$ & $1.32\pm0.14$           & 0.360 \\
  807&   J1625.9+4349  & 246.48 &  43.81  & 87GB 162418.8+435342&      7.78 & BZQ & $12.13_{-1.90}^{+2.30}$ & $2.04_{-0.42}^{+0.50}$  & 1.048 \\
  808&   J1626.5-2951  & 246.55 & -29.85  & PKS 1622-29&               5.07 & BZQ & $15.54_{-2.89}^{+3.55}$ & $1.32\pm0.66$           & 0.815 \\
  829&   J1643.1+3951  & 250.76 &  39.81  & 3C 345&                    7.05 & BZQ & $20.71_{-3.11}^{+3.30}$ & $1.17_{-0.35}^{+0.35}$  & 0.592 \\
  843&   J1654.0+3946  & 253.47 &  39.75  & Mrk 501&                   50.65 & BZB & $71.58_{-2.29}^{+2.22}$ & $2.39\pm0.07$          & 0.033 \\
  1492&  J1658.5+0518  & 254.62 &   5.29  & RX J1658.5+0515&          5.54 & BZQ & $12.75_{-2.81}^{+3.62}$ & $1.79_{-0.61}^{+0.71}$  & 0.879 \\
  1510&  J1740.7+5197  & 265.16 &  51.97  & 1RXS J174036.3+521155&    5.4 & BZQ & $8.70_{-3.11}^{+2.29}$ & $1.90_{-0.58}^{+0.73}$    & 1.375 \\
  1524&  J1759.6+7846  & 269.88 &  78.46  & [HB89] 1803+784&          5.26 & BZQ & $9.08_{-2.15}^{+2.35}$ & $1.93_{-0.55}^{+0.67}$   & 0.680 \\
  1530&  J1809.6-4585  & 272.41 & -45.85  & ICRF J180957.8-455241&    5.43 & BZQ & $12.68_{-2.17}^{+3.15}$ & $1.80_{-0.46}^{+0.53}$  & 0.069 \\
  1531&  J1810.0-6554  & 272.47 & -65.92  & PMN J1809-6556&           6.22 & BZQ & $13.21_{-2.08}^{+3.02}$ & $1.84_{-0.53}^{+0.60}$  & 0.180 \\
  980&   J1829.4+4846  & 277.40 &  48.73  & 3C 380&                    6.69 & BZQ & $15.23_{-2.37}^{+2.72}$ & $1.52_{-0.38}^{+0.39}$  & 0.692 \\
  1551&  J1848.5+6704  & 282.12 &  67.04  & 8C 1849+670&              6.39 & BZQ & $6.39_{-1.90}^{+1.54}$ & $2.72_{-0.72}^{+1.06}$   & 0.657 \\
  1564&  J1924.8+5531  & 291.19 &  55.51  & 87GB[BWE91] 1923+5523&    5.28 & BZQ & $9.96_{-3.10}^{+3.14}$ & $1.75_{-0.52}^{+0.59}$   & 0.345 \\
  1565&  J1924.9-2918  & 291.21 & -29.18  & [HB89] 1921-293&          7.75 & BZG & $16.22_{-3.42}^{+3.47}$ & $2.04_{-0.41}^{+0.46}$  & 0.352 \\
  1038&  J1928.0+7356  & 292.10 &  73.94  & 4C+73.18&                 8.65 & BZQ & $11.04_{-1.47}^{+1.80}$ & $2.50_{-0.42}^{+0.53}$  & 0.302 \\
  1570&  J1941.3-6216  & 295.32 & -62.16  & PKS 1936-623&             6.32 & BZQ & $17.84_{-3.51}^{+3.25}$ & $1.32_{-0.50}^{+0.48}$  & 2.480 \\
  1573&  J1948.4-7975  & 297.10 & -79.75  & 6dF J1949458-794523&      6.5 & BZQ & $9.95_{-2.04}^{+2.19}$ & $2.18_{-0.48}^{+0.58}$    & 1.127 \\
  1058&  J1959.6+6507  & 299.97 &  65.15  & QSO B1959+650&             23.05 & BZB & $29.03_{-1.83}^{+1.72}$ & $2.67_{-0.18}^{+0.17}$ & 0.047 \\
  1066&  J2010.6-2521  & 302.66 & -25.34  & 1RXS J201020.0-252356&     5.0 & BZQ & $10.83_{-4.11}^{+3.67}$ & $1.77_{-0.64}^{+0.73}$   & 0.824 \\
  1068&  J2011.5-1544  & 302.83 & -15.74  & PKS 2008-159&              7.15 & BZQ & $12.6_{-3.27}^{+2.92}$ & $2.41_{-0.55}^{+0.72}$   & 1.180 \\
  1082&  J2033.4+2147  & 308.38 &  21.76  & 4C+21.55&                 15.34 & BZQ & $30.81_{-2.88}^{+2.85}$ & $2.01_{-0.18}^{+0.21}$ & 0.173 \\
  1093&  J2056.0-4713  & 313.82 & -47.16  & PKS 2052-47&               12.76 & BZQ & $18.27_{-2.82}^{+2.23}$ & $2.19_{-0.36}^{+0.40}$ & 1.489 \\
  1112&  J2129.1-1538  & 322.31 & -15.60  & PKS 2126-15&               8.14 & BZQ & $20.05_{-2.56}^{+3.13}$ & $1.79_{-0.43}^{+0.48}$  & 3.268 \\
  1126&  J2148.0+0657  & 327.03 &   6.93  & PKS 2145+06&               8.09 & BZQ & $17.37_{-3.50}^{+2.63}$ & $1.90_{-0.37}^{+0.43}$  & 0.990 \\
  1605&  J2148.4-7557  & 327.10 & -75.57  & [HB89] 2142-758&          6.04 & BZQ & $14.59_{-2.23}^{+2.78}$ & $1.41_{-0.47}^{+0.48}$  & 1.139 \\
  1129&  J2152.0-3030  & 327.98 & -30.46  & PKS 2149-306&              33.32 & BZQ & $89.3_{-4.13}^{+3.09}$ & $1.61_{-0.09}^{+0.08}$  & 2.345 \\
  1608&  J2157.4-3316  & 329.35 & -33.26  & SWIFT J2157.4-3316&       5.86 & BZQ & $15.12_{-3.53}^{+4.08}$ & $1.63_{-0.43}^{+0.47}$  & 1.671 \\
  1136&  J2202.8+4218  & 330.68 &  42.26  & BLLac&                    16.87 & BZB & $34.91_{-3.25}^{+2.06}$ & $1.76_{-0.17}^{+0.18}$ & 0.068 \\
  1137&  J2203.0+3146  & 330.75 &  31.75  & 4C+31.63&                 8.16 & BZQ & $15.56_{-3.08}^{+1.97}$ & $1.92_{-0.39}^{+0.44}$  & 0.295 \\
  1143&  J2211.7+1843  & 332.95 &  18.70  & IIZw171&                  7.79 & BZQ & $15.72_{-3.33}^{+2.71}$ & $1.88_{-0.37}^{+0.39}$  & 0.070 \\
  1149&  J2219.7+2614  & 334.96 &  26.25  & 2MASX J22194971+2613277&   8.62 & BZQ & $15.13_{-2.80}^{+3.07}$ & $2.10_{-0.35}^{+0.41}$  & 0.085 \\
  1154&  J2229.7-0831  & 337.47 &  -8.49  & PKS 2227-088&              6.0 & BZQ & $17.08_{-3.17}^{+3.68}$ & $1.46_{-0.51}^{+0.54}$   & 1.559 \\
  1155&  J2232.5+1141  & 338.17 &  11.71  & [HB89] 2230+114&           10.78 & BZQ & $30.05_{-3.14}^{+3.36}$ & $1.49_{-0.25}^{+0.26}$ & 1.037 \\
  1616&  J2233.9+1007  & 338.48 &  10.16  & MG1 J223400+1008&         6.28 & BZQ & $16.29_{-2.81}^{+3.58}$ & $1.65_{-0.43}^{0.48}$   & 1.854 \\
  1620&  J2246.7-5208  & 341.66 & -52.11  & RBS 1895&                 5.16 & BZG & $6.97_{-1.57}^{+1.76}$ & $2.51_{0.70}^{+1.07}$    & 0.194 \\
  1170&  J2251.8-3210  & 342.94 & -32.09  & 1RXS J225146.9-320614&     7.27 & BZB & $13.98_{-1.91}^{+2.47}$ & $2.01_{-0.43}^{+0.50}$  & 0.246 \\
  1169&  J2251.9+2215  & 342.95 &  22.30  & MG3 J225155+2217&          5.89 & BZQ & $9.61_{-1.61}^{+2.07}$ & $2.36_{-0.50}^{+0.67}$   & 3.668 \\
  1171&  J2253.9+1608  & 343.48 &  16.14  & 3C 454.3&                  61.98 & BZQ & $158.36_{-2.78}^{+2.80}$ & $1.50\pm0.05$         & 0.859 \\
  1196&  J2327.4+1525  & 351.80 &  15.42  & 2MASX J23272195+1524375&   7.4 & BZQ & $10.77_{-2.46}^{+1.85}$ & $2.58_{-0.55}^{+0.76}$   & 0.045 \\
  1197&  J2327.5+0938  & 351.88 &   9.66  & PKS 2325+093&              10.39 & BZQ & $29.73_{-3.85}^{+2.85}$ & $1.40_{-0.28}^{+0.29}$ & 1.843 \\
  1200&  J2333.9-2342  & 353.47 & -23.69  & PKS 2331-240&              6.2 & BZQ & $16.1_{-3.39}^{+3.83}$ & $1.40_{-0.48}^{+0.50}$    & 0.047 \\
  1209&  J2359.0-3038  & 359.77 & -30.57  & H2356-309&                10.06 & BZB & $14.88_{-1.85}^{+1.97}$ & $2.28_{-0.37}^{+0.44}$ & 0.165 \\                        
\hline

\multicolumn{10}{l}{\begin{minipage}{\textwidth}
 \tablenotetext{a}{BAT number as provided in the BAT 105 catalog \citep{BAT_105}.}
 \tablenotetext{b}{BAT name.}
 \tablenotetext{c}{BAT coordinates.}
 \tablenotetext{d}{Updated associated counterparts from the BASS spectroscopic campaign.}
  \tablenotetext{e}{Signal to noise ratio of BAT detection}.
 \tablenotetext{f}{Updated source type from the BASS DR2 spectroscopic classification.}
 \tablenotetext{g}{Source $14-195\,\rm keV$ flux and associated 90\% error as listed in the BAT 105 catalog.}
 \tablenotetext{h}{The BAT spectral index, computed from a power-law ﬁt to the eight-band BAT data and reported the BAT 105 catalog.}
 \tablenotetext{i}{Redshift as reported in the BASS DR2 catalog \citep{Kossa_2022,Kossb_2022}}
\end{minipage}
}
\end{longtable}

\endgroup

\bibliographystyle{aasjournal}

\bibliography{bat}

\end{document}